\documentclass[12pt,epsf]{article}

\begin{document}
\baselineskip=0.7cm
\newcommand{\EQ}{\begin{equation}}
\newcommand{\EN}{\end{equation}}
\newcommand{\EQA}{\begin{eqnarray}}
\newcommand{\EQN}{\end{eqnarray}}
\newcommand{\EQAN}{\begin{eqnarray*}}
\newcommand{\EQNN}{\end{eqnarray*}}
\newcommand{\e}{{\rm e}}
\newcommand{\Sp}{{\rm Sp}}
\renewcommand{\theequation}{\arabic{section}.\arabic{equation}}
\newcommand{\Tr}{{\rm Tr}}
\newcommand{\lpartial}{\buildrel \leftarrow \over \partial}
\newcommand{\rpartial}{\buildrel \rightarrow \over 
\partial}
\renewcommand{\thesection}{\arabic{section}.}
\renewcommand{\thesubsection}{\arabic{section}.\arabic{subsection}}
\makeatletter
\def\section{\@startsection{section}{1}{\z@}{-3.5ex plus -1ex minus 
 -.2ex}{2.3ex plus .2ex}{\large}} 
\def\subsection{\@startsection{subsection}{2}{\z@}{-3.25ex plus -1ex minus 
 -.2ex}{1.5ex plus .2ex}{\normalsize\it}}
\makeatother
\makeatletter
\def\lesim{\mathrel{\mathpalette\gl@align<}}
\def\gtsim{\mathrel{\mathpalette\gl@align>}}
\def\gl@align#1#2{\lower.7ex\vbox{\baselineskip\z@skip\lineskip.2ex%
  \ialign{$\m@th#1\hfil##\hfil$\crcr#2\crcr\sim\crcr}}}
\makeatother

\newcommand{\Nt}{\widetilde{N}}
\newcommand{\bra}[1]{\langle #1\vert}
\newcommand{\ket}[1]{\vert #1\rangle}
\newcommand{\braket}[2]{\langle #1\vert #2\rangle}
\newcommand{\bbbk}[4]{{}_1\langle #1|{}_2\langle #2|
                      {}_3\langle #3|#4\rangle_{123}}
\newcommand{\C}{C_{\rm vac}}
\newcommand{\HI}{H_{\rm 3}}
\newcommand{\E}{E}
\renewcommand{\P}{{\bf P}}
\newcommand{\PSV}{P_{SV{\rm -}PS}}

\newcommand{\HSV}{\ket{\HI}_{SV{\rm -}PS}}
\newcommand{\HD}{\ket{\HI}_{DPPRT}}
\newcommand{\HCK}{\ket{\HI}_{CK}}

\newcommand{\vac}{{\rm vac}}
\renewcommand{\v}{{\rm v}}

\renewcommand{\a}{\alpha}
\renewcommand{\b}{\beta}
\newcommand{\inI}{2}
\newcommand{\inII}{3}
\newcommand{\out}{1}
\newcommand{\ainI}{\a_{(\inI)}}
\newcommand{\ainII}{\a_{(\inII)}}
\newcommand{\aout}{\a_{(\out)}}
\newcommand{\ar}{\a_{(r)}}
\newcommand{\as}{\a_{(s)}}
\newcommand{\asp}{\a_{(s')}}
\newcommand{\arp}{\a_{(r')}}
\newcommand{\atp}{\a_{(t')}}
\newcommand{\XI}{X_{\rm I}}
\newcommand{\XII}{X_{\rm II}}
\newcommand{\YI}{Y}
\newcommand{\yI}{Y_{\rm I}^{(1)}}
\newcommand{\YII}{Z}
\newcommand{\aDim}{\Delta^{\langle1\rangle}}
\newcommand{\Ctree}{C^{\langle0\rangle}}
\def\thefootnote{\fnsymbol{footnote}}

\begin{flushright}
hep-th/0406225\\
UT-KOMABA/04-8\\
June 2004
\end{flushright}
\vspace{0.3cm}
\begin{center}
\Large
Resolving the Holography in the Plane-Wave Limit \\
of AdS/CFT Correspondence

\vspace{0.7cm}

\normalsize
 \vspace{0.4cm}
Suguru {\sc Dobashi}
\footnote{e-mail address:\ \ {\tt doba@hep1.c.u-tokyo.ac.jp}}
and 
Tamiaki {\sc  Yoneya}
\footnote{
e-mail address:\ \ {\tt tam@hep1.c.u-tokyo.ac.jp}}

\vspace{0.3cm}

{\it Institute of Physics, University of Tokyo\\
Komaba, Meguro-ku, Tokyo 153-8902}

\vspace{1cm}
Abstract
\end{center}
\vspace{0.4cm}
The issue of holographic mapping between bulk and 
boundary 
in the plane-wave limit of AdS/SYM correspondence 
is reexamined from the viewpoint of 
correlation functions. We first study  the limit of large angular 
momentum for  the so-called GKP-W relation in supergeravity 
approximation, 
connecting directly the effective action in the bulk and the generating 
functional of correlation functions on the boundary. 
The spacetime tunneling picture which has been proposed 
in our previous works naturally emerges.  This gives not only 
a justification of our 
previous proposal, with some important refinements,  on the mapping 
between bulk effective interaction and the 
CFT coefficients on the boundary in the plane-wave limit, but also 
implies various insights on the interpretation 
of holography in the plane-wave limit. Based on this 
result, we construct a new `holographic' string 
field theory. We confirm for several nontrivial 
examples that this gives the 
CFT coefficients derived by perturbation theory 
on the gauge-theory side. 
 Our results are useful for understanding how 
apparently different duality maps proposed from different 
standpoints are 
consistent with each other and with our definite spacetime 
picture for the AdS holography in the plane-wave limit.

\newpage
\section{Introduction}
An impressive amount of computations have been done 
following the  BMN conjecture \cite{bmn} as to the 
identification of stringy operators in AdS/CFT 
correspondence.  In spite of all  those important  works of two years, 
however, it seems that 
the question of  holographic correspondence 
of correlation functions for the BMN operators 
still has not been appropriately understood. 
In the case of the original AdS$_5$/SYM$_4$ correspondence, 
the relation between the bulk fields $\{ \phi_i(z, \vec{x}) \}$ and the 
gauge-invariant operators $\{O_i(\vec{x})\}$ of 4D Yang-Mills theory 
has been concretely 
formulated as the famous GKP-W relation \cite{gkpw} 
\EQ
Z[\phi] \sim \langle \exp \Big[\int d^4 x \sum_i\phi_i(\vec{x})O_i(\vec{x})\Big]
\rangle , 
\EN
which connects the boundary values $ \lim_{z\rightarrow 0} z^{\Delta_i-4} \phi_i(z, \vec{x})=\phi_i(\vec{x}) $ of the bulk fields to the source fields coupled with $\{ {\cal O}_i(\vec{x})\}$ at the 
conformal boundary of AdS spacetime. 
If one naively followed the Penrose limit in the bulk 
of AdS spacetime in obtaining a plane-wave 
 approximation, one would end up in a puzzling 
situation that the plane-wave geometry corresponding to the 
large angular momentum along a direction 
of $S^5$ cannot be 
related to the conformal boundary, since 
the null trajectory adopted by the BMN proposal never reaches the conformal boundary. 
Because of this difficulty, some different 
ways of comparing both sides without relying directly 
upon the GKP-W relation have been discussed in the literature \cite{reviews}.  
However, lacking for  more direct links to 
physical observables,  
these proposals seem to be still regarded as 
phenomenological data towards a better understanding of 
holography.  
It is very important to resolve the issue of holographic 
correspondence from the viewpoint of correlation 
functions, since it would be a crucial 
basis in addressing  physically more  
relevant  questions related to the duality 
between closed string theories and gauge theories.

In previous works, we have presented basic ideas on a possible 
reconciliation of the BMN proposal with the GKP-W relation. 
In ref. \cite{dsy}\cite{yy}, we proposed to interpret the 
GKP-W relation in the plane-wave limit as a 
consequence of the tunneling propagation of the BMN states of 
closed strings  from boundary to boundary. 
The motivation for our proposal was puzzles 
which arises in connection of holography when we adopt seemingly familiar 
premises in the literature, especially, the identification of the 
global AdS time (or light-cone time) with the time of 
radial quantization on the CFT side.\footnote{
In our opinion, the origin of a 
 puzzle discussed in \cite{polman} 
is also related to this identification. 
For a list of other approaches on the plane-wave holography, we would like to refer the reader to several review articles cited in \cite{reviews}. 
}
We have argued that these puzzles are resolved, 
by considering the tunneling trajectory connecting 
AdS boundary to AdS boundary.  Since the role of 
time parameter is played by the affine 
parameter along the tunneling trajectory 
which is orthogonal to the conformal boundary, 
we cannot identify 
it with radial time directly, and consequently all the 
puzzles are naturally resolved. 
In subsequent works  \cite{asy},  
our ideas have been extended successfully 
to a more general nonconformal case of D$p$-brane 
backgrounds, by deriving 
the generalized correspondence \cite{sy} obtained 
previously 
between nonconformal 
SYM theories and D$p$-brane backgrounds. 

However,  our original argument
in \cite{dsy} has not yet been completely satisfactory,  since 
it involves some ambiguities with respect to normalization and 
short-distance cutoff when we discuss 3-point and higher 
correlation functions. 
One of the purposes of the present work 
is to reexamine our basic ideas from an equivalent but a more systematic 
standpoint by studying the plane-wave limit 
directly on the GKP-W relation, and to 
strengthen our picture by presenting further supports 
and extensions. 
In our first work \cite{dsy}, 
we have not started from  
the GKP-W relation. Instead, we have treated the 
bulk field equation in the WKB approximation 
and then proposed a natural ansatz for relating 
correlation functions and the {\it Euclidean} S-matrix 
in the same spirit as the GKP-W relation.  
In the present work,  we study 
the limit of large R-charge ($J$) for the GKP-W relation 
directly and confirm that our 
original picture emerges automatically within the 
supergravity approximation.  Then, 
on the basis of the known relation \cite{lmrs} 
for chiral operators between supergravity and 
SYM gauge theory, we  establish a definite 
relation between 3-point correlators and the bulk 
effective interactions which are consistent with our 
original picture, including precise normalization and  cutoff.   

It turns out that the resultant 
supergravity effective theory in the plane-wave limit 
cannot be obtained from any versions of previously known string field theories. Our result can be adopted as a strong constraint in constructing a string field theory describing higher stringy modes in accordance with our picture.  
We propose a new `holographic' 
string field theory, which reduces to the derived 
effective action when restricted in the supergravity sector,
  and simultaneously gives the correct 3-point 
correlation functions on the SYM side derived by 
perturbation theory, via our holographic mapping. 
As a byproduct, we also give some 
clarifications on the 
relation of our picture with  other approaches    
which have been discussed in some of recent works 
for mapping the known 
versions of string field theories to gauge theory. 
We believe that our results not only resolve holography 
in the plane-wave limit,  but also lay a foundation 
for understanding relation among different proposals and for 
investigating further extensions, 
on the basis of holography for correlation functions. 
In particular, we give a definite prediction for impurity non-preserving cases which have not 
been treated in the literature before.

The present paper is organized as follows. 
In section 2, we analyze the large $J$-limit 
directly on the usual diagrammatic rules (`Witten diagrams') 
for computing 3-point functions from the 
viewpoint of bulk theory. We demonstrate how our tunneling 
picture arises in the WKB approximation and establish the validity 
of this picture by comparing with  exact computations. 
In section 3,  we derive directly the effective action in the 
plane-wave limit  from the bulk supergravity effective  
action given previously in ref. \cite{lmrs}. 
We then formulate the holographic relation 
which should be valid for (non BPS) stringy operators. 
In section 4, we construct the  `holographic' 
string field theory which is 
consistent with our effective action for supergravity 
sector. 
In section 5, we clarify the relation of our results with other 
approaches. In section 6, we confirm our general 
discussion by explicit examples. 
 We conclude the paper in section 7 by giving 
further remarks.  Two appendices A and B 
are devoted to some details of calculation and  to a summary 
of the properties of string-interaction vertices 
of string field theory, respectively.

%\newpage
\section{The direct large $J$-limit of GKP-W relation}
\setcounter{equation}{0}

Let us start from briefly recalling 
the standard 
perturbative computation of correlation functions 
from bulk supergravity theory. For simplicity, suppose that 
the bulk theory is effectively described by  scalar 
fields  $\phi_i(x)$ in the AdS$_5$ background with action, 
\EQ
S_{\phi}=\int d^5x \sqrt{g}\, [\sum_{i=1}^3{1\over 2}(\partial \phi_i)^2 
+ \sum_{i=1}^3{m_i^2\over 2}\phi_i^2 + g
\phi_1\phi_2\phi_3
] , \quad m_i^2=J_i(J_i+4)/R^2. 
\label{scalaraction}
\EN
We assume the Euclidean AdS metric ($R^4 =4\pi g_s N (\alpha')^2
$), 
\EQ
ds^2={R^2 \over z^2}(dz^2 + d\vec{x}^2) 
\EN
using the Poincar\'{e} coordinate. According to the GKP-W 
dictionary,  
  a 3-point correlation function 
of three operators ${\cal O}_i(\vec{x}_i)$ (conformal 
dimension = $\Delta_i=J_i+\tilde{k}_i=k_i$) 
corresponding to the bulk fields $\phi_i$ 
 is given, up to an overall normalization factor and 
to the lowest order with respect to the 
coupling constant $g$, as 
\EQ
\langle 
{\cal O}_1(\vec{x}_1) {\cal O}_2(\vec{x}_2)
{\cal O}_3(\vec{x}_3) \rangle
=\int {d^4\vec{x} dz 
\over z^5} \, 
K_{\Delta_1}(z, \vec{x};\vec{x}_1)
K_{\Delta_2}(z, \vec{x};\vec{x}_2)
K_{\Delta_3}(z, \vec{x};\vec{x}_3) 
\label{3-point}
\EN 
where  the bulk-to-boundary propagator 
\EQ
K_{\Delta}(z, \vec{x};\vec{y})
={\Gamma(\Delta)\over 
\pi^2 \Gamma(\Delta-2)}
\Big(
{z \over z^2+ (\vec{x}-\vec{y})^2}
\Big)^{\Delta}
\label{bbpropagator}
\EN satisfies 
\EQ
\Big[z^5{\partial \over \partial z}\Big(z^{-3}{\partial \over \partial z}\Big)
+z^2 {\partial^2 \over \partial \vec{x}^2}-m^2 \Big]
K_{\Delta}(z, \vec{x};\vec{y})=0, \quad \mbox{for} \, \, 
z>0
\EN
and 
\EQ
\lim_{z\rightarrow 0} z^{\Delta -4}K_{\Delta}(z, \vec{x};\vec{y})
=\delta(\vec{x}-\vec{y}). 
\label{bcond}
\EN 
The PP-wave limit amounts to taking the 
limit where $J_i, \, R \rightarrow \infty$ with $J_i\alpha'/R^2 $ being kept fixed and $\tilde{k}_i \sim 
O(1)$.  The angular momentum 
which comes from an SO(2) part of 
the SO(6) R-symmetry must be assumed to be 
conserved, say, 
$
J_1=J_2 +J_3. 
$
Obviously, since $\Delta_i \sim J_i \rightarrow \infty$,  
the integral in the expression eq. (\ref{3-point}) can be studied by saddle-point methods.\footnote{
A preliminary discussion on the approach  
of the present paper has been 
given by one of the present authors in a talk at the 
Strings 2003 conference \cite{y-st04}. 
}
For our purpose it is useful to do a 
warm-up in the case of two-point functions. 
In what follows until stated otherwise explicitly, 
it is convenient to adopt the unit such that the 
AdS$_5$ ($S^5$) radius to be one, $R=1$, since  
$R$ is the only length scale characterizing this system 
in the supergravity limit $\alpha'\rightarrow 0$ with 
fixed $R$. 

\subsection{Two-point functions}
Consider a 2-point function of the following form 
\EQ
G_2(\vec{x}_1, \vec{x}_2) \equiv \int {d^4\vec{x} dz 
\over z^5} \, 
K_{\Delta}(z, \vec{x};\vec{x}_1)
K_{\Delta}(z, \vec{x};\vec{x}_2)z^{\epsilon}, 
\label{defG2}
\EN
where $\epsilon \rightarrow 0+$ is a 
parameter for regularization. 
The saddle-point equations are 
\EQ
{\partial \over \partial z}
\Big(
\ln [{z \over z^2 + (\vec{x}-\vec{x}_1)^2 }]
+
\ln [{z \over z^2 + (\vec{x}-\vec{x}_2)^2 }]
\Big) =0, 
\EN
\EQ
{\partial \over \partial \vec{x}}
\Big(
\ln [{z \over z^2 + (\vec{x}-\vec{x}_1)^2 }]
+
\ln [{z \over z^2 + (\vec{x}-\vec{x}_2)^2 }]
\Big) =0.
\EN
The general solution is 
\EQ
\vec{x}_0={1\over 2}(\vec{x}_1+\vec{x}_2)
-{1\over 2}(\vec{x}_1-\vec{x}_2) \tanh \tau, \quad 
z_0={|\vec{x}_1-\vec{x}_2|\over 2\cosh\tau}
\EN
with $\tau$ being an undermined integration constant.  
Thus the integral can be approximated as a one dimensional 
integral over the `collective' coordinate $\tau$. 
In conformity with our previous works, 
the solution  describes a tunneling process 
from one boundary point $\vec{x}_1$ at $\tau \sim -\infty$ 
to another  boundary point 
$\vec{x}_2$ at $\tau\sim +\infty$. 
Thus we have naturally arrived at the same 
picture for the PP-wave holography as we have proposed 
in previous works. 

Following the standard method of semi-classical 
path-integrals,  the integration measure 
in  (\ref{3-point}) is replaced by 
\EQ
{dz d^4\vec{x} \over z^5} \Rightarrow d\tau
d\tilde{z}d^3\vec{\tilde{x}}_{\perp}J(\tau)
\label{measure}
\EN
for the collective coordinate $\tau$ and the fluctuating coordinates  $\tilde{z}, \, \vec{\tilde{x}}$, which 
are defined by the following 
shift of the bulk coordinates, 
\EQ
z=z_0(\tau) + \tilde{z}, \quad 
\vec{x}=\vec{x}_0(\tau) + \vec{\tilde{x}}
\EN
with the orthogonality constraint
\EQ
{dz_0\over d\tau}\delta\tilde{z}
+{d\vec{x}_0\over d\tau}\cdot \delta\vec{\tilde{x}}=0.
\label{constraint}
\EN
The effective metric for the fluctuations is 
found to be, to the lowest 
nontrivial order in the 
fluctuations,
\EQ
{1\over z^2}\Big((dz_0 + d\delta \tilde{z})^2 
+(d\vec{x}_0+ d\vec{\delta \tilde{x}})^2\Big)
\Rightarrow
(d\tau)^2
+ {4 \cosh^2\tau\over |\vec{x}_1-\vec{x}_2|^2}
\Big(\cosh^2\tau(d\delta\tilde{z})^2 
+ (d\delta\vec{\tilde{x}}_{\perp})^2\Big)
\label{effmetric}
\EN
where we have used the solution for the 
constraint (\ref{constraint}), 
\EQ
\delta \vec{\tilde{x}}
=-\vec{n} \sinh \tau \delta{\tilde{z}}
+ \delta \vec{\tilde{x}}_{\perp}, \quad \vec{n}\cdot
\delta \vec{\tilde{x}}_{\perp}=0
\EN
with $\vec{n}$ being the unit vector along the 
direction of the vector $\vec{x}_1-\vec{x}_2$, which  
connects two points $\vec{x}_1, \vec{x}_2$ on the boundary. 
Thus, the Jacobian 
is given as 
\EQ
J(\tau)=\sqrt{\Big( {4 \cosh^2\tau\over |\vec{x}_1-\vec{x}_2|^2}
\Big)^4 \cosh^2 \tau} . 
\EN

On the other hand, the effective second-order action 
for the saddle-point integral 
is 
\EQ
S_{eff}^{(2)}
={4\Delta \over |\vec{x}_1-\vec{x}_2|^2}
\Big(
\cosh^4 \tau \, \tilde{z}^2 + \cosh^2 \tau \, (\vec{\tilde{x}}_{\perp})^2
\Big) + \epsilon\mbox{-dependent factor}.
\EN
The classical part of the action gives only a 
factor which is independent of the collective 
coordinate $\tau$ and has the 
correct dependence on the distance 
of boundary points, as 
can be checked by using 
\EQ
K_{\Delta}(z_0(\tau), \vec{x}_0(\tau); 
\vec{x}_{1,2})
={\Gamma(\Delta)\over 
\pi^2 \Gamma(\Delta-2)}|\vec{x}_1-\vec{x}_2|^{-\Delta}
\e^{\pm \Delta \tau}
\label{Kalongtraj}
\EN
where the sign on the exponentials depends on 
the points $\vec{x}_1$ or $\vec{x}_2$, respectively. 
It is now evident that the Jacobian factor 
is canceled by the integrations  
over the fluctuating coordinates, up to a $\Delta$-dependent 
proportionality constant. 
Consequently, the two-point function 
in the large $\Delta$ limit is given simply by 
\EQ
 G_2(\vec{x}_1, \vec{x}_2)\Rightarrow 
{\Delta^2\over \pi^2}|\vec{x}_1-\vec{x}_2|^{-2(\Delta-\epsilon)}
\int_{-\infty}^{+\infty} d\tau \, (2\cosh \tau)^{-\epsilon} ,
\EN  
where the $\epsilon$ dependent contributions 
come from the factor $z^{\epsilon}$ in the 
defining expression (\ref{defG2}). 
Thus, in the limit $\epsilon\rightarrow 0+$, 
we reproduce the correct behavior for two-point 
correlators for conformal operators, up to the 
pole singularity
\EQ
\int_{-\infty}^{+\infty} d\tau \, (2\cosh \tau)^{-\epsilon}
\sim {2 \over \epsilon} .
\EN
We can compare this result with that of exact integration: 
\EQ
G_2(\vec{x}_1, \vec{x}_2)=\Big({\Gamma(\Delta)\over 
\pi^2 \Gamma(\Delta-2)}\Big)^2\times 
{\pi^2 \over 2}
{\Gamma(\Delta -2)
\Gamma(\epsilon/2)^2
\over 
\Gamma(\Delta) \Gamma(\epsilon)}
|\vec{x}_1-\vec{x}_2|^{-2(\Delta-\epsilon)}
\sim {2\Delta^2 \over \pi^2}{1\over \epsilon}|\vec{x}_1-\vec{x}_2|^{-2\Delta}
\EN
Here we have used the general formula for this type of 
integral \cite{dfmmr},
\[
\int_0^{\infty} dz \int d^D\vec{x}
{z^a 
\over (z^2 + (\vec{x}-\vec{x}_1)^2)^b
(z^2 + (\vec{x}-\vec{x}_2)^2)^c}=
|\vec{x}_1-\vec{x}_2|^{1+a + D -2b -2c}
I(a, b, c, D), 
\]
\EQ
I(a, b, c, D)\equiv 
{\pi^{D/2}\over 2}
{\Gamma({a\over 2}+ {1\over 2})
\Gamma(b+c -{D\over 2} -{a\over 2} -{1\over 2})
\Gamma({1\over 2} +{a\over 2} +{D\over 2} -b)
\Gamma({1\over 2} +{a\over 2} +{D\over 2} -c)
\over 
\Gamma(b)\Gamma(c)\Gamma(1+a+D -b -c)}. 
\EN

In reality,  for the 
effective theory described by (\ref{scalaraction}), 
it is more appropriate to consider 
\EQ
\int {d^4\vec{x} dz 
\over z^{5-\epsilon}} \, 
\Big[g^{\mu\nu}(z,\vec{x}) \partial_{\mu} K_{\Delta}(z, \vec{x};\vec{x}_1)\partial_{\nu}
K_{\Delta}(z, \vec{x};\vec{x}_2)
+m^2 K_{\Delta}(z, \vec{x};\vec{x}_1)
K_{\Delta}(z, \vec{x};\vec{x}_2)\Big]
\EN
than (\ref{defG2}). 
It is easy to repeat the above 
calculation for this case. Only difference  of the final result from the case of $G_2$  is the multiplication of 
$(m^2 -\Delta^2)/m^2 \sim -4/\Delta$. 

\subsection{3-point functions} 
Armed by this exercise, we now go back to the 3-point 
function (\ref{3-point}). The saddle-point equations  
are 
\EQ
\sum_{i=1}^3 \Delta_i\Big(
{1\over z} - {2z \over z^2 + (\vec{x}_i-\vec{x})^2}
\Big)=0 ,
\EN
\EQ
\sum_{i=1}^3 \Delta_i{\vec{x}_i -\vec{x} 
\over z^2 + (\vec{x}_i-\vec{x})^2}=0. 
\EN
It is easy to convince oneself
 that, for generic configurations of three 
boundary points, there is no solution to these 
equations. However, if we take a limit where 
two of the boundary points, say, $\vec{x}_2$ and 
$\vec{x}_3$ approach sufficiently to 
one point $\vec{x}_c=(\vec{x}_2 + \vec{x}_3)/2$, 
the same trajectory 
connecting $\vec{x}_1$ and $\vec{x}_c$ 
as we have discussed in the previous subsection 
can be regarded as an approximate solution.  
Therefore, let us try to reduce the integral to the one 
along the following trajectory, 
\EQ
\vec{x}_0(\tau) ={1\over 2}(\vec{x}_1+\vec{x}_c)
-{1\over 2}(\vec{x}_1-\vec{x}_c) \tanh \tau, \quad 
z_0(\tau) ={|\vec{x}_1-\vec{x}_c| \over 2\cosh\tau}
\EN
with the fluctuations, $\tilde{\vec{x}}=
\vec{x}-\vec{x}_0$ and $\tilde{z}=
z-z_0$. For our purpose it is sufficient to 
evaluate the integral to the leading order 
in the short-distance limit, 
\[
\vec{\delta} \equiv \vec{x}_2-\vec{x}_c= -
\vec{x}_3+\vec{x}_c\rightarrow 0 .
\]
In order to avoid unnecessary complications,   
we assume that all three points $\vec{x}_1, \vec{x}_2, 
\vec{x}_3$ are along a single line on the boundary. 

The effective action for this computation 
\EQ
S_{eff}=\sum_{i=1}^3\Delta_i \ln 
{z^2 + (\vec{x}_i-\vec{x})^2 \over z} 
\EN
is rewritten as 
\[
{\Delta_1+ \Delta_2 +\Delta_3 \over 2}\Big( \ln {z^2 + (\vec{x}_1-\vec{x})^2 \over z}
+{z^2 + (\vec{x}_c-\vec{x})^2 \over z}\Big)
\]
\[
+\, {\Delta_1-\Delta_2 -\Delta_3\over 2}
\Big(\ln 
{z^2 + (\vec{x}_1-\vec{x})^2 \over z}
-\ln 
{z^2 + (\vec{x}_c-\vec{x})^2 \over z}
\Big)
\]
\EQ
+\Delta_2 \ln \Big(1+ {2\vec{\delta}\cdot(\vec{x}_c-\vec{x})+\vec{\delta}^2
\over z^2 + (\vec{x}_c-\vec{x})^2 }\Big)
+\Delta_3 \ln \Big(1+ {-2\vec{\delta}\cdot(\vec{x}_c-\vec{x})+\vec{\delta}^2
\over z^2 + (\vec{x}_c-\vec{x})^2 }\Big).
\label{3paction}
\EN
The first line of (\ref{3paction}) can 
be treated in exactly the same way as for the 2-point 
case, by replacing $\Delta$ by 
$(\Delta_1+ \Delta_2 + \Delta_3)/2  \, (\sim J_1 
\rightarrow \infty) $. 
Next, since 
\EQ
\Delta_1-\Delta_2-\Delta_3 \sim O(1), 
\EN
the second line can be approximated by 
its value on the classical trajectory
\EQ
{\Delta_1-\Delta_2 -\Delta_3\over 2}
\Big(\ln 
{z_0^2 + (\vec{x}_1-\vec{x}_0)^2 \over z_0}
-\ln 
{z_0^2 + (\vec{x}_2-\vec{x}_0)^2 \over z_0}\Big)
=(\Delta_1-\Delta_2 -\Delta_3)\tau
\EN 
 
The third line gives, to the zero-th order with respect to the 
fluctuations, 
\EQ
-2(\Delta_2-\Delta_3){\delta \over |\vec{x}_1-\vec{x}_c|}
-(\Delta_2+ \Delta_3){\delta^2 \over |\vec{x}_1-\vec{x}_c|^2} 
+
(\Delta_2+ \Delta_3){\delta^2 \over |\vec{x}_1-\vec{x}_c|^2}
\e^{2\tau} + O(\delta^3). 
\label{zerofluc}
\EN
Here we have kept the third term which is of second order 
in $\delta\rightarrow 0$, since it shows that for sufficiently 
large $\tau$ there is a natural 
cutoff for the range of the affine parameter
 for arbitrary small  $\delta=|\vec{\delta}|  \sim 0$. 
The other terms which are independent of $\tau$ can be 
ignored in the limit $\delta\rightarrow 0$. 

As for the contribution of fluctuating coordinates in the 
third line, 
it is easy to convince ourselves that 
keeping only the first order term with respect both 
to the fluctuations and to the short-distance cutoff $\delta$ 
is sufficient for 
our purpose. 
The relevant terms are arranged as 
\[
-4(\Delta_2-\Delta_3)\Big[{\vec{\delta}\cdot (\vec{x}_c-\tilde{\vec{x}})
(\tilde{z}\cdot z_0 -\tilde{\vec{x}}\cdot 
(\vec{x}_c-\vec{x}_0))
\over 
(z_0^2 + (\vec{x}_c -\vec{x}_0)^2)^2}
+{1\over 2}{\vec{\delta}\cdot\tilde{\vec{x}} \over 
 z_0^2 + (\vec{x}_c -\vec{x}_0)^2}
\Big]
\] 
\EQ
=\tilde{z} {4\delta(\Delta_2-\Delta_3)
\over  |\vec{x}_1-\vec{x}_c|^2}\e^\tau \cosh^2 \tau, 
\EN
which involves only the $\tilde{z}$-fluctuation. 
Combining this with the relevant part of the Gaussian 
factor coming from the first line, 
the integral with respect to
 $\tilde{z}$  is 
\[
\int d\tilde{z} \, [\mbox{measure}]\, 
\exp \Big[
-\tilde{z}^2{4J_1 \cosh^4 \tau \over |\vec{x}_1-\vec{x}_2|^2 }
-\tilde{z}{4\delta(\Delta_2-\Delta_3)
\over  |\vec{x}_1-\vec{x}_c|^2}\e^\tau \cosh^2 \tau
\Big]
\]
\EQ=
\exp\Big[
{\delta^2 (\Delta_2-\Delta_3)^2 \e^{2\tau} 
\over J_1|\vec{x}_1-\vec{x}_2|^2} 
\Big] 
\times 
\int d\tilde{z} \, [\mbox{measure}]\, 
\exp \Big[
-\tilde{z}^2{4J_1 \cosh^4 \tau \over |\vec{x}_1-\vec{x}_2|^2 }
\Big]
\EN
to the present order of approximation. The prefactor 
here implies that, together with the contribution from the 3rd term 
of (\ref{zerofluc}), the total factor which is 
responsible for the cutoff at large $\tau$ region is 
\EQ
\exp\Big[
-\e^{2\tau}{\delta^2 (\Delta_2 +\Delta_3) \over 
 |\vec{x}_1-\vec{x}_2|^2}
+ \e^{2\tau} {(\Delta_2-\Delta_2)^2 \over J_1}
{\delta^2 \over  |\vec{x}_1-\vec{x}_2|^2} 
\Big]
\sim \exp\Big[-4{\delta^2 \over
 |\vec{x}_1-\vec{x}_2|^2} {J_2J_3 \over J_1} \e^{2\tau}
\Big]. 
\EN

Putting all of these results together and remembering that 
the Jacobian and the Gaussian integral 
cancel, the integral of the 3-point function 
now takes the form
\EQ
{\pi^2 \over J_1^2}
|\vec{x}_1 -\vec{x}_c|^{-(\Delta_1 + \Delta_2 
+\Delta_3)}
\int_{-\infty}^{+\infty} d\tau \, 
\e^{-(\Delta_1-\Delta_2-\Delta_3)\tau}
\exp[-{J_2J_3\over  J_1}{(2\delta)^2\over 
|\vec{x}_1-\vec{x}_c|^2}\e^{2\tau}]
\label{gammaintegral} 
\EN
in the limit of large $J_i$ and small $\delta$. 
This expression shows that the precise form of the 
cutoff mentioned above is $\tau \lesim \tau_c$ with
\EQ
e^{-\tau_c}= \sqrt{{J_2J_3 \over J_1}}{2\delta \over 
|\vec{x}_1-\vec{x}_c|}. 
\label{cutoff}
\EN
Thus,  the 3-point integral 
can be expressed by a Gamma function and leads to 
\EQ
{\pi^2 \over J_1^2} \, |\vec{x}_1-\vec{x}_c|^{-2\Delta_1}
(2\delta)^{-(\Delta_2+\Delta_3-\Delta_1)}
\Big({J_2J_3 \over J_1}\Big)^{-(\Delta_2+\Delta_3-\Delta_1)/2}
{\Gamma({\Delta_2+\Delta_3-\Delta_1\over 2}+1)
\over -\Delta_1+\Delta_2+\Delta_3} \lambda . 
\label{3psaddle}
\EN
After fixing the convention for normalization suitably, 
this give the 
correct short-distance 
limit for the 3-point correlator, from which we can 
identify the CFT coefficient by 
equating the expression with 
\EQ
\lim_{\vec{\delta}\rightarrow 0}
{C_{123} \over 
|\vec{x}_1-\vec{x}_2|^{2\alpha_3}
|\vec{x}_2-\vec{x}_3|^{2\alpha_1}
|\vec{x}_3-\vec{x}_1|^{2\alpha_2}}
\sim {C_{123}\over |\vec{x}_1-\vec{x}_c|^{2\Delta_1}
|2\vec{\delta}|^{2\alpha_1}}
\EN
where $\alpha_1=(\Delta_2+\Delta_3 -\Delta_1)/2$ etc. 
The precise normalization will be discussed in the next 
section.

Let us compare this result with the exact computation. 
The same integral formula cited before gives
\EQ
\int {d^4 \vec{x}dz \over z^5}
\prod_{i=1}^3 \Big(
{z \over z^2+ (\vec{x}-\vec{x}_i)^2}
\Big)^{\Delta_i}
={c(\Delta_1, \Delta_2, \Delta_3) 
\over |\vec{x}_1-\vec{x}_2|^{\Delta_1+\Delta_2 -\Delta_3}
|\vec{x}_2-\vec{x}_3|^{\Delta_2 + \Delta_3-\Delta_1}
|\vec{x}_3-\vec{x}_1|^{\Delta_3 + \Delta_1 -\Delta_2}}, 
\label{3pintegral}
\EN
\[
c(\Delta_1, \Delta_2, \Delta_3) ={\pi^2 \over 2}
{\Gamma({1\over 2}(\Delta_1+\Delta_2 -\Delta_3))
\Gamma({1\over 2}(\Delta_2+\Delta_3 -\Delta_1))
\Gamma({1\over 2}(\Delta_3+\Delta_1 -\Delta_2))
\over \Gamma(\Delta_1)\Gamma(\Delta_2)\Gamma(\Delta_3)}
\]
\EQ
\times \Gamma({1\over 2}(\Delta_1+\Delta_2 +\Delta_3-4)) .
\EN
By taking the large $J_i$ limit using the Stirling formula 
$\lim_{J\rightarrow \infty} \Gamma(J+\tilde{k}) \sim 
\sqrt{2\pi} \exp((J+\tilde{k} -{1\over 2})\ln J -J)$, 
we obtain 
\EQ
{\pi^2 \over J_1^2}|\vec{x}_1-\vec{x}_c|^{-2\Delta_1}
(2\delta)^{-(\Delta_2+\Delta_3-\Delta_1)}
\Big({J_2J_3\over J_1}\Big)^{-(\Delta_2+\Delta_3-\Delta_1)/2}
{\Gamma({\Delta_2+\Delta_3-\Delta_1\over 2}+1)
\over -\Delta_1+\Delta_2+\Delta_3}, 
\label{3plimit}
\EN
which exactly matches (\ref{3psaddle}).  

Thus we have confirmed that the 3-point correlators 
 in the short distance limit $\vec{\delta} \rightarrow 0$  
can be computed effectively as a process 
in the bulk occurring along a single 
tunneling trajectory connecting boundary to boundary, 
whose amplitude essentially takes the general form, 
\EQ
{\epsilon^{\Delta_1-\Delta_2-\Delta_3}
\over -\Delta_1+\Delta_2+\Delta_3}\tilde{g}, 
\label{originalform} 
\EN
\EQ
\tilde{g}\sim g \, 
\Big({J_2J_3\over J_1}\Big)^{{\Delta_1-\Delta_2-\Delta_3\over 2}}\Gamma({-\Delta_1+\Delta_2+\Delta_3\over 2}+1)
\label{3pcorrelator} , \quad 
\epsilon =2\delta , 
\label{couplingcftrelation}
\EN
with a suitable normalization convention, 
apart from  the usual spacetime dependent factor  $|\vec{x}_1-\vec{x}_c|^{-2\Delta_1}$. 

The form (\ref{originalform}) 
 is  consistent with what we have 
proposed as the general structure 
for 3-point correlators which is expected from the 
spacetime picture for holography in the PP-wave limit. 
The parameter $\epsilon \sim \e^{-\tau_c}$ is 
the cutoff related to the distance of the 
operators ${\cal O}_2(\vec{x}_2)$ and 
${\cal O}_3(\vec{x}_3)$ 
at the boundary, as (\ref{cutoff}). 
In our first original discussion, 
the emergence of this particular form 
is originated from the formal integral 
$\int d\tau \exp[(\Delta_1 -\Delta_2 -\Delta_3)\tau]$ 
which appears in perturbation theory along the 
tunneling trajectory. This is actually ill-defined as 
it stands.  
To justify the expression 
(\ref{originalform}), we had to invoke
a  wave-packet picture which is rather subtle in the 
Euclidean tunneling as is alluded to 
in \cite{dsy}. 
According to the present 
argument, the ill-defined integral must be replaced by 
(\ref{gammaintegral}) which can be defined unambiguously 
 by analytic continuation, due to 
the presence of the natural cutoff for large $\tau$ region. 
Intuitively,  the large $\tau$ cutoff corresponds to the 
limitation of the picture using the single tunneling trajectory 
in discussing 3-point correlators for small but 
nonzero $\delta$. 
Note that if we take the limit 
$\delta \rightarrow 0$ naively inside the integral, this factor 
would have simply disappeared.   In fact, the additional factor 
$\Big({J_2J_3\over J_1}\Big)^{(\Delta_1-\Delta_2-\Delta_3)/2}\Gamma({-\Delta_1+\Delta_2+\Delta_3\over 2}+1)
$ was missing in the previous discussions, owing to 
this ambiguity. 

 It is known that the coupling constants 
of supergravity modes in the bulk 
  vanish for the so-called `extremal' case 
with $\Delta_1-\Delta_2-\Delta_3=0$. However, as is 
well known, the extremal correlation functions themselves 
 do {\it not} vanish \cite{dfmmr2}. 
This remarkable fact  corresponds to 
the presence of the singular denominator in 
(\ref{originalform}).  
  We  propose that the holographic correspondence relating the bulk 3-point couplings and the CFT coefficients 
should obey the above general relation. Of course, 
the computation of this section is restricted to 
the supergravity approximation. Below, we will 
argue that this should generalize to non-BPS stringy 
modes too, with slight corrections.   
Actually, if we only consider a restricted set of the 
correlators  in which the numbers of `impurities' are  conserved,   
as has been assumed in the literature,  
the correction factor 
can be ignored  in the large $\mu$ limit 
to the leading order in making comparison with 
perturbative computation 
on the Yang-Mills side, since 
$\Delta_1-\Delta_2 -\Delta_3 \sim O(1/\mu^2)$ 
for such cases.

\section{Effective action along the 
tunneling trajectory} 
\setcounter{equation}{0}

Since we have established that the 
computation of correlation functions (or 
CFT coefficients)  can be 
reduced to the processes along a 
single tunneling trajectory connecting 
AdS boundary to AdS boundary, we can now 
proceed to derive an effective action 
along the trajectory directly from a more general 
effective action of supergravity in the bulk. 

\subsection{Bulk effective action and the CFT 
coefficients}
We study the bulk effective action for general chiral 
primary operators consisting of the 
SO(6) scalar fields $\phi_i(x)$
\EQ
O^I=\kappa {\cal C}_{i_1i_2\cdots i_k}^I
\Tr( \phi^{i_1}\phi^{i_2} \cdots \phi^{i_k}). 
\label{so6ope}
\EN 
Here, ${\cal C}$ is a totally symmetric traceless tensor whose 
contraction is 
normalized as 
\EQ
\langle {\cal C} ^{I_1}
{\cal C}^{I_2} \rangle 
={\cal C}_{i_1i_2 \cdots i_k}^{I_1}
{\cal C}^{I_2, \, i_1 i_2 \cdots i_k}
=\delta^{I_1I_2}, 
\label{so6norm}
\EN
and $\kappa=(2\pi)^k/(g_{YM}^2N)^{k/2}\sqrt{k}$ is the normalization factor such that 
the 2-point functions takes the form 
($\vec{x}_{12}=\vec{x}_1- \vec{x}_2$) ,
\EQ
\langle O^{I_1}(\vec{x}_1)O^{I_2}(\vec{x}_2)\rangle =
{\delta^{I_1I_2} \over |\vec{x}_{12}|^{2k}}.
\EN 
Since the BMN operators in the supergravity sector are essentially contained 
in this set of local operators, it should be possible to 
derive the effective action along the 
tunneling trajectory starting from an appropriate effective action for these operators. 
The BMN  supergravity 
modes are a subset of (\ref{so6ope}),  expressed by using the 
complex basis for the SO(2) directions 
$i=5, 6$ with $Z={1\over \sqrt{2}}
(\phi_5+i \phi_6)$:
\EQ
O^I=\tilde{\kappa} \tilde{{\cal C}}_{i_1i_2\cdots i_{\tilde{k}}}^I
\Tr(Z^J \phi^{i_1}\phi^{i_2} \cdots \phi^{i_{\tilde{k}}} 
+ \mbox{permutations}) , \quad k=J + \tilde{k}, 
\quad 
J\rightarrow \infty, 
\label{so4ope}
\EN
where 
$\tilde{{\cal C}}_{i_1i_2\cdots i_{\tilde{k}}}^I
$ 
are now completely symmetrized traceless SO(4) 
tensors with unit normalization 
(similarly as (\ref{so6norm})) under contraction. 
The normalization constant is related to that of the 
original set as
\EQ
\tilde{\kappa}=\kappa \pmatrix{k \cr J \cr}^{-1/2} .
\EN
Note that in the large $J$-limit the operators 
with `vector' excitations $D_i Z$ are regarded essentially as 
the derivatives of (\ref{so4ope}), 
and hence are included in this set defined by (\ref{so4ope}) if 
we suitably take into account the variation of the boundary 
coordinates $\vec{x}$. In fact, as we will argue later on 
we have to take a special care in the interpretation of 
vector excitations. 

The bulk effective action for the SO(6) operators has been derived in 
ref. \cite{lmrs}:  
\EQ
S_{bulk}={4N^2\over (2\pi)^5}\int d^5 x \sqrt{g}
\Big[\sum_{I} {1\over 2}(\nabla\phi_I)^2 
+ {1\over 2}k(k-4)\phi_I^2 
-{1\over 3}\sum_{I_1,I_2, I_3}
{{\cal G}_{I_1I_2I_3}\over 
\sqrt{A_{I_1}A_{I_2}A_{I_3}}}
\phi_{I_1}\phi_{I_2}\phi_{I_3}\Big]
\label{bulkaction}
\EN
where we used Euclidean metric and assumed 
a particular normalization for the scalar fields:
\EQ
A_I=2^{6-k}\pi^3 {k(k-1)\over (k+1)^2} .
\EN
This is the effective action in the sense that it reproduces 
correlation functions through the Witten diagrams 
as studied in the previous section. Therefore, the freedom of field redefinition on the bulk side and correspondingly the choice of the 
basis of chiral operators on the boundary side 
are both fixed already. 
Remarkably, the set of bulk fields $\{\phi_I\}$ corresponding to the set $\{O^I \}$ are effectively treated as 
scalar fields propagating in the AdS$_5$ background 
without derivative coupling. The derivative 
interactions are removed  by 
making a particular field redefinition in the derivation 
of this action.\footnote{
This does not mean that the  interactions containing derivatives along the 
$S^5$ directions are removed. 
} Actually, 
the effective action (\ref{bulkaction}) was obtained from the 
equation of motion of IIB supergravity. 
It was argued that the result must be 
correct  including the normalization of the CFT coefficients 
by relating them to 
the R-symmetry currents which have been 
known to be given exactly by the free gauge theory. 
We refer the reader to  \cite{lmrs} and the references therein 
for more details on this effective action.  Also for a 
summary on the non-renormalization properties 
of 2-and 3-point functions of 
chiral operators, see {\it e.g.} \cite{dhoker} and 
the references therein. 

The CFT coefficients $C^{I_1I_2I_3}$ of the above operators defined by 
\EQ
\langle O^{I_1}(\vec{x}_1) O^{I_2}(\vec{x}_2) 
O^{I_3}(\vec{x}_3) \rangle
={C^{I_1I_2 I_3}
\over |\vec{x}_{12}|^{2\alpha_3}|\vec{x}_{23}|^{2\alpha_1}
|\vec{x}_{31}|^{2\alpha_2}}
\label{gene3pointcorr}
\EN
with 
\EQ
\alpha_1 ={\Delta_2 + \Delta_3 -\Delta_1 \over 2} ,\quad etc
\EN
is related to the 3-point interaction 
in the effective action (\ref{bulkaction}) by 
\EQ
C^{I_1I_2 I_3}={\sqrt{k_1k_2 k_3} \over N}\langle 
{\cal C}^{I_1}{\cal C}^{I_2} {\cal C}^{I_3}\rangle
\EN
\EQ
\langle 
{\cal C}^{I_1}{\cal C}^{I_2} {\cal C}^{I_3}\rangle
=
{{\cal G}_{I_1I_2I_3}\over 
a(k_1,k_2,k_3)} 
{(k_1+1)(k_2+1)(k_3+1)
\over 
2^7 \Sigma ((\Sigma/2)^2-1)((\Sigma/2)^2-4)
\alpha_1\alpha_2\alpha_3}
\EN
\EQ
\Sigma=k_1 + k_2 + k_3
\EN
\EQ
a(k_1,k_2,k_3)={\pi^3 \over ({\Sigma\over 2}+2)! 2^{(\Sigma-2)/2}}
{k_1!k_2!k_3!\over \alpha_1!\alpha_2! \alpha_3!}
\EN
The first expression in this list is nothing but the 
result of free field computation of correlators 
on the Yang-Mills side, with $
\langle 
{\cal C}^{I_1}{\cal C}^{I_2} {\cal C}^{I_3}\rangle$ 
being all the possible contractions among 
the SO(6) indices of ${\cal C}$-tensors. 
The second expression is its representation 
in terms of the bulk quantities, in which 
the factor $a(k_1,k_2,k_3)$ arises 
as the coefficient relating 
$
\langle 
{\cal C}^{I_1}{\cal C}^{I_2} {\cal C}^{I_3}\rangle$ 
to the overlap integral of $S^5$ harmonics $Y^I$,  
which appears in reducing 10 D theory to 
5D effective theory on AdS$_5$ backgrounds,  as
\EQ
\int_{S^5} Y^{I_1}Y^{I_2}Y^{I_3}
=a(k_1, k_2, k_3)
\langle 
{\cal C}^{I_1}{\cal C}^{I_2} {\cal C}^{I_3}\rangle
\EN 
The definitions of the $S^5$ 
harmonics and their integrals are summarized in
 the Appendix A, to which we refer the reader for more 
details. The  fact that the supergravity effective action 
is exactly matched to the free-field results on the gauge-theory 
side is interpreted as a consequence of non-renormalization 
properties of 3-point functions of chiral operators. 
This is quite remarkable 
since the supergravity limit $\alpha'\rightarrow 
0$ is nothing but a strong-coupling limit 
$g_{{\rm YM}}^2N =R^4/(\alpha')^2 \rightarrow \infty$ 
on the gauge-theory side. 

Before going to our main task of this section, 
let us here check that the relation between the above 
effective action and the CFT coefficients is  
consistent with the prediction 
of previous section.  
Taking the limit $J_1(=J_2+J_3), J_2, J_3 
\rightarrow \infty$, 
we find 
\EQ
C^{I_1I_2I_3}={1\over N}{2^{J_1 +{\tilde{\Sigma}\over 2} -9}\over \pi^3}
{\sqrt{J_1J_2J_3}\over J_1^2} \Big({J_1 \over J_2 J_3}\Big)^{\alpha_1}\, 
{\alpha_1! \over \alpha_1}\,  {\cal G}_{I_1I_2 I_3} .
\label{couplingcft2}
\EN
Apart from the normalization factor which is independent 
of $\alpha_1=(\Delta_2+ \Delta_3 -\Delta_1)/2 
=(\tilde{k}_2+\tilde{k}_3-\tilde{k}_1)/2 \equiv \tilde{\alpha}_1$, 
this indeed coincides with the result obtained in the previous section, as it should. 
The difference of the normalization is 
owing to the fact that we have not yet fixed the precise normalization 
of two-point functions in the previous section.  

The CFT coefficients can be expressed in terms of the 
SO(4) basis (\ref{so4ope}) 
of the BMN operators, using the 
relation 
\EQ
\langle{\cal C}^{\overline{I}_1}{\cal C}^{I_2}
{\cal C}^{I_3}\rangle 
= \alpha_1!
\Big({J_1\over J_2 J_3}\Big)^{\tilde{\alpha}_1}
\Big({J_2\over J_1}\Big)^{\tilde{k}_2/2}
\Big({J_3 \over J_1}\Big)^{\tilde{k}_3/2}
 {\sqrt{\tilde{k}_1!\tilde{k}_2!\tilde{k}_3!}\over 
\tilde{\alpha}_1!\tilde{\alpha}_2!\tilde{\alpha}_3!}
\langle \tilde{{\cal C}}^{\overline{I}_1}
\tilde{{\cal C}}^{I_2}\tilde{{\cal C}}^{I_3}\rangle, 
\EN
which is valid in the large $J_i$ limit. 
For a derivation of this relation, see the Appendix A. 
Thus the 3-point coupling coefficient in the above 
effective action is 
\EQ
{\cal G}_{\overline{I}_1I_2 I_3}=\alpha_1 \, 
\pi^32^{-J_1 -{\tilde{\Sigma}\over 2} +9}J_1^2 
\Big({J_2\over J_1}\Big)^{\tilde{k}_2/2}
\Big({J_3 \over J_1}\Big)^{\tilde{k}_3/2}
{\sqrt{\tilde{k}_1!\tilde{k}_2!\tilde{k}_3!}\over 
\tilde{\alpha}_1!\tilde{\alpha}_2!\tilde{\alpha}_3!}
\langle \tilde{{\cal C}}^{\overline{I}_1}
\tilde{{\cal C}}^{I_2}\tilde{{\cal C}}^{I_3}\rangle
\label{scalarvertex}
\EN
in terms of the SO(4) contractions.

\subsection{Effective $0+1$ dimensional action}
With these preparations, we are now in the position of deriving the effective 
action along the tunneling trajectory starting from 
(\ref{bulkaction}).  Let us parametrize 
the coordinates near the trajectory as in the 
previous section, 
\EQ
z=z_0(\tau) + \tilde{z}, \quad \vec{x}=\vec{x}_0 (\tau)+ \tilde{\vec{x}}
\label{fluczx}
\EN
 with 
\EQ
\tilde{\vec{x}}=-\vec{n} \tilde{z} \sinh \tau
+ \tilde{\vec{x}}_{\perp}. 
\label{flucx}
\EN
Since the fluctuations around the trajectory 
can be assumed to be of order $1/\sqrt{J}$ from the 
discussion of the previous section, we  derive the effective 
metric which is correct to the second order 
in the fluctuations. 
The result is 
\[
ds^2 = (1 + \tilde{z}^2 + \tilde{\vec{x}}_{\perp}^2) 
d\tau^2 +
d\tilde{z}^2 + d\tilde{\vec{x}}_{\perp}^2 
+ \mbox{higher order terms}
\]
after rescaling as 
\[
\tilde{z} \rightarrow {|\vec{x}_1 -\vec{x}_c| 
\over 2 \cosh^2  \tau}\tilde{z}, 
\quad 
\tilde{\vec{x}}_{\perp} \rightarrow 
{|\vec{x}_1 -\vec{x}_c| 
\over 2 \cosh \tau}\tilde{\vec{x}}_{\perp}
\]
and then making the shift of the time coordinate, 
\[
\tau \rightarrow \tau + {\sinh \tau \over 2\cosh \tau}
(  \tilde{z}^2 
+   \tilde{\vec{x}}^2_{\perp}).
\]
 
To avoid notational confusions, 
we denote the rescaled fluctuations $(\tilde{\vec{x}}_{\perp}, \tilde{z})$ by a four vector 
$\vec{y}=(y_1, y_2, y_3, y_4)$. Thus the original fluctuating  four-vector $\tilde{\vec{x}}$ in the bulk is now expressed as 
\EQ
(\tilde{\vec{x}}_{\perp}, 
\vec{n}\cdot \tilde{\vec{x}})={|\vec{x}_1 -\vec{x}_c| 
\over 2 \cosh  \tilde{\tau}}(y_1, y_2, y_3, -{\sinh {\tilde{\tau}} \over \cosh \tilde{\tau}}y_4) 
\label{defy}
\EN
with the time parameter being redefined as 
\EQ
\tau = \tilde{\tau} + {\sinh \tilde{\tau} \over 2\cosh \tilde{\tau}}\vec{y}^{\,2} .
\label{taushift}
 \EN
The effective metric 
\EQ
ds^2 = (1 + \vec{y}^{\, 2}) 
d\tilde{\tau}^{2} +d\vec{y}^{\, 2}
\label{bulkmetric}
\EN
 has an SO(4) symmetry with respect to the rotation 
of $\vec{y}$. We also note that, since 
 the order of magnitude 
of $\vec{y}$ is supposed to be 
constant in $\tau$ on the basis of 
this $\tau$-independent metric, 
(\ref{defy}) implies that the 
original fluctuating coordinates  
$\tilde{\vec{x}}$ decrease 
as  $\e^{-|\tau|}$, when we approach the boundary 
$\tau \rightarrow \infty$.  

Using this metric, the quadratic terms of the bulk effective 
action (\ref{bulkaction}) are given, to the accuracy of second order in $\vec{y}$, as 
\EQ{4N^2\over (2\pi)^5}
\int d\tilde{\tau} d^4\vec{y} 
\Big[
(1- {1\over 2}y^2) 
\partial_{\tilde{\tau}}\overline{\phi_I}
 \partial_{\tilde{\tau}}\phi_I 
+\partial_y\overline{\phi_I}\partial_y\phi_I
+(1+ {1\over 2}y^2) 
(J+ \tilde{k})(J+\tilde{k}-4)
\overline{\phi_I}\phi_I
\Big]
\EN
Note that we have changed the SO(6) index to the complex SO(2)$\times$SO(4) index. 

The assumption 
that the magnitude
 of the fluctuating coordinates is of order $1/\sqrt{J}$ is 
justified  in the language of the effective action as follows. 
Redefine the fields by 
\EQ
\phi(\tau, \vec{y})
=\e^{-J\tilde{\tau}}\phi_0^{(J)}(\vec{y}) 
\psi(\tau)
\label{reduction1}, 
\EN
\EQ
\overline{\phi}(\tau,\vec{y})
=\e^{+J\tilde{\tau}}\phi_0^{(J)}(\vec{y}) 
\overline{\psi}(\tau)
\label{reduction2}, 
\EN
with 
\EQ
\phi_0^{(J)}(\vec{y})=
(J/\pi)^2 
\exp[-{1\over 2}J\vec{y}^{\,2}]
\EN
being the ground state wave function for the kinetic 
operator for the fluctuating coordinates, 
\EQ
h=-\partial_{y}^2 
+ J^2 \vec{y}^{\,2}. 
\label{harmonicop}
\EN
The hermiticity condition for our Euclidean 
field theory is \cite{dsy}
\EQ
\overline{\phi}(\tau, \vec{y})
=\phi(-\tau,  \vec{y})
\EN
which requires the choice of signs in the 
definition (\ref{reduction2}) on the exponential.  
Here and in what follows, we rewrite $\tilde{\tau}$ 
and $\tilde{z}$ by $\tau$ and $z$ without tilde again for notational brevity. 

In terms of the reduced field $\psi(\tau), \overline{\psi}(\tau)$, 
the leading term of the free action is then of order O($J$), taking the familiar nonrelativistic form in 
0+1 dimensions as 
\EQ
{4N^2\over (2\pi)^5}
\int d\tau   \, J 
\Big[
\overline{\psi}\partial_{\tau}\psi-
\partial_{\tau}\overline{\psi} \psi
+2\tilde{k}\overline{\psi}\psi
\Big]. 
\EN
The order $O(J^2)$ terms of the form $J^2 \overline{\psi}\psi$ are canceled, and 
the zero-point energy of the operator (\ref{harmonicop}) 
is responsible for cancelling a part of the 
order $O(J)$ term in the last term 
in the lagrangian density. 
Note that the term $|\partial_{\tilde{\tau}}\psi|^2$ is of order one and hence can be ignored in the large $J$-limit. 
Absorbing the normalization factor 
for the field, the free Hamiltonian is seen to be $\tilde{k}$ 
which  correctly reproduces the $O(1)$ part of the 
`energy' $\Delta =J+\tilde{k}$.

By performing the above redefinition for the 
cubic interaction term and rescaling,  
$\psi \rightarrow {(2\pi)^{5/2}\over 2N}{1\over \sqrt{2J}}\, 
\psi$, 
to renormalize the quadratic terms in the standard form, 
the total effective action is 
\EQ
\int d\tau\sum_{I}\Big[
{1\over 2}(\overline{\psi}_I\partial_{\tau}\psi_I-
\partial_{\tau}\overline{\psi}_I \psi_I
)+ \tilde{k}_I\, \overline{\psi}_I\psi_I
\Big]
+{1\over 2}\int d\tau \sum_{I_1,I_2,I_3}
\lambda_{\overline{I}_1,
I_2, I_3}(
\overline{\psi}_{I_1}\psi_{I_2}\psi_{I_3} + 
h.c.) , 
\label{totalactionscalar}
\EN
\EQ
\lambda_{\overline{I}_1,I_2,I_3}=
{1\over \pi^3N}
2^{-8+J_1}{2^{\tilde{\Sigma}/2}\over J_1^2}
\sqrt{J_1J_2J_3}\, {\cal G}_{\overline{I}_1,
I_2, I_3}, 
\EN
with $J_1=J_2+J_3$. 
Comparing this result with the known 
relation (\ref{couplingcft2}) connecting 
the CFT coefficients to 
the 3-point coupling constant, we confirm that the relation 
is again precisely the one discussed in (\ref{originalform}) and 
(\ref{couplingcftrelation}).  This establishes 
that the effective theory along the tunneling 
trajectory for the BMN operators with only scalar 
excitations is (\ref{totalactionscalar}). 

Now it is very natural to extend this action by 
including the higher excited states with respect to 
the Schr\"{o}dinger operator (\ref{harmonicop}). Since the 
fluctuating coordinates near the boundary 
$z_0 \rightarrow 0 \, \,(\tau\sim \pm \infty)$ are essentially 
the four vector $\vec{x}$ of the 4D base space up to an 
exponential factor $|\vec{x}_1-\vec{x}_c| \e^{-|\tau|}$,  
taking these excited states into account seems to correspond 
to the  inclusion of  the vector excitations $D_i Z(\overline{Z})$ 
of the BPS-BMN operators. 
Here we write down this extension and will give a precise discussion 
of this correspondence in the next section.  
The  decomposition (\ref{reduction1}) is generalized to the 
infinite expansion over the complete set of excited states, 
\EQ
\phi_I(\tau, \vec{y})
={(2\pi)^{5/2}\over 2N}{1\over \sqrt{2J}}\e^{-J\tau}\sum_n\phi_n^{(J)}(\vec{y}) 
\psi_n(\tau) ,
\label{reduction11}
\EN
\EQ
\overline{\phi}_I(\tau,  \vec{y})
={(2\pi)^{5/2}\over 2N}{1\over \sqrt{2J}}\e^{+J\tau}\sum_n\phi_n^{(J)}(\vec{y}) 
\overline{\psi}_n(\tau) ,
\label{reduction22} 
\EN
where 
\EQ
\phi_n^{(J)}(\vec{y}) 
=\prod_{i=1}^4 \Big({J\over \pi}\Big)^{1/4}
{2^{-n_i/2}\over \sqrt{n_i!}}H_{n_i}(\sqrt{J} \vec{y})
\e^{-Jy^2/2}
\EN
are the normalized eigenfunctions.  The kinetic term is then extended to 
\EQ
\sum_{I, n}\Big[{1\over 2}(\overline{\psi}_{I,n}\partial_{\tau}\psi_{I,n}-
\partial_{\tau}\overline{\psi}_{I,n} \psi_{I,n}
)+ (\tilde{k}_I
+ \sum_{i=1}^4 n_i) \, \overline{\psi}_{I,n}\psi_{I,n}\Big]
\label{freeactionvector}
\EN
and similarly for the interaction term
\EQ
{1\over 2}\sum_{I_1, n^{(1)}, I_2, n^{(2)}, I_3, n^{(3)}}
{1\over \pi^3N}
2^{-8+J_1}{2^{\tilde{\Sigma}/2}\over J_1^2}
\sqrt{J_1J_2J_3}({\cal G}_{\overline{I}_1,
I_2, I_3}^{n^{(1)}, n^{(2)}, n^{(3)}}
\overline{\psi}_{I_1,n^{(1)}}\psi_{I_2,n^{(2)}}\psi_{I_3,n^{(3)}} + 
h.c.) .
\label{intactionvector}
\EN
with
\EQ
{\cal G}_{\overline{I}_1,
I_2, I_3}^{n^{(1)}, n^{(2)}, n^{(3)}}
={\cal G}_{\overline{I}_1,
I_2, I_3} {\pi J_1 \over J_2J_3}
\int d^4\vec{y}\, \phi^{(J_1)}_{n^{(1)}}(\vec{y})
\phi^{(J_2)}_{n^{(2)}}(\vec{y})\phi^{(J_3)}_{n^{(3)}}
(\vec{y})
\label{vertexwithvector}
\EN

\subsection{Vector excitations}
The form of the effective action (\ref{intactionvector}) 
and (\ref{vertexwithvector}) 
including higher excited states 
with respect to the fluctuations $\vec{y}$ indicates that 
the vector excitations of  BMN operators in the bulk 
can be treated in the same manner as the 
scalar excitations. On the boundary, 
 BMN operators with vector excitations restricted to 
supergravity modes  
can  be regarded  essentially as derivatives 
of those without vector excitations;
for sufficiently large $J$, 
\[
\tilde{C}^{I}_{i_1i_2\cdots i_{\tilde{k}}}
K^{L}_{j_1j_2 \cdots j_{\ell}} \partial_{j_1}
\partial_{j_2}\cdots \partial_{j_{\ell}}
\Tr (Z^J\phi^{i_1}\phi^{i_2} \cdots \phi^{i_{\tilde{k}}}
+ \mbox{permutations}) 
\] 
\EQ
\sim
\tilde{C}^{I}_{i_1i_2\cdots i_{\tilde{k}}}
K^{L}_{j_1j_2 \cdots j_{\ell}}
\Tr(
Z^{J-\ell}D_{j_1}ZD_{j_2}Z\cdots D_{j_{\ell}}Z
\phi^{i_1}\phi^{i_2} \cdots \phi^{i_{\tilde{k}}}
+ \mbox{permutations})
\label{boundaryvectorope}
\EN
where $K^{L}_{j_1j_2 \cdots j_{\ell}}$ 
is again a completely symmetric 
and traceless tensor\footnote{Strictly speaking, we should 
include the derivatives of the scalar impurity fields and also higher derivatives appropriately 
in the right-hand side of 
(\ref{boundaryvectorope}). 
In the present section, we do not treat the trace part 
for simplicity. They will be partly considered in 
section 5 after we construct the holographic string 
field theory. 
} along the base-space directions 
on the 4D boundary, with  the normalization condition 
of the same type as for $\tilde{C}^I$.   
The `permutation' of the second line 
 means a summation over all 
possibilities of different orderings of  operators 
including $D_jZ$.    We expect that the excitations with respect to $y_j$ 
components correspond to the action of 
 $\partial_j$ on the boundary, 
as suggested in the original BMN proposal. 
However, we encounter two puzzles  
with this naive expectation. 

Firstly, we have stressed in 
subsection 3.2 that the fluctuations with respect to 
the original 4-coordinate $\vec{x}$ vanish 
exponentially $\e^{-|\tau|}$ as we approach the 
boundary. We then naively expect that 
the higher excited states do not affect the boundary theory. 
However, this is actually as it should be  in accordance with  
our S-matrix picture \cite{dsy} for the PP-wave holography. 
When an asymptotic state is an excited state, we 
have to supply an additional energy-factor 
$\e^{\ell|\tau|}$ to the wave function, by the 
definition of the S-matrix, which would just 
cancel the decreasing exponential 
associated with $\vec{y}$ fluctuations. 
In the case of scalar excitations discussed in section 2, 
the boundary condition (\ref{bcond}) of the bulk-boundary 
propagator already takes this into account, as is visible in 
the boundary condition (\ref{bcond}).  

Secondly, the effective 0+1 dimensional action 
clearly demands that the 2-point functions 
must satisfy an orthogonality condition, 
since the quadratic terms are diagonalized. 
To the contrary, 
the two-point functions obtained 
from the standard form 
${1\over |\vec{x}_1- \vec{x}_2|^{2\Delta}}$  
by acting derivative 
do not satisfy othogonality: for instance, 
\EQ
\partial_{1,j}\ln |\vec{x}_1- \vec{x}_2|=
{n_j\over |\vec{x}_1- \vec{x}_2|}, 
\label{1der}
\EN
\EQ
\partial_{1,j_1}\partial_{2,j_2}\ln |\vec{x}_1- \vec{x}_2|=
-{1 \over |\vec{x}_1- \vec{x}_2|^2}
\Big(
\delta_{j_1j_2} -2n_{j_1}n_{j_2}\Big)
\EN
with $n_j=(x_1-x_2)_j/|\vec{x}_1-\vec{x}_2|$. 
The appearance of the tensor factor $\delta_{j_1j_2} -2n_{j_1}n_{j_2}\equiv C_{j_1j_2}$ 
is  consistent with the 
relation (\ref{defy}) 
of fluctuating coordinates with the vector $\vec{y}$, 
requiring that the directions of $\vec{y}$ along $\vec{n}$ 
are actually opposite at two asymptotic regions 
$\tau \rightarrow \mp \infty$, corresponding to 
$\vec{x}_1$ and $\vec{x}_2$, respectively, at the boundary. 
The tensor $C_{ij}$ is 
usually associated with the conformal inversion 
$\vec{x} \rightarrow \vec{x}/ |x|^2$. In our case, this 
is an automatic consequence of the bulk-boundary 
correspondence. 
However, (\ref{1der})  evidently leads to an inconsistency with 
the orthogonality between scalar and vector excitations. 
In the literature (see {\it e.g.} \cite{GKT}), a particular way out  of this difficulty has been discussed without a definite physical picture 
for holography. 
The suggested procedure recovers orthogonality only after 
taking the limit $|\vec{x}_1-\vec{x}_2|\rightarrow \infty$. 
Since our picture for holography must be valid 
for any finite $|\vec{x}_1-\vec{x}_2|$, 
it is not satisfactory from our standpoint. 

  From our viewpoint of a direct correspondence 
between bulk and boundary as clarified 
up to this point,  the origin of this apparent 
discrepancy lies in an important difference  
with respect to the SO(4) symmetry on both sides.   
For the bulk, we use the metric (\ref{bulkmetric}) 
which characterizes the geometry close to the 
tunneling trajectory traversing from a fixed boundary 
point to another fixed boundary point. In particular, 
the trajectory is asymptotically 
orthogonal to the boundary spacetime, and 
  the fluctuations near the boundary have an SO(4) symmetry 
under global rotations of $\vec{y}$. 
The SO(4) rotations must be performed simultaneously at 
all points along the trajectory, and consequently 
at both ends on the boundary. It is important to 
recognize that this SO(4) symmetry 
cannot be identified with the SO(4) isometry 
of the AdS metric with respect to $\vec{x}$, as 
is indicated by the relation (\ref{defy}). Since the latter 
isometry should correspond to the SO(4) symmetry 
of the boundary theory, the rotations of $\vec{y}$ cannot be 
directly related to the usual SO(4) symmetry 
at the boundary. 

 For the boundary theory, 
we are using the flat Euclidean metric along the 
boundary space. 
The change of a distance caused by a small variation 
of points   in the
 flat Euclidean metric is  not of 
second order with respect to the variations 
of coordinates. 
\EQ
(\vec{x}_1+\delta \vec{x}_1-\vec{x}_2-\delta \vec{x}_2)^2 
=|\vec{x}_1-\vec{x}_2|^2\Big[1  
+ 2\Big(n \cdot {\delta\vec{x}_1 -\delta \vec{x}_2
\over |\vec{x}_1-\vec{x}_2| }\Big)
+\Big({|\delta\vec{x}_1-\delta\vec{x}_2|
\over |\vec{x}_1-\vec{x}_2|}\Big)^2 \Big]. 
\label{ordvariation}
\EN
This is not SO(4) symmetric under the 
rotation of $\vec{y}$. Note that 
the SO(4) transformations of $\vec{y}$ 
must be done with fixed fiducial points, $\vec{x}_1$ and 
$\vec{x}_2$, in such a way that the vicinities around 
them are simultaneously rotated. Obviously, such an  
SO(4) symmetry {\it cannot} be identified
 with the usual global SO(4) 
isometry of the original AdS metric. 
One might wonder that this is then a 
self-contradiction. But that is not so. 
As we have emphasized, the magnitude of fluctuations 
around the trajectory 
vanishes exponentially as we approach the boundary, and 
hence there is no contradiction. This is also 
consistent with the fact that the tunneling trajectory 
is a classical solution under the restriction 
that the fluctuations satisfy the Dirichlet boundary 
condition at $z=0$. 
The Euclidean S-matrix, however, requires 
us to `blow up' the vanishing fluctuations 
near the boundary at the vicinities of the 
fiducial points by multiplying the  
powers of $\e^{|\tau|} =\e^{T} 
\rightarrow \infty$ in such a way that the SO(4) symmetry 
with respect to $\vec{y}$ is kept. The process of 
blowing-up amounts to introducing a particular UV cutoff 
for the boundary theory. 

Therefore, the naive identification of the 
vector excitations in the bulk with the derivations 
$D_iZ$ on the boundary is not precise.  
We have to check from the bulk point 
of view how the small variations 
near the boundary are treated  by deriving two-point functions 
for them following the approach of section 2. 
Suppose that external lines are general vector states corresponding to 
the operators (\ref{boundaryvectorope}) with 
two traceless symmetric tensors $K^{L_1}$ and $K^{L_2}$.  
This amounts to extracting the parts proportional to  
\[
K^{L_1}_{j_1j_2\cdots j_{\ell_1}}y^{j_1}y^{j_2}
\cdots y^{j_{\ell_1}} , \quad 
K^{L_2}_{j_1'j_2'\cdots j_{\ell_2}'}y^{j_1'}y^{j_2'}
\cdots y^{j_{\ell_2}'}
\]
from the bulk-boundary propagators in the integrand 
(\ref{defG2}),  
\[
\Big({z\over z^2 + (\vec{x}+\delta \vec{x}-
\vec{x}_1-\delta \vec{x}_1)^2}\Big)^{J+\tilde{k}}, 
\quad 
\Big({z\over z^2 + (\vec{x}+\delta \vec{x}-
\vec{x}_2-\delta \vec{x}_2)^2}\Big)^{J+\tilde{k}}, 
\]
respectively. 
They are given by
\EQ
2^{\ell_1}J^{\ell_1}{x^{j_1}x^{j_2}\cdots x^{j_{\ell_1}}
\over z^{\ell_1}}
\Big({z\over z^2 + (\vec{x}-
\vec{x}_1)^2}\Big)^{J+\tilde{k}+\ell_1}, 
\EN
\EQ
2^{\ell_2}J^{\ell_2}{\overline{x}^{j_1'}\overline{x}^{j_2'}\cdots \overline{x}^{j_{\ell_2'}}
\over z^{\ell_2}}
\Big({z\over z^2 + (\vec{x}-
\vec{x}_2)^2}\Big)^{J+\tilde{k}+\ell_2}
\EN
respectively, in the large $J_i$-limit, where 
\EQ
\overline{x}^i=(\delta^{ij}-2n^i n^j)x^j. 
\EN
Note that the use of redefined $\overline{x}^i$ is 
required by the change of sign in the 
relation (\ref{defy}) between the fluctuating coordinates 
with $\vec{y}$ in the limit $\tau =\pm T \, (T\rightarrow \infty)$. 
Although the 4-th components actually 
give the powers of $|\tanh \tau|$ in converting from 
$ \vec{x}/z$ (or $\overline{\vec{x}}/z$)
 to $\vec{y}$, they can be replaced by 
one in the leading singular part with respect to the 
regularization $\epsilon$. 
Then, the result of saddle-point integral is 
\EQ
\delta^{L_1L_2}{J^2\over \pi^2}
(2J)^{\ell_1}\ell_1!
|\vec{x}_1-\vec{x}_2|^{-2(J+\tilde{k}+\ell_1)}
\delta^{L_1L_2}{2 \over \epsilon}  .
\EN 
The Kronecker $\delta^{L_1L_2}$ and the prefactor 
arises by the Gaussian integral 
\EQ
(2J)^{\ell_1+\ell_2}K^{L_1}_{j_1j_2\cdots j_{\ell_1}}K^{L_2}_{j_1'j_2'\cdots j_{\ell_2}'}{\int dy^4 
y^{j_1}y^{j_2}
\cdots y^{j_{\ell_1}}y^{j_1'}y^{j_2'}
\cdots y^{j_{\ell_2}'}\e^{-Jy^2}\over 
\int dy^4 \e^{-Jy^2}}
=\delta^{L_1L_2}\ell_1! (2J)^{\ell_1} .
\EN 

In terms of derivatives with respect to $(\vec{x}_1, \vec{x}_2)$, 
this result is expressed as  
\EQ
K^{L_1}_{j_1j_2 \cdots j_{\ell_1}}\partial_{1, j_1}
\partial_{1, j_2} \cdots \partial_{1, j_{\ell_1}}
K^{L_2}_{j_1'j_2'\cdots j'_{\ell_2}}\overline{\partial}_{2, j_1'}
\overline{\partial}_{2, j_2'}\cdots \overline{\partial}_{2, j'_{\ell_2}}{1\over |\vec{x}_{12}|_g^{2(J+\tilde{k})}}=
{
(2J)^{\ell}\ell_1! \over |\vec{x}_1- \vec{x}_2|^
{2(J+\tilde{k}+\ell_1)}
}
\delta^{L_1L_2}
\label{2pointortho}
\EN
with 
\EQ
\overline{\partial}_i =(\delta_{ij}-2n_in_j)\partial_j
\EN
being the derivatives with respect to $\overline{x}^i$. 
Here the subscript $g$ for the distance $|\vec{x}_{12}|$ indicates that the 
derivatives are computed by {\it assuming} that the 
small variations with respect to the boundary positions are defined by
\EQ
|\vec{x}_{12}|^2 \rightarrow 
(\vec{x}_1-\vec{x}_2)^2 \Big(1
- 2{\delta\vec{x}_1\cdot \overline{\delta\vec{x}}_2\over 
|\vec{x}_1-\vec{x}_2|^2}\Big)
\label{bulkvariation}
\EN
instead of the naive expression (\ref{ordvariation}).  
Apart from the terms $\delta\vec{x}_1^2, \delta\vec{x}_2^2$ 
which do not contribute owing to the traceless condition for 
the tensors $K^{L_1}$ and  $K^{L_2}$, 
this amounts to dropping all terms which violates the 
SO(4) symmetry in (\ref{ordvariation}). Strictly speaking, 
we should have denoted the derivatives $\partial_i, \overline{\partial}_i$ 
by using different notations to indicate the situation. 
Effectively, our prescription is equivalent to 
dropping terms which violates the SO(4) symmetry,  
and is consistent with  the prescriptions adopted in other 
approaches. Practically, the following rule is valid: 
First we consider 
only the directions orthogonal to $\vec{n}$ to 
obtain SO(3) symmetric answer. Then, we extend the 
results formally to SO(4) symmetric ones.  

The variation indicated by (\ref{bulkvariation}) can 
be interpreted as the variation 
of the distances measured using the Plane-wave metric 
(\ref{bulkmetric}). The minimal distance 
defined by the distance functional  
\EQ
\ell(\vec{y}_1,\vec{y}_2;T) 
=\int d\tau \sqrt{1 + \vec{y}^2 + 
\Big({d\vec{y}\over d\tau} \Big)^2} 
\sim \int d\tau (1 +{1\over 2}\Big(
\vec{y}^2 + 
\Big({d\vec{y}\over d\tau} \Big)^2\Big)
\EN
with the boundary condition
\EQ
\vec{y}(T)=y_1, \quad \vec{y}(-T)=y_2
\EN
behaves  as 
\EQA
\ell(\vec{y}_1,\vec{y}_2;T)&=&
2T + {1\over 2}(\vec{y}_1^2+\vec{y}_2^2) \tanh 2T
-{\vec{y}_1\cdot \vec{y}_2\over \sinh 2T} \nonumber \\
&\sim& 2T + {1\over 2}(\vec{y}_1^2+\vec{y}_2^2) 
-2\vec{y}_1\cdot \vec{y}_2 \e^{-2T}+ O(\e^{-4T})
\label{bulkdisvariation}
\EQN
Thus the transition amplitude is 
\EQ
\e^{-(J+\tilde{k})\ell(\vec{y}_1,\vec{y}_2;T)}
\sim \e^{-2(J+\tilde{k})T}\Big(1
- 2{\delta\vec{x}_1\cdot \overline{\delta\vec{x}_2}\over 
|\vec{x}_1-\vec{x}_2|^2}\Big)^{J+\tilde{k}}
\EN
with 
\EQ
\delta\vec{x}_1\sim {\e^{-T}\vec{y}_1\over |\vec{x}_1-\vec{x}_2|}, \quad 
\overline{\delta\vec{x}_2}\sim {\e^{-T}
\vec{y}_2\over |\vec{x}_1-\vec{x}_2|}
\EN
which are required by the relation (\ref{defy}) for 
large $|\tau|\rightarrow T$. 
Note that we can ignore the term ${1\over 2}(\vec{y}_1^2+\vec{y}_2^2)$ in (\ref{bulkdisvariation}), 
which corresponds to the redefinition (\ref{taushift}) and 
does not contribute owing to the traceless condition. 

We can  extend the above argument to
3-point correlators with vector excitations. 
By an obvious simplification of notations, we are lead 
to the integral 
\[
 {\Big(
\Pi_{i=1}^3(2J_i)^{\ell_i}K^{L_i}
\Big)
\over \int dy^4 \e^{-J_1y^2}}\int dy^4
y^{j_1}y^{j_2}
\cdots y^{j_{\ell_1}}y^{j_1'}y^{j_2'}
\cdots y^{j_{\ell_2}'}
y^{j_1''}y^{j_2''}
\cdots y^{j_{\ell_3''}}\e^{-J_1y^2}
\]
\[
\times (\mbox{ scalar integral  
with  } \Delta_i \rightarrow 
J_i + \tilde{k}_i + \ell_i) 
\]
\[
=2^{(\ell_1+\ell_2+\ell_3)/2}J_2^{\beta_3}J_3^{\beta_2}
{\ell_1!\ell_2!\ell_3! \over \beta_1! \beta_2! \beta_3!}
{(\alpha_1 + \beta_1-1)! \over (\alpha_1-1)!}
\langle K^{L_1}K^{L_2}K^{L_3}\rangle
\]
\EQ
\times  
{C^{I_1I_2I_3}
 \over |\vec{x}_{1c}|^{2(J_1+\tilde{k}_1 +\ell_1)}|2\delta|^{2
(\alpha_1+\beta_1)}}
\EN
where
\EQ
\beta_1=(\ell_2 +\ell_3 -\ell_1)/2, \quad \mbox{etc} . 
\EN
 Taking into account the normalization factor, 
this leads to the CFT coefficient
\[
C^{I_1I_2I_3, L_1L_2L_3}=
\Big({J_ 1 \over  J_2 J_3}\Big)^{\beta_1}
\Big({J_2 \over J_1}\Big)^{\ell_2/2}
\Big({J_3 \over J_1}\Big)^{\ell_3/2}
\]
\EQ \times
{\sqrt{\ell_1!\ell_2!\ell_3!} \over \beta_1! \beta_2! \beta_3!}
{(\alpha_1 +\beta_1-1)! \over (\alpha_1-1)!}
C^{I_1I_2I_3}\langle K^{L_1}K^{L_2}K^{L_3}\rangle
\EN

From the viewpoint of boundary, this result can again be formulated as the consequence 
of the replacements, 
\EQ
|\vec{x}_{12}|^2 \rightarrow 
(\vec{x}_1-\vec{x}_0)^2 \Big(1
- 2{\delta\vec{x}_1\cdot \overline{\delta\vec{x}}_2\over 
|\vec{x}_1-\vec{x}_0|^2}\Big), 
\quad 
|\vec{x}_{13}|^2 \rightarrow 
(\vec{x}_1-\vec{x}_0)^2 \Big(1
- 2{\delta\vec{x}_1\cdot \overline{\delta\vec{x}}_3\over 
|\vec{x}_1-\vec{x}_0|^2}\Big)
\label{bulkvariation23}
\EN
\EQ
|2\delta|^2 \rightarrow 
(2\delta)^2 \Big(1
- 2{\overline{\delta\vec{x}_2}\cdot \overline{\delta\vec{x}}_3\over 
|2\delta|^2}\Big)
\label{bulkvariation0}
\EN
in acting the derivation  
\[
K^{L_1}_{j_1j_2 \cdots j_{\ell_1}}\partial_{1, j_1}
\partial_{1, j_2} \cdots \partial_{1, j_{\ell_1}}
K^{L_2}_{j_1'j_2'\cdots j'_{\ell_2}}\overline{\partial}_{2, j_1'}
\overline{\partial}_{2, j_2'}\cdots \overline{\partial}_{2, j'_{\ell_2}}
K^{L_3}_{j_1"j_2'\cdots j"_{\ell_3}}\overline{\partial}_{3, j_1"}
\overline{\partial}_{3, j_3"}\cdots \overline{\partial}_{3, j"_{\ell_3}}
\]
directly to the general form (\ref{gene3pointcorr}) of the 3-point correlator without 
vector excitations.
The prescriptions  (\ref{bulkvariation23}) are equivalent to (\ref{bulkvariation}) to the  leading 
order in $\delta$, while (\ref{bulkvariation0}) 
is necessary to be consistent with the 
SO(4) symmetry. 
 
Combining  with the CFT 
coefficient of pure scalars, we obtain
\[
C^{I_1I_2I_3, L_1L_2L_3}={1\over N}{2^{J_1 +{\tilde{\Sigma}\over 2} -9}\over \pi^3}
{\sqrt{J_1J_2J_3}\over J_1^2} \Big({J_1 \over J_2 J_3}\Big)^{\alpha_1+\beta_1}\, 
{(\alpha_1+\beta_1)! \over \alpha_1+\beta_1}\,  {\cal G}_{I_1I_2 I_3} 
\]
\EQ
\times \Big({J_2 \over J_1}\Big)^{\ell_2/2}
\Big({J_3 \over J_1}\Big)^{\ell_3/2}
{\sqrt{\ell_1!\ell_2!\ell_3!} \over \beta_1! \beta_2! \beta_3!}
\langle K^{L_1}K^{L_2}K^{L_3}\rangle
\label{cftvector1}
\EN
in terms of the pure-scalar 3-point coupling, or 
using 
the expression for the scalar vertex (\ref{scalarvertex})
\[
C^{\overline{I}_1I_2I_3, L_1L_2L_3}={1\over N} {\sqrt{J_1J_2J_3}\over J_1^2} \Big({J_1 \over J_2 J_3}\Big)^{\alpha_1+\beta_1}(\alpha_1+\beta_1)! 
{\alpha_1\over \alpha_1+\beta_1}
\]
\EQ
\times \Big({J_2\over J_1}\Big)^{(\tilde{k}_2+\ell_2)/2}
\Big({J_3 \over J_1}\Big)^{(\tilde{k}_3+\ell_3)/2}
{\sqrt{\tilde{k}_1!\tilde{k}_2!\tilde{k}_3!}\over 
\tilde{\alpha}_1!\tilde{\alpha}_2!\tilde{\alpha}_3!}
\langle \tilde{{\cal C}}^{\overline{I}_1}
\tilde{{\cal C}}^{I_2}\tilde{{\cal C}}^{I_3}\rangle
{\sqrt{\ell_1!\ell_2!\ell_3!} \over \beta_1! \beta_2! \beta_3!}
\langle K^{L_1}K^{L_2}K^{L_3}\rangle
\label{cftvector2}
\EN
in terms of the SO(4) tensors. On the other hand, 
the effective interaction (\ref{vertexwithvector}) 
including the vector excitations shows that 
the 3-point coupling for the external lines in consideration is 
given by replacing the coefficient ${\cal G}_{I_1I_2I_3}$ 
for purely scalar excitations as 
\EQ
{\cal G}_{I_1I_2I_3}^{L_1L_2L_3}
={\cal G}_{I_1I_2I_3}\Big({J_2 \over J_1}\Big)^{\ell_2/2}
\Big({J_3 \over J_1}\Big)^{\ell_3/2}
{\sqrt{\ell_1!\ell_2!\ell_3!} \over \beta_1! \beta_2! \beta_3!}
\langle K^{L_1}K^{L_2}K^{L_3}\rangle. 
\label{gcouplingvector}
\EN 
Comparing this with (\ref{cftvector1}) taking account the 
normalization factor in the 3-point interaction term 
of the effective action, 
we can conclude that  
the general form, 
as derived in section 2  
for the holographic correspondence of scalar supergravity excitations, extends 
to a more general case including the vector excitations, 
\EQ
C_{I_1I_2I_3}^{L_1L_2L_3}=
{\tilde{\lambda}_{I_1I_2I_3}^{L_1L_2L_3}
\over \Delta_2+\Delta_3-\Delta_1}
\EN
\EQ
\tilde{\lambda}_{I_1I_2I_3}^{L_1L_2L_3}
=\Big({J_2J_3\over J_1}\Big)^{(\Delta_1-\Delta_2-\Delta_3)/2}\Gamma({\Delta_2+\Delta_3-\Delta_1 \over 2}+1)
\lambda_{I_1I_2I_3}^{L_1L_2L_3}
\EN
with  
$\Delta_i=k_i+ \ell_i=J_i +\tilde{k}_i +\ell_i$. 

Supersymmetry demands that this should extend 
further to fermionic excitations.  Since we treat the full supersymmetric 
string field theory later, we do not discuss the supersymmetrization 
of our arguments explicitly. 
We only mention that to discuss 
orthogonality including fermionic excitations, the 
spinor equivalent to the inversion tensor $\delta_{j_1j_2}-2n_{j_1}n_{J_2}$ 
transforming spinor indices between 
two boundary points is the SO(4) 
gamma matrix 
$i\gamma^i n^i$. In computing 
correlation functions, we have to 
extract SO(4) invariant pieces in similar manner 
as for bosonic excitations  by taking into account the 
transformation of spinor indices between 
two boundary points.  This is effectively equivalent 
with the procedures adopted in other works.

\subsection{Operator representation and the $\alpha'$ corrections}

On the basis of our results, we are now prepared to proceed to  the construction of a  full `holographic' string field theory.  
Before that, it is  
convenient to first rewrite the foregoing 
results using oscillator representation. 
By substituting the 
expression for the general 3-point coupling, the 
effective action is written as 
\EQ
S_{eff}=S_2 +S_3
\EN
\EQ
S_2=\sum_{I, L}\int d\tau \Big[
{1\over 2}(\overline{\psi}_{I,L}\partial_{\tau}\psi_{I,L}-
\partial_{\tau}\overline{\psi}_{I,L} \psi_{I,L}
)+ {1\over R}(\tilde{k}_I+ \ell_L)\, \overline{\psi}_{I,L}\psi_{I,L}
\Big], 
\EN
\[
S_3 = 
{1\over NR}\int d\tau \sum_{I_1,I_2,I_3}
\alpha_1
\sqrt{{J_1J_2J_3}} 
\, \, \overline{\psi}_{I_1,L_1}\psi_{I_2,L_2}\psi_{I_3,L_3}
\Big({J_2\over J_1}\Big)^{(\tilde{k}_2+\ell_2)/2}
\Big({J_3 \over J_1}\Big)^{(\tilde{k}_3+\ell_3)/2}
\]
\EQ
\times 
{\sqrt{\tilde{k}_1!\tilde{k}_2!\tilde{k}_3}\over 
\tilde{\alpha}_1!\tilde{\alpha}_2!\tilde{\alpha}_3!}
{\sqrt{\ell_1!\ell_2!\ell_3!} \over \beta_1! \beta_2! \beta_3!}
\langle \tilde{{\cal C}}^{\overline{I}_1}
\tilde{{\cal C}}^{I_2}\tilde{{\cal C}}^{I_3}\rangle 
\langle K^{L_1}K^{L_2}K^{L_3}\rangle  + 
h.c. , 
\EN
where we have recovered the length dimension. 
The integrals over 
the angular momentum $J$ should be understood implicitly.  
We introduce the bra-ket notation for the field 
by using the 8 dimensional oscillator algebra of 
$a_i, a^{\dagger}_i$;$ \,  a_i|0\rangle =0 , \, 
[a_i, a^{\dagger}_j]=\delta_{ij} \, \, (i=1, 2, \ldots, 8)$ 
where the first 4 directions correspond to 
the SO(4) indices of the vector directions $\vec{y}$, and 
the remaining 4 directions $(i=5, \ldots, 8)$ represents the 
SO(4) indices of the scalar excitations.  
\EQ
\{\psi_I(\tau) \} \rightarrow 
|\psi(\tau)\rangle =
\sum_{I,L}\psi_{I,L}(\tau) 
|I, L \rangle ,
\EN
\EQ
|I ,L\rangle=
\tilde{C}^I_{i_1 i_2 \cdots i_{\tilde{k}}}K^L_{j_1j_2\cdots j_{\ell}}
{1\over \sqrt{\tilde{k}!\ell!}}a^{\dagger}_{i_1}a^{\dagger}_{i_2}
\cdots a^{\dagger}_{i_{\tilde{k}}}
a^{\dagger}_{j_1}a^{\dagger}_{j_2}
\cdots a^{\dagger}_{j_{\ell}}\,  |0\rangle .
\label{oscstate}
\EN
The free term of the action is then
\EQ
S_2=\int d\tau \Big[{1\over 2}(\langle \overline{\psi}| \partial_{\tau}|\psi\rangle 
-{1\over 2}(\partial_{\tau}\langle \overline{\psi}|)|\psi\rangle +
\langle \overline{\psi}| h_{sv} |\psi\rangle \Big]
\EN
with 
\EQ
h_{sv}=h_s + h_v, \quad h_s={1\over R}\sum_{i=1}^4a_i^{\dagger}a_i, 
\quad h_v={1\over R}\sum_{j=5}^8 a_j^{\dagger}a_j. 
\label{freehsugra}
\EN
The parameters familiar in string field theory literature are 
$R^4/J_1^2(\alpha')^2 =\lambda'=1/(\mu p^+ \alpha')^2=
g_{{YM}}^2N/J_1^2, g_2=J_1^2/N$. In terms of the string-length 
parameter, $\alpha_{(r)}=\alpha'p^+_r$, 
$\mu |\alpha_{(r)}|=\alpha'J_r/R^2$. Note that 
$\mu \sim 1/R$ since $|p^+_r| \sim J_r/R$ and also that 
 $\alpha_{(1)} <0, \alpha_{(2)}>0, \alpha_{(3)} 
>0$ in the present convention. 
The supergravity limit corresponds to the limit 
$\alpha'\rightarrow 0$ with fixed $R=(g_{{\rm YM}}^2N)^{1/4}\alpha'$.

%%%%%%% from here %%%%
The interaction term is expressed in terms 
of the overlap state defined by
\begin{equation}
|v_0\rangle =
\exp [-{1\over 2}\sum_{r,s=1}^3\, \Big(
\sum_{i=1}^4
a^{\dagger}_{i(r)}n_{00}^{rs}
a_{i(s)}^{\dagger} 
+\sum_{j=5}^8
a^{\dagger}_{j(r)}n_{00}^{rs}
a_{j(s)}^{\dagger}\Big)
]|0\rangle
\end{equation}
with the supergravity part of the familiar Neumann 
functions in the zero-slop limit $\alpha'\rightarrow 0$. 
\begin{equation}
n_{00}^{rs}=
\delta^{rs}-\sqrt{{J_rJ_s\over J_1^2}} , 
\quad 
n_{00}^{r1 }=n_{00}^{1r}=
-\sqrt{{J_r\over J_1}}
\quad \quad \mbox{for}  
\quad r,s =2, 3 \quad \mbox{and} \quad 
n_{00}^{11}=0
\end{equation}
The unusual minus sign on the exponent 
is owing to our phase convention 
for the creation-annihilation operators. 
As is well known, the overlap state is the ground-state 
solution for the continuity conditions satisfying 
\begin{equation}
\sum_{r=1}^3 p_{(r)}|v_0\rangle =0, \quad 
(x_{(3)} -{J_2\over J_1}x_{(2)} -{J_3\over J_1}x_{(3)}) |v_0\rangle =0
\label{conticond}
\end{equation}
where 
\begin{equation}
x_{(r)}={R\over \sqrt{2J_r}}
(a_{(r)}+a^{\dagger}_{(r)}), \quad 
p_{(r)}=-i\sqrt{J_r\over 2R^2}(a_{(r)} - a_{(r)}^{\dagger})
\end{equation}
for all 8 directions.  The `ground state' means that 
it also satisfy the locality condition of the form 
\begin{equation}
(x_{(1)}-x_{(2)})|v_0\rangle =
(x_{(1)}-x_{(3)})|v_0\rangle =0. 
\label{locality}
\end{equation}
We find that 
\[
_{(1)}\langle \overline{I}_1,L_1|_{(2)}\langle I_2,L_2|
_{(3)}\langle I_3, L_3| 
|v_0\rangle =
\Big({J_2\over J_1}\Big)^{{(\tilde{k}_2+\ell_2)\over 2}}
\Big({J_3\over J_1}\Big)^{{(\tilde{k}_3+\ell_3)\over 2}}
\]
\begin{equation} \times
{\sqrt{\tilde{k}_1! \tilde{k}_2! \tilde{k}_3!}
\over \tilde{\alpha}_1! \tilde{\alpha}_2! \tilde{\alpha}_3!}
\langle \tilde{C}^{\overline{I}_1}
\tilde{C}^{I_2}\tilde{C}^{I_3}\rangle
{\sqrt{\ell_1!\ell_2!\ell_3!} \over \beta_1! \beta_2! \beta_3!}
\langle K^{L_1}K^{L_2}K^{L_3}\rangle .
\end{equation}
This shows that the 3-point interaction part of the 
effective action takes the following simple form
\begin{equation}
S_3=
{1\over 2N}\int d\tau  \,  _{(1)}\langle 
\overline{\psi}|   _{(2)}\langle \psi|  _{(3)}\langle \psi| \sqrt{J_1J_2J_3}
(h^{(2)}_s+h^{(3)}_s -h^{(1)}_s)|v_0\rangle 
+ h.c. \, .
\label{s3zero}
\end{equation}
The integral with respect to the angular momentum under 
the conservation condition $J_1=J_2+J_3$ should 
be understood implicitly as before. 
The most important characteristic of this expression is that 
the so-called `prefactor' involves only the energies 
of scalar excitations.  As we have stressed in previous works, the holographic relation of our type 
does not necessarily 
demand that the prefactor itself is the difference of 
the free Hamiltonian $h_{sv}^{(2)}+h_{sv}^{(3)}-h_{sv}^{(1)}$.  
As a consequence of this remarkable property, the 3-point 
vertices vanish whenever the scalar energies are preserved. 
This implies that, except for special cases when the vector energies are 
simultaneously preserved, the 3-point correlators 
vanish, as has been explicitly exhibited 
in the expression of the CFT coefficient (\ref{cftvector2}). Since we have  constraints
$\tilde{k}_1 \le \tilde{k}_2 + \tilde{k}_3$ and 
$\ell_1 \le \ell_2 + \ell_3$, the conservation of both 
scalar and vector energies 
is possible only when the total energies are 
conserved, namely, when $\Delta_1=\Delta_2 + \Delta_3$. 
Note that in this extremal cases we have finite results 
for the 3-point correlators, owing to the 
vanishing denominator in our holographic relation. 

Of course, the above action for supergravity modes is 
valid in the limit $\alpha'\rightarrow 0$ with fixed 
$R$.  The limit $\alpha'\rightarrow 0$ corresponds to the 
imposition of the locality conditions  (\ref{locality}) 
for the overlap state. However, 
when we consider a finite $\alpha'$, the latter conditions 
cannot be preserved, as it should be for general  string states 
which have nonzero extension.  Then only the more general continuity conditions  
(\ref{conticond}) are satisfied. Correspondingly, the 
Neumann functions $n_{00}^{rs}$ for the zero modes 
must be replaced by  $N^{rs}_{00}$, defined by
\begin{equation}
N_{00}^{rs}=fn^{rs}_{00}=f(
\delta^{rs}-\sqrt{{J_rJ_s\over J_1^2}}) , 
\quad 
N_{00}^{r1 }=n_{00}^{r1}=
-\sqrt{{J_r\over J_1}}
\quad \quad \mbox{for}  
\quad r,s =2, 3 \quad \mbox{and} \quad 
N_{00}^{11}=0
\end{equation}
where $f=1-4\mu\alpha_{(1)}\alpha_{(2)}\alpha_{(3)} K$ is a 
nontrivial function of $\mu\alpha_{(r)} \, (r=1,2,3)$. 
For the definition of the quantity $K$, 
we refer the reader to  the Appendix B. 
In the supergravity limit  $\alpha'\rightarrow 0$, keeping $R$ 
and $J_i$ fixed,  we have $f\rightarrow 1$.  
In the opposite large $\mu  $ limit, it is known that 
\begin{equation}
f\rightarrow {|\alpha_{(1)}| \over 4\pi \mu \alpha_{(2)} \alpha_{(3)}}
={R^2 J_1\over 4\pi \alpha'J_2J_3}. 
\end{equation}
We denote the overlap state with this modification 
for finite $\alpha'$  by 
$|V_0\rangle $. 
\begin{equation}
|V_0\rangle =
\exp [-{1\over 2}\sum_{r,s=1}^3\, \Big(
\sum_{i=1}^4
a^{\dagger}_{i(r)}N_{00}^{rs}
a_{i(s)}^{\dagger} 
+\sum_{j=5}^8
a^{\dagger}_{j(r)}N_{00}^{rs}
a_{j(s)}^{\dagger}\Big) 
]|0\rangle
\end{equation}
The use of the $\alpha'$-corrected 3-point vertex which is 
given by the same form as (\ref{s3zero}) 
but $|v_0\rangle$ being replaced by $|V_0\rangle$  leads to a correction factor 
$f^{\alpha_1}$ for the matrix elements, multiplying the original local form.

On the other hand, the normalization of the 
CFT coefficients for the BPS operators used 
in the foregoing subsections 
is believed to be exact to all orders in the gauge coupling  
$\lambda=g_{{\rm YM}}^2N$, because of the 
non-renormalization property of 3-point 
functions of chiral operators.  
Indeed, as we stressed before, the CFT coefficients 
are nothing but the results from the free field theory, 
which corresponds to the large $\mu$ limit. 
This implies that for 
nonzero $\alpha'$, the rule of mapping between 
the matrix elements of the 
bulk 3-point vertex and the 
CFT coefficients must be extended  as a natural ansatz to 
\begin{equation}
C_{I_1I_2I_3}^{L_1L_2L_3}=
{\tilde{\lambda}_{I_1I_2I_3}^{L_1L_2L_3}
\over \Delta_2+\Delta_3-\Delta_1} , 
\label{clambdarelation1}
\end{equation}
\begin{equation}
\tilde{\lambda}_{I_1I_2I_3}^{L_1L_2L_3}
=\Big(f {J_2J_3\over J_1}\Big)^{-(\Delta_2+\Delta_3-\Delta_1)/2}\Gamma({\Delta_2+\Delta_3-\Delta_1 \over 2}+1)
\lambda_{I_1I_2I_3}^{L_1L_2L_3}
\label{clambdarelation2}
\end{equation}
with 
\begin{equation}
\lambda_{I_1I_2I_3}^{L_1L_2L_3}
=\,  _{(1)}\langle 
I_1, L_1|   _{(2)}\langle I_2, L_2|  _{(3)}\langle I_3, L_3| {\sqrt{J_1J_2J_3}\over N}\, 
R(h^{(2)}_s+h^{(3)}_s -h^{(1)}_s)|\tilde{V}_0\rangle , 
\label{clambdarelation3}
\end{equation}
and the 3-point interaction term of the action is 
\begin{equation}
S_3=
{1\over 2}\int d\tau  \,  _{(1)}\langle 
\overline{\psi}|   _{(2)}\langle \psi|  _{(3)}\langle \psi| {\sqrt{J_1J_2J_3}\over N}
(h^{(2)}_s+h^{(3)}_s -h^{(1)}_s)|V_0\rangle 
+ h.c. \, .
\label{s3zero1}
\end{equation}
This is the main prediction of the present paper. 
By  construction, this gives the same CFT 
coefficients for the purely supergravity modes. 
But for higher stringy modes, the correction factor  would 
play an important role for impurity non-preserving 
processes. The appearance of the factor $f$ 
with non-trivial $\alpha'$ dependence 
is not surprising if we recall that the origin of the factor  
$\Big({J_2J_3\over J_1}\Big)^{-(\Delta_2+\Delta_3-\Delta_1)/2}$ 
is the relation between the length $\delta$ and 
the cutoff with respect to the $\tau$ integration, as 
exhibited in the integral (\ref{gammaintegral}). 
Our prediction indicates that the nonlocality of the vertex 
is essentially represented by a rescaling 
$\delta \rightarrow \delta \sqrt{f}$. 
 Finally, we also mention that this interaction vertex 
is not SO(8) symmetric, 
contrary to a possibility suggested in  \cite{dsy}.  
The origin of this phenomenon is  the asymmetric roles 
   played by scalar and vector excitations 
from the viewpoint of boundary theory, as we have analyzed 
in the previous subsection.

%\newpage
\section{Holographic string field theory}
\setcounter{equation}{0}
In our first work \cite{dsy} on the PP-wave holography, 
we have emphasized that 
string field theory describing the plane-wave background 
cannot be constructed uniquely by the requirement 
of supersymmetry alone. Our claim was that 
there should exist a unique `holographic' string field theory 
which realizes our holographic relation. 
We now have a more concrete constraint 
for the holographic string field that it should 
have the  3-point vertex (\ref{s3zero1}) when restricted to the 
bosonic supergravity sector. 

There are basically two 
different proposals for 3-point vertex, conforming to the 
requirement of supersymmetry
 algebra $\{Q_{a},Q_{\dot a}^{\dagger} \}=2\delta_{a\dot a}H$ up 
to the first order with respect to the string coupling.  
Let us first briefly review them. The most familiar is the 
one which was first proposed in \cite{SV1,SV2}
as a generalization of the light-cone string field theory 
in flat spacetime, as constructed by Brink, Green, and Schwarz long time ago. This proposal was corrected and established in subsequent works \cite{Pan1,PS1,Pan2,PS2}. We denote this vertex by 
\begin{equation}
|H_3\rangle_{SV}=P_{SV}|E\rangle
\label{SV vertex}
\end{equation}
where $|E\rangle$ is the overlap vertex including both 
bosonic and fermionic oscillators. The explicit form 
of the prefactor $P_{SV}$ is given in the Appendix B, where 
all of the other necessary definitions and formulas of 
string field theory are summarized, in order
 to make the present 
paper reasonably self-contained.  
It has been shown that this vertex has a 
nontrivial relation with the matrix of operator mixing 
which appears in the perturbative computation 
of anomalous dimensions on the gauge-theory side. 
We will come back to this point later.   

Another version of the 3-point vertex, which was proposed \cite{DPPRT} later in connection with 
the duality relation of our type, takes the following form
\begin{equation}
|H_3\rangle_D =(H_2^{(2)}+H_2^{(3)}-H_2^{(1)})|E\rangle
\label{dvertex}
\end{equation}
where 
\begin{equation}
H_2^{(r)}=\frac{1}{|\alpha_{(r)}|}\sum_{n=-\infty}^{\infty}
\omega_n^{(r)}(a_n^{(r)\dagger}a_n^{(r) }
+b_n^{(r)\dagger}b_n^{(r)}) \quad \mbox{with} \quad  \omega^{(r)}_n=
\sqrt{n^2 + (\mu\alpha_{(r)})^2}
\end{equation}
are the free Hamiltonian operators. Here the string-length 
parameters are chosen such that $\alpha_{(1)}+\alpha_{(2)}+\alpha_{(3)}=0$ with $\alpha_{(2)}, \alpha_{(3)}>0$ and $\alpha_{(1)}<0$ 
to be consistent with our conventions used in 
the foregoing analyses. Note that this Hamiltonian 
coincides with the total free Hamiltonian $h_{sv}$, (\ref{freehsugra}),  of the previous section 
when it is restricted to 
supergravity modes with the identification 
$\mu =1/R$. 
 The form (\ref{dvertex}) is the simplest realization 
of the way which was discussed in our first work \cite{dsy} to obtain susy-compatible interaction vertices.  In 
particular, it satisfies the SO(8) symmetry suggested there.  
The authors of \cite{DPPRT}  has  shown that 
this vertex gives nontrivial matrix elements in the 
leading order of the $1/\mu$-expansion for impurity 
preserving processes, 
following the duality relation of our type 
\begin{eqnarray}\label{duality}
 (\Delta_\inI+\Delta_\inII-\Delta_\out)C_{123}
={\sqrt{J_1J_2J_3}\over N}\frac{1}{\mu} \braket{1,2,3}{\HI},
\end{eqnarray} 
when the mixing of gauge-theory operators is {\it ignored}. 
This duality relation itself for impurity-preserving 
processes was first proposed in 
\cite{Constable et. al. 1} from  a 
viewpoint which is entirely different from ours, and was 
actually abandoned later since it turned out that the 
relation was not appropriate for the original purpose 
after correcting a sign error in some earlier 
versions of works on PP-wave string field theory.  
This relation is  in fact obtained from our 
holographic relation in the limit of large $\mu$ for arbitrary 
3-point processes when the 
number of impurities  is preserved.  We stress that 
in the light of our works, the more general 
relation summarized in (\ref{clambdarelation1})$\sim$
(\ref{clambdarelation3}) has  a firm physical foundation from  the holographic principle. We should also recall 
that we are using the Euclidean picture 
of the tunneling and hence this form of the 
interaction proportional to energy difference 
must not be thrown away by a possible nonlinear unitary 
transformation: The `energies' are not 
conserved for the Euclidean S-matrix.  For a detailed 
discussion of our S-matrix picture, we invite the 
reader to \cite{dsy}. 

These two candidates for the 3-point vertex conform to 
different duality relations connecting bulk and boundary.   
The first type $|H_3\rangle_{SV}$ does not fit to our 
holographic relation. The second type 
$|H_3\rangle_D$ does not either, though motivated by (\ref{duality}), since for all of our arguments 
from section 2 and 3,  
 the gauge-theory operators have to be definite conformal operators of 
which 2-point functions are diagonalized and  3-point 
correlators take the 
standard conformal form (\ref{gene3pointcorr}). 
However, it is important here to recall that
 the requirement of 
supersymmetry algebra to the first order 
in the string coupling puts only a linear constraint.  
Hence, any linear combination 
of two possible vertices with some global coefficients 
depending only on  string length parameters and 
the curvature parameter $\sqrt{\alpha'}\mu \sim \sqrt{\alpha'}/R$ can be 
an allowed candidate for the 3-point vertex. 

We can now  study whether it is possible to 
obtain our effective action (\ref{s3zero1}) 
for supergravity modes 
from these vertices.  
Let us first examine how the above two 
vertices behave when they are 
restricted to purely supergravity modes. 
For purely bosonic external states, they take the 
following forms
\begin{eqnarray}
|H_3\rangle_{SV}&\Rightarrow &\frac{1}{2}
\left[
\sum_{i=5}^{8}
\left(\left(\XI^i\right)^2-\left(\XII^{i}\right)^2\right)
-\sum_{\mu=1}^{4}
\left(\left(\XI^{\mu}\right)^2-\left(\XII^{\mu}\right)^2\right]
\right)\ket{\E_a}  \nonumber \\
&=&\sum_{r=1}^3\sum_{i=5}^8
\left(\sum_{m=0}^{\infty}
\frac{\omega^{(r)i}_{m}}{\alpha_{(r)}}
a^{(r)i\dagger}_{m}a^{(r)i}_{m}
-\sum_{m=1}^{\infty}
\frac{\omega^{(r)i}_{m}}{\alpha_{(r)}}
a^{(r)i\dagger}_{-m}a^{(r)i}_{-m}
\right)\ket{\E_a} \nonumber \\
&-&\sum_{r=1}^3\sum_{i=1}^4
\left(\sum_{m=0}^{\infty}
\frac{\omega^{(r)}_{m}}{\alpha_{(r)}}
a^{(r)i\dagger}_{m}a^{(r)i}_{m}
-\sum_{m=1}^{\infty}
\frac{\omega^{(r)i}_{m}}{\alpha_{(r)}}
a^{(r)i\dagger}_{-m}a^{(r)i}_{-m}
\right)\ket{\E_a}, 
\end{eqnarray}

\begin{eqnarray}
|H_3\rangle_D
&\Rightarrow& \frac{1}{2}\left[
\sum_{i=5}^{8}
\left(\left(\XI^i\right)^2+\left(\XII^{i}\right)^2\right)
+\sum_{\mu=1}^{4}
\left(\left(\XI^{\mu}\right)^2+\left(\XII^{\mu}\right)^2\right)
\right]\ket{\E_a} \nonumber \\
&=&\sum_{r=1}^3\sum_{i=1}^8
\left(\sum_{m=0}^{\infty}
\frac{\omega^{(r)}_{m}}{\alpha_{(r)}}
a^{(r)i\dagger}_{m}a^{(r)i}_{m}
+\sum_{m=1}^{\infty}
\frac{\omega^{(r)}_{m}}{\alpha_{(r)}}
a^{(r)i\dagger}_{-m}a^{(r)i}_{-m}
\right)\ket{\E_a}.
\end{eqnarray}
The state $\ket{E_a}$ is the bosonic part of the 
overlap state. 
The expression $\XI$ and $\XII$, defined in the Appendix B,  contain 
the oscillators of only $\cos$ and $\sin$ modes, respectively, 
which are represented by the creation-and-annihilation 
operator of positive ($a_n^{(r)}$) and negative 
($a_{-n}^{(r)}) $ indices, respectively. 
As briefly discussed there, the factorization formula for
the Neumann matrices \cite{Schwarz,Pan1} allows us 
to derive the above simple expressions in terms 
of world-sheet energies $\omega_m^{r}$. 

By restricting to the zero modes, we see that neither 
of them separately reduces to the effective action of the previous section. However, it is now evident that 
if we combine them with equal weights we can get 
the desired form
\EQA
|H_3\rangle_h\equiv {1\over 2}(|H_3\rangle_{SV}+|H_3\rangle_D)&\Rightarrow&
\left(
\sum_{i=5}^8 (\XI^i)^2 + 
\sum_{i=1}^4(\XII^i)^2
\right)|E_a\rangle 
\nonumber \\
&&\hspace{-3.8cm}=
\sum_{r=1}^3
\left(\sum_{i=5}^8\sum_{m=0}^{\infty}
\frac{\omega^{(r)}_{m}}{\alpha_{(r)}}
a^{(r)i\dagger}_{m}a^{(r)i}_{m}
+\sum_{i=1}^4\sum_{m=1}^{\infty}
\frac{\omega^{(r)}_{m}}{\alpha_{(r)}}
a^{(r)i\dagger}_{-m}a^{(r)i}_{-m}
\right)\ket{\E_a}, 
\label{bosonicpart}
\EQN
which in the zero-mode sector reduces to the form
$
(h_s^{(2)}+h_s^{(3)}-h_s^{(1)})|V_0\rangle, 
$
involving only the scalar oscillators in the prefactor, as derived 
in the previous section. Apart from the overall 
normalization fixed by comparing with (\ref{s3zero1}), 
this combination is unique. Therefore, we have arrived at 
the uniquely possible string field 3-point interaction term 
which is consistent with the conclusion that we have 
reached by foregoing discussions. 
 
We note that when restricted  to the scalar modes 
the form (\ref{bosonicpart}) coincides with the one 
suggested previously in \cite{CK} as the possible prefactor 
which is compatible with the holographic relation (\ref{duality}).  Namely, they have shown that 
this gives the correct CFT coefficients after taking 
into account the operator mixing for scalar operators with 
two impurities.  The result (\ref{bosonicpart}) shows that 
the scalar prefactor indeed consists only of the $\cos$ modes 
as they have proposed, 
while actually the vector prefactor must consist only of the $\sin$ modes, in order to be consistent with supersymmetry.  

Thus the conclusion of this section is that 
the holographic string field theory is given in the 
following form 
up to the first order in the string coupling 
\EQ
S_{h}=S_{h,2} + S_{h,3} ,
\EN
\EQ
S_{h,2}=\int d\tau \Big[{1\over 2}(\langle \overline{\psi}| \partial_{\tau}|\psi\rangle 
-{1\over 2}(\partial_{\tau}\langle \overline{\psi}|)|\psi\rangle +
\langle \overline{\psi}| H_2 |\psi\rangle \Big] , 
\EN
\EQ
S_{h,3}={1\over 2N}\int d\tau  \,  _{(1)}\, \langle 
\overline{\psi}|   _{(2)}\langle \psi|  _{(3)}\langle \psi| 
\sqrt{J_1J_2J_3} |H_3\rangle_h 
+ h.c. 
\EN
The integral over $J_i$ under the condition $J_1=J_2+J_3$ is implicit as before. 

This action has no SO(8) symmetry, nor 
even the $Z_2$ symmetry.  We also note that the holographic 
string field theory does not directly reduce to the 
familiar Green-Schwarz form in the flat limit 
$R\rightarrow \infty$. This was the case already in the 
supergravity sector. Since the AdS/CFT correspondence 
between bulk gravity (closed string theory) and CFT 
on the boundary is a concept which requires 
a global consideration of the whole AdS spacetime, 
the fact that the flat limit is not direct is not surprising. 

\section{Relation with other possible duality maps}
\setcounter{equation}{0}

In the present section, we discuss the relation of our results 
to other proposals for duality maps, especially the one 
advocated in refs. \cite{GMP1}. As we have already mentioned, it has been shown that 
the 3-point vertex $|H_3\rangle_{SV}$ of the first type 
has a close connection with the matrix of operator mixing 
in the perturbative computation of anomalous dimensions. 
It is indeed reasonable to 
assume that there exists a quantity in 
the bulk which plays the role of the  
operator mixing.  The operator mixing is necessarily associated 
with perturbative renormalization  procedure 
in higher orders with respect both to  
gauge coupling and to genus expansion in our 
situation where there is 
 a large degeneracy with 
respect to conformal dimensions at the lowest order.  

The first indication for the necessity of taking into account the 
operator mixing comes from the behavior of 3-point functions 
at the first order in $\lambda'\equiv \lambda/J^2$. 
Take  for instance the simplest case of the 3-point function 
of operators with two different impurities ($i \ne j$), 
\[
\hspace{-8cm}
\langle 
\bar{O}^{J}_{ij, n}(x_\out)O^{y J}_{ij, m}(x_\inI)
O_{\vac }^{(1-y)J}(x_\inII)
\rangle =
\]
\begin{eqnarray}
\label{bare three point function}g_2 C_{n;my}[1-
\lambda'\left(a_{n;my}\log(x_{\out\inI}\Lambda)^2
 +b_{n;my}\log\left(\frac{x_{\inII\out}x_{\out\inI}
\Lambda}{x_{\inI\inII}}\right)\right)],
\end{eqnarray}
where 
\[
O^{J}_{ij,n}=\frac{1}{\sqrt{JN^{J+2}}}
\sum_{l=0}^{J}\e^{\frac{2\pi i n l}{J}}\Tr
\left(\phi_i Z^{l} \phi_{j} Z^{J-l}\right)
\]
and $O_{\vac }^{J}$ is just the BMN operator corresponding 
to the ground state, with 
\EQ
a_{n;my}=m^2/y^2,  \quad b_{n;my}=n(n-m/y) , \quad 
 \mbox{and} \quad C_{n;my}=\sqrt{{1-y\over yJ}}
{\sin^2 (\pi n y)\over \pi^2(n-{m\over y})^2} .
\EN
For notational brevity, in the present section, 
we suppress the classical part 
(namely, the zero-th order in $\lambda'$) of 
spacetime dependence for gauge-theory correlation functions. 
Note that $a_{n;my}$ is noting but the anomalous dimension
of $O^{y J}_{m}$ since it comes from 
Feynman diagrams 
where the interaction occurs only  between $O_n^J(x_\out)$ and
$O_m^{yJ}(x_\inI)$,  
while $b_{n;my}$ comes from graphs
where the interaction occurs among all three operators
$O^{J}_n(x_\out)$, $O_m^{yJ}(x_\inI)$ and $O_{\vac}^{(1-y)J}(x_\inII)$.
This form, however, does not take the standard form 
of a 3-point correlation function of conformal operators. 
In other words, to this first order in $\lambda'$, 
the BMN operator $O^{J}_{ij,n}$ cannot be regarded as 
a conformal operator characterized by the standard 
conformal transformation property. 

As is well known now, 
this difficulty is resolved if we take into account the 
operator mixing with multi-trace operators. 
Consider a 2$\times$2 matrix of two-point functions 
of operators, 
 \begin{equation}
O^{J}_{ij,n}=\frac{1}{\sqrt{JN^{J+2}}}
\sum_{l=0}^{J}\e^{\frac{2\pi i n l}{J}}\Tr
\left(\phi_i Z^{l} \phi_{j} Z^{J-l}\right)\quad\mbox{and}\quad
T_{ij,n}^{J,y}=:O^{y\cdot J}_{ij,n}\,O^{(1-y)\cdot J}_{\vac}: \, . 
\end{equation}
With respect to the genus-expansion parameter 
$g_2\equiv  J^2/N$, the diagonal elements begin from the 
zero-th order, while the off-diagonal elements, 
being interpreted as a 2-body to 1-body (or 1-body to 2-body) process, 
starts from the first order. Therefore, we can express the 
general structure of this matrix to the 
first order both in $\lambda'$ and $g_2$ as \cite{Kristjansen et. al., Constable et. al. 1,
Beisert et. al., Constable et. al. 2}
\begin{eqnarray}
\label{two point function matrix}
\langle \bar  O_A(0) O_B(x) \rangle
\equiv G_{AB}+\lambda'\Gamma_{AB}\log(x\Lambda)^{-2}, 
\end{eqnarray}
with 
\begin{eqnarray}
 \hspace{-1cm} G_{AB}=&& \hspace{-0.6cm}
 \left(
 \begin{array}{@{\,}cc@{\,}}
  \delta_{mn} & 0\\
  0 & \delta_{mn}\delta_{yz}\\
 \end{array}
 \right)
+g_2 \left(
 \begin{array}{@{\,}cc@{\,}}
  0 &  C_{n;my}\\
  C_{n;my} & 0\\
 \end{array}
 \right),
\label{GAB}\\
%%%%%%%%%%%%%%
\hspace{-1cm} \Gamma_{AB}=&&\hspace{-0.6cm}
 \left(
 \begin{array}{@{\,}cc@{\,}}
  n^2\delta_{nm} & 0\\
  0 & \frac{m^2}{y^2}\delta_{nm}\delta_{yz}\\
 \end{array}
 \right)
+g_2 \left(
 \begin{array}{@{\,}cc@{\,}}
  0 & (a_{n;my}+b_{n;my}) C_{n;my}\\
  (a_{n;my}+b_{n;my}) C_{n;my}& 0\\
 \end{array}
 \right).\label{GammaAB}
\end{eqnarray}  
Note hat the 2-point functions between a single-trace and 
a double-trace operator are obtained by taking the limit
$x_{23} \sim 1/\Lambda 
\rightarrow 0 $, and also that $1/\Lambda$ is identified with the  short-distance cutoff parameter associated with loops in the sense of Feynman diagrams. 

The specific ansatz proposed in \cite{GMP1} in order 
to relate this mixing matrix  to string-field theory 
vertex $|H_3\rangle_{SV}$ is as follows. 
We first transform the basis for these operators by
$\widetilde O_A=\widetilde U_{AB}O_B$
such that 
\begin{eqnarray}
 \langle \bar{\widetilde O}_{A}(0)\widetilde {O}_{B}(x)\rangle
= \delta_{AB}+{\cal O}(\lambda^{'}) .
\end{eqnarray}
However, this requirement alone does not determine
the basis  completely. To restrict the basis further
the authors demand that
the transformation should be symmetric and real.
Then, the transformation matrix $\tilde{U}$
can be fixed  to the present order ${\cal O}(g_2)$ as
\begin{eqnarray}
 \tilde{U}=1-\frac{g_2}{2}G^{(1)},
\end{eqnarray}
where $G^{(1)}$ is the ${\cal O}(g_2)$ part of $G_{AB}$.
By this change of the basis,
the matrix  $\Gamma_{AB}$
is transformed to
\begin{eqnarray}\label{off-diagonal}
 \Gamma'=\Gamma^{(0)}+ g_2\Gamma^{(1)}-\frac{g_2}{2}\{\Gamma^{(0)},G^{(1)}\} 
\end{eqnarray}
to the present order of approximation. 
The observation of \cite{GMP1}  is  that
the off-diagonal part of this matrix
coincides with the matrix elements 
of the interaction vertex  of the first type, namely, 
$|H_3\rangle_{SV}$. This has been confirmed 
in several cases, for instance, 
for scalar impurities in \cite{GMP1,GMP2,GKT}, and 
for other cases including vectors and fermions in 
\cite{GKT, GT}.  

For the supergravity 
chiral operators, the mixing matrix $\Gamma_{AB}$ 
vanishes. In connection with this, we note that 
in this case the matrix $G_{AB}$ should not be interpreted as representing  operator mixing, as we will touch briefly below.  This is as it should be, since the 3-point functions of the chiral supergravity 
operators are 
not renormalized and hence take 
the standard conformally-invariant 
form even after including the 
higher-order effects. 

 We now show that this ansatz  can be understood 
as a consequence of our holographic string field theory,   
given the observation of \cite{DPPRT}. 
In other words, all existing duality relations are 
actually compatible.  Since our results are 
based on a clear spacetime picture which is lacking 
unfortunately in other approaches, 
checking the consistency 
with more formal correspondences  
seems to provide a useful guide for obtaining 
a unified viewpoint on the holography in the 
plane-wave limit. 

Let us consider a general  class of three general operators
$O_\out(x_\out)$, $O_\inI(x_\inI)$ and $O_\inII(x_\inII)$,
which have the classical conformal dimensions $\Delta^{\langle0\rangle}_r$ satisfying the condition of degeneracy 
$\Delta^{\langle0\rangle}_\out=\Delta^{\langle0\rangle}_\inI+\Delta^{\langle0\rangle}_\inII$. 
The number of impurities contained in $O_\out$
is thus equal to the sum of those in $O_{\inI}$ and $O_{\inII}$. 
Here and in what follows, we use the subscript $\langle0\rangle, \langle 
1\rangle, etc $ to denote the order with respect to 
$\lambda'$. 
The coupling constant $g_2$ with respect to which the 
order is denoted  by the usual subscript $(0), (1), etc$ is 
factored out in order to make clear the order of 
 the quantities we are dealing with. 
The 3-point function takes the following general  form
\begin{eqnarray}\label{general three point function}
&&\hspace{-1.5cm}\langle 
\bar{O}_{\out}(x_\out)O_{\inI}(x_\inI)
O_{\inII}(x_\inII)
\rangle\nonumber\\
&&\hspace{-1cm}= g_2 \Ctree_{123}\left[1-
\lambda'\left\{\aDim_{\inI}\log(x_{\out\inI}\Lambda)^2
+\aDim_{\inII}\log(x_{\out\inII}\Lambda)^2
 +b_{123}\log\left(\frac{x_{\inII\out}x_{\out\inI}
\Lambda}{x_{\inI\inII}}\right)\right\}\right],
\end{eqnarray}
where $C_{123}^{\langle0\rangle}$ 
denotes the 3-point coefficient of the free gauge theory 
and $\aDim_r$ is the anomalous dimension of the
operator $O_{r}(x_r)$ to the first order in $\lambda'$. 
As above, this expression in general 
 does not conform to 
the standard form of the 3-point correlation functions 
of conformal operators. However, we can easily 
check that this is the most general form which would lead 
to the standard form ((\ref{standard}) below)  after taking into account the operator mixing.  
In the ${\cal O}(\lambda')$ part
on the right hand side of  (\ref{general three point function}), 
the first term $\aDim_{\inI}\log(x_{\out\inI}\Lambda)^2$ 
comes from a class of Feynman diagrams
in which the interaction occurs only
between $O_\inI(x_\inI)$ and $O_\out(x_\out)$. Similarly, 
the second term $\Delta^{\langle0\rangle}_3\log(x_{13}\Lambda)^2$ 
comes from the ones where the interaction occurs only
between $O_\inII(x_\inII)$ and $O_\out(x_\out)$,
and the third term $b_{123}\log((x_{\inII\out}x_{\out\inI}
\Lambda)/x_{\inI\inII})$ from the remaining ones,
where the interaction occurs among all three operators.
Supposing that the double-trace operator in consideration 
is obtained from the product of $O_\inI(x_\inI)$ and
$O_\inII(x_\inII)$ by taking the limit $x_{23} \sim 1/\Lambda$, the matrices $G_{AB}$ and $\Gamma_{AB}$ 
in the subspace of operators $O_\out(0)$ and 
$:\hspace{-3pt}O_\inI(x)O_\inII(x)\hspace{-3pt}:$
takes the form
\begin{equation}
\hspace{-3.8cm} G_{AB}=
 \left(
 \begin{array}{@{\,}cc@{\,}}
  1 & 0\\
  0 & 1\\
 \end{array}
 \right)
+g_2 \left(
 \begin{array}{@{\,}cc@{\,}}
  0 &  \Ctree_{123}\\
  \Ctree_{123} & 0\\
 \end{array}
 \right),
\label{general GAB}
\end{equation}
%%%%%%%%%%%%%%
\[
\hspace{-5cm} 
\Gamma_{AB}=
 \left(
 \begin{array}{@{\,}cc@{\,}}
  \aDim_{\out} & 0\\
  0 & \aDim_{\inI}+\aDim_{\inII}\\
 \end{array}
 \right)
\]
\begin{equation}
\hspace{5cm} +g_2 \left(
 \begin{array}{@{\,}cc@{\,}}
  0 & (\Delta_{\inI}^{\langle1\rangle}+\Delta_{\inII}^{\langle1\rangle}+b_{123})C_{123}^{\langle0\rangle}\\
  (\Delta_{\inI}^{\langle1\rangle}+\Delta_{\inII}^{\langle1\rangle}+b_{123})C_{123}^{\langle0\rangle}& 0\\
 \end{array}
 \right),\label{general GammaAB}
\end{equation}
where we have taken into account the fact that the ${\cal O}(g_2^0)$ part
of $\Gamma_{AB}$ gives the anomalous dimensions of
$O_1$ and $:O_2O_3:$.
In the case of two different scalar impurities discussed above,
$\aDim_\inI=m^2/y^2$, $\aDim_\inII=0$, $\aDim_\out=n^2$,
$b_{123}=b_{n;my}=n(n-m/y)$, and $\Ctree_{123}=C_{n;my}$.

Now, in order to extract the correct CFT coefficient 
with the operator mixing being taken into account,  we 
 introduce the transformation matrix $U_{AB}$ which diagonalizes 
the matrix $\langle \bar O_A(0)O_B(x)\rangle$ as a whole. 
Namely, contrary to the previous $\tilde{U}$,  both of the matrices $G_{AB}$ and $\Gamma_{AB}$ are simultaneously diagonalized. It takes the form
\begin{eqnarray}
& U=
 \left(
 \begin{array}{@{\,}cc@{\,}}
  1 & 0\\
  0 & 1\\
 \end{array}
 \right)
+g_2 \left(
 \begin{array}{@{\,}cc@{\,}}
  0 &  D_{123}\\
  E_{123} & 0\\
 \end{array}
 \right),
\end{eqnarray}
where
\begin{equation}
D_{123}
 = \frac{\aDim_{\inI}+\aDim_{\inII}-\aDim_{\out}+b_{123}}
{\aDim_\out-\aDim_\inI-\aDim_\inII}
\,\Ctree_{123},
\qquad E_{123}
 = - \frac{b_{123}}{\aDim_\out-\aDim_\inI-\aDim_\inII}\,\Ctree_{123}.
\end{equation}
Then the 3-point function of the 
operators $O'_A=U_{AB} O_B$ in the new basis  
is given by
\begin{eqnarray}\label{transformed three point function}
%& \langle 
%\bar{O}^{'J}_{n}(x_\out)
%O^{'y J}_{m}(x_\inI)
%O_{\vac }^{(1-y)J}(x_\inII)
%\rangle\nonumber\\
%&= 
%\langle 
%\bar{O}{}^{J}_{n}(x_\out)
%O^{y J}_{m}(x_\inI)
%O_{\vac }^{(1-y)J}(x_\inII)
%\rangle
%+g_2
%\sum_{m',y'}D_{n;m'y'}\langle 
%\bar{T}^{J}_{m'y'}(x_\out)
%O^{y J}_{m}(x_\inI)
%O_{\vac }^{(1-y)J}(x_\inII)
%\rangle+{\cal O}(g_2^2)\nonumber\\[-6pt]
%&=
&& \langle 
\bar{O}'_{\out}(x_\out)
O'_{\inI}(x_\inI)
O'_{\inII}(x_\inII)
\rangle\nonumber\\
&&= 
\langle 
\bar{O}_{\out}(x_\out)
O_{\inI}(x_\inI)
O_{\inII}(x_\inII)
\rangle
+g_2
D_{12'3'}\langle 
\overline{:O_{\inI'} O_{\inII'}:}(x_\out)
O_{\inI}(x_\inI)
O_{\inII}(x_\inII)
\rangle+{\cal O}(g_2^2)\nonumber\\[-6pt]
&&=
g_2( \Ctree_{123}+D_{123})\nonumber\\
&&\quad-\lambda'g_2
\left[
 \left\{2\aDim_\inI(\Ctree_{123}+D_{123})+b_{123}\Ctree_{123}
\right\}\log(x_{\out\inI}\Lambda)\right.\\
&& \hspace{2cm}+\left.\left\{2\aDim_\inII(\Ctree_{123}+D_{123})+b_{123}\Ctree_{123}
\right\}\log(x_{\inII\out}\Lambda)
 -b_{123}\Ctree_{123}\log(x_{\inI\inII}\Lambda)
\right].\nonumber
\end{eqnarray}
Here, 
the mixing effect
 does not affect the third term $\log(x_{\inI\inII}\Lambda)$,
since a correlation function
which contains a double trace operator
at $x_\inI$ or $x_\inII$
gives a ${\cal O}(g_2)$ contribution by itself.
Taking into account the definition of $D_{123}$,
we then obtain the following result 
regardless of the expression of $b_{123}$, which is 
consistent with the canonical form of the 3-point function
for operators with  (anomalous) conformal dimensions
$\aDim_\out$, $\aDim_\inI$, and $\aDim_\inII$, respectively:
\begin{eqnarray}\label{canonical three point function}
%&\langle 
%\bar{O}^{'J}_{n}(x_\out)
%O^{'y J}_{m}(x_\inI)
%O_{\vac }^{(1-y)J}(x_\inII)\rangle\nonumber\\
&&\langle 
\bar{O}'_{\out}(x_\out)
O'_{\inI}(x_\inI)
O'_{\inII}(x_\inII)
\rangle\nonumber\\
&&=g_2 C_{123}
\left[
1-
\lambda'
\left\{
 \left(\aDim_\out+\aDim_\inI-\aDim_\inII\right)\log(x_{\out\inI}\Lambda)
\right.\right. \label{standard}\\
&&\hspace{3cm}\left.\left.
+\left(\aDim_\inII+\aDim_\out-\aDim_\inI\right)\log(x_{\inII\out}\Lambda)
 +\left(\aDim_\inI+\aDim_\inII-\aDim_\out\right)\log(x_{\inI\inII}\Lambda)
\right\}
\right],\nonumber
\end{eqnarray}
where the true CFT coefficient 
$C_{123}=C_{123}^{\langle0\rangle}+D_{123}$ is expressed in terms of $\Ctree_{123}$ and $b_{123}$ as 
\begin{eqnarray}\label{CC}
 C_{123}%= \frac{b_{123} \Ctree_{123}}{\aDim_\out-\aDim_\inI-\aDim_\inII}
= \frac{\lambda' b_{123} \Ctree_{123}}
{\Delta_\out-\Delta_\inI-\Delta_\inII}.
\label{cftcoeff}
\end{eqnarray}

Here, we have used the relation 
$\Delta_\out-\Delta_\inI-\Delta_\inII=\lambda'(\aDim_\out-\aDim_\inI-\aDim_\inII)$,
which is valid for impurity-preserving processes. 
Though we have taken into account the operator 
mixing to higher order with respect to both $g_2$ and $\lambda'$, 
the correction to the 
CFT coefficients thus starts from the lowest order, namely 
the same order as $C^{\langle0\rangle}_{123}$.   

Note that this  argument can be applied
for any kind of impurities, except for the case of 
pure chiral operators where $a_{123}, b_{123}$ and $\Delta_i^{\langle0\rangle}$ all vanish: 
The relation (\ref{CC}) holds 
with different $b_{n;my}$'s depending on impurities \cite{GKT,GT}. Also, for the chiral operators of supergravity, this procedure would lead to a nonsensical result $C_{123}=0$, 
indicating that for the supergravity  BMN operators 
 the matrix $G_{AB}$ cannot be regarded as the  mixing matrix.  

By identifying the true CFT coefficient with the 
3-point vertex $|H_3\rangle_h$  in accordance with
 our result for the holographic string field theory, 
we must have 
\begin{eqnarray}\label{step1}
{\sqrt{J_1J_2J_3}\over N} \bra{1,2,3}\left(
\frac{1}{2\mu}|H_3\rangle_{D}
+\frac{1}{2\mu}|H_3\rangle_{SV}
\right)
=- g_2{\lambda'}\,b_{123} C_{123}^{\langle 0 \rangle}.
\label{cftdsv}
\end{eqnarray}
On the other hand, the result of \cite{DPPRT} 
indicates that 
\begin{eqnarray}\label{step2}
{\sqrt{J_1J_2J_3}\over N} \frac{1}{\mu}\braket{1,2,3}{\HI}_D
&=g_2\left(\Delta_\inI+\Delta_\inII-\Delta_\out\right)\Ctree_{123}. 
%&=-\frac{1}{\mu}\,\left(\aDim_\out-\aDim_\inI-\aDim_\inII\right)\Ctree_{123}
\label{cftbare}
\end{eqnarray}
The off-diagonal 
matrix elements of the $\Gamma$-matrix in the 
particular basis which makes the partial diagonalization 
are given in the form
\begin{eqnarray}
\lambda'\Gamma'_{\rm off}\label{step3}
=\frac{1}{2}\left(\Delta_\inI+\Delta_\inII-\Delta_\out\right)\Ctree_{123}
+\lambda' b_{123}\Ctree_{123} .
\end{eqnarray}
Using the equations (\ref{cftcoeff}), (\ref{cftdsv}) and 
(\ref{cftbare}), we finally find  
\begin{equation}
g_2\lambda'\Gamma'_{\rm off}=-{\sqrt{J_1J_2J_3}\over N}
\frac{1}{2\mu}\braket{1,2,3}{\HI}_{SV}. 
\end{equation}
This is nothing but the claim made in \cite{GMP1,GMP2},
except for the overall factor $-1/2$.
This factor  can be understood 
from the difference of normalization. 
Our  convention differs just 
by this factor from the one adopted in the literature discussing 
this subject;  see, for example, (B.14)  and (B.9) in \cite{GT}. 

It should be remarked that the above relation of 
the 3-point vertex 
$|H_3\rangle_{SV}$  
with the operator mixing in 
perturbation theory at the boundary is 
intrinsically restricted to the 
processes where the numbers of impurities are 
conserved.    For this particular class of processes,  
our argument clarified why 
the correct interaction vertex of the 
holographic string field theory must be the particular  combination 
of the two types of string interaction vertices: 
Roughly speaking, the part $|H_3\rangle_{D}$ describes 
the `bare' part of the interaction of BMN operators, while 
the part $|H_3\rangle_{SV}$ describes the mixing among them.  Both are necessary for  describing the 
processes in the bulk, corresponding to a 
propagation of them from boundary to boundary 
along the tunneling null geodesic.  
Note that the observation of \cite{DPPRT} that the string 
overlap vertex, the bare part of the interaction, 
 in the large $\mu$ limit 
precisely corresponds to the 
free-field contraction is also quite natural 
for impurity preserving processes. 
Through  the holographic string field theory, this natural 
property is related to the specific 
ansatz relating $|H_3\rangle_{SV}$ to the 
matrix of operator mixing, which singles out the 
particular basis of the gauge-theory operators. 

We warn the reader, however, that the above intuitive 
interpretation on the different roles of $|H_3\rangle_{SV}$ and 
$|H_3\rangle_{D}$ does not apply to more general 
processes in which the numbers of impurities are not 
conserved. It is difficult to extend the above argument 
directly to such cases.

\section{Explicit examples} 
The purpose of this section is to present  
 some concrete computations in order 
to confirm our general discussions. In the present work, 
for simplicity we restrict ourselves 
to the cases of two (conserved) impurities, for which 
we can utilize many results by other authors on the gauge theory side. 
It is sufficient to 
focus on the cases of vector, mixed scalar-vector, and  fermionic impurities, since  as we have 
mentioned before the case of pure scalar impurities 
has been practically treated in \cite{CK} 
and was shown to be 
consistent with our holographic relation. We are planning to study more 
general cases, especially the cases where the numbers of 
impurities are not conserved, in a forthcoming work. 
In the present section, 
we denote the first 4-vector indices ($i=1, \ldots, 4$) by Greek letters ($\mu, \nu$) for a clear discrimination between 
the vector and scalar directions.

\subsection{Vector impurities}

Let us begin  from the BMN operators 
with two vector impurities. 
The CFT coefficients on the gauge theory side 
have been already computed  in ref.\cite{GKT}.
Non-trivial processes are listed bellow  
for $(\Delta_\inI+\Delta_\inII-\Delta_\out)C_{123}$:
\begin{eqnarray}
 &&\mu_{m}\nu_{-m}+\vac\to\mu_n\nu_{-n}:
 \qquad
 -\lambda'\C \frac{\sin^2(\pi n y)}
 {y  \pi^2}\frac{m/y}{n-m/y}\label{vector1}\\
 &&\nu_{m}\mu_{-m}+\vac\to\mu_n\nu_{-n}:
 \qquad
 \lambda'\C \frac{\sin^2(\pi n y)}
 { y  \pi^2}\frac{m/y}{n+m/y}\label{vector2}\\
 &&\mu_{m}\mu_{-m}+\vac\to\nu_n\nu_{-n}:
 \qquad
 -\frac{\lambda'}{2}\C \frac{\sin^2(\pi n y)}
 {y  \pi^2}\label{vector3}\\
 &&\mu_{m}\mu_{-m}+\vac\to\mu_n\mu_{-n}:
 \qquad
 -\frac{\lambda'}{2}\C \frac{\sin(\pi n y)}{y \pi^2}
 \frac{n^2+3m^2/y^2}{n^2-m^2/y^2},\label{vector4}
\end{eqnarray}
where the left hand side represents symbolically
the processes with two vector impurities 
with $\mu$ and $\nu$ supposed to denote different vector indices. 
We have already used the fact that 
the difference of the conformal dimensions, 
$\Delta_\inI+\Delta_\inII-\Delta_\out$, 
is given by $\lambda'(m^2/y^2-n^2)$ with $y\equiv J_\inI/J_\out$
at the leading order of small 
$\lambda'=1/(\mu \aout)^2$.
The mode numbers $n$ and $m$ are supposed to satisfy 
$n>0$ and $m\ge0$, respectively.
The overall numerical constant $\C$ is  
the 3-point function for vacuum BMN operators,
$O^{J_i}_{\rm vac}\equiv(J_{i}N^{J_{i}})^{-1/2}{\rm Tr}(Z^{J_i})$
with $J_\inI+J_\inII=J_\out$:
\begin{eqnarray}\label{Normalization}
\C=\frac{\sqrt{J_1J_2J_3}}{N}=g_2\frac{\sqrt{y(1-y)}}{\sqrt{J_\out}}\,.
\end{eqnarray}
Note that the normalization constant of the 
BMN operator has always been
chosen such that  two-point functions  
take the form 
$\langle \bar O^{J}(x) O^J(0)\rangle=
1/|x|^{2J}$. This overall factor coincides with the corresponding  
factor for the 3-point vertex of the holographic 
string field theory, 
as determined in the previous section on the basis 
of the comparison with the supergravity analysis. 

The 3-point vertex on the string-theory side which should be 
compared with is the one with the prefactor with only 
$\sin$ modes, since the scalar part $X_{{\rm I}}^2$ vanishes 
for these cases, 
 \begin{eqnarray}\label{XII operation on E}
\XII^2\ket{\E}
%&=\sum_{r=1}^3\sum_{m=1}^\infty\frac{\omega_m^{(r)}}{\ar}
%a_{-m}^{(r)\dagger} a_{-m}^{(r)}\ket{\E}\\
&=&%\frac{1}{2} 
-\sum_{r,s=1}^{3}\sum_{m,n=1}^{\infty}
\frac{\omega_m^{(r)}}{\ar}
\left(\widetilde N^{rs}_{mn}-\widetilde N^{rs}_{m-n}\right)
\left(\a_{m}^{(r)\dagger} \a_{n}^{(s)\dagger}
+\a_{-m}^{(r)\dagger} \a_{-n}^{(s)\dagger}\right)\ket{\E}
\nonumber \\
&&%\frac{1}{2}
+\sum_{r,s=1}^{3}\sum_{m,n=1}^{\infty}
\left(\frac{\omega_m^{(r)}}{\ar}+\frac{\omega_n^{(s)}}{\as}\right)
\left(\widetilde N^{rs}_{mn}-\widetilde N^{rs}_{m-n}\right)
\a_{m}^{(r)\dagger} \a_{-n}^{(s)\dagger}\ket{\E},
\end{eqnarray}
where we have expressed the formula in terms of the 
exponential basis defined as
\begin{eqnarray}\label{sin/cos-exponential}
 \a_0=a_0,\quad
\a_{n}=\frac{1}{\sqrt{2}}(a_n-ia_{-n}),\quad
\a_{-n}=\frac{1}{\sqrt{2}}(a_n+ia_{-n}),\quad
\end{eqnarray}
which  directly corresponds to the momentum basis of  the BMN operators on the boundary. 

For the first process (\ref{vector1}), assuming  $m$ and $n$ are non-zero, we obtain the following matrix element 
\begin{eqnarray}\label{case1}
&&\hspace{-0.8cm}_{123}\bra{v}
\a_{n}^{(\out)\mu}\a_{-n}^{(\out)\nu}
\a_{m}^{(\inI)\mu}\a_{-m}^{(\inI)\nu}
\,\frac{1}{2\mu}\left(\XII^2\right)\ket{\E}\nonumber\\
&=& 
\frac{1}{2\mu}
\left(
 \frac{\omega_n^{(\out)}}{\aout}(\Nt_{nm}^{\out\inI}-\Nt_{n-m}^{\out\inI})
+\frac{\omega_m^{(\inI)}}{\ainI}(\Nt_{mn}^{\inI\out}-\Nt_{m-n}^{\inI\out})
\right)
\Nt_{-n-m}^{\out\inI}\nonumber \\
&&+\frac{1}{2\mu}
\left(
 \frac{\omega_n^{(\out)}}{\aout}(\Nt_{nm}^{\out\inI}-\Nt_{n-m}^{\out\inI})
+\frac{\omega_m^{(\inI)}}{\ainI}(\Nt_{mn}^{\inI\out}-\Nt_{m-n}^{\inI\out})
\right)
\Nt_{nm}^{\out\inI},
\end{eqnarray}
where the second line comes form 
the case where 
the oscillators $\a_n^{(\out)\mu}$ and $\a_m^{(\inI)\mu}$ 
are contracted through the prefactor $\XII^2$ 
and the operators $\a_{-n}^{(\out)\nu}$ and $\a_{-m}^{(\inI)\nu}$
through the overlap $\ket{E}$,
while the third line comes form the opposite case.
The net results is 
\begin{eqnarray}
\frac{1}{\mu}\left(\frac{\omega_n^{(\out)}}{\aout}+
 \frac{\omega_m^{(\inI)}}{\ainI}\right)
(\Nt_{n-m}^{\out\inI}-\Nt_{nm}^{\out\inI})\Nt_{nm}^{\out\inI}.
\end{eqnarray}
Using the explicit form of the Neumann coefficients 
in the large $\mu$ limit
presented in (\ref{Neumann for large mu}),
we can confirm that
this reduces in this limit to the gauge theory result (\ref{vector1}),
after including the above overall factor. 
 As for the special case $m=0$, 
the absence of zero-mode oscillators in the prefactor $\XII$ 
leads to the vanishing result, which matches the gauge
theory. 

The second case (\ref{vector2}) is related 
by a change $m\to -m$ to the first one.  
On the string theory side, the large $\mu$ behavior 
of Neumann functions $\widetilde N^{rs}_{mn}$ satisfies the 
same relation with respect to  this sign change,  as exhibited 
in  (\ref{Neumann for large mu}). 
Thus the first case ensures the correct matching in 
the second case. 

For the third case (\ref{vector3}) ,  we obtain 
for $n\neq0$ and $m\neq0$, 
\begin{eqnarray}
&&{}_{123}\bra{v}
\a_{n}^{(\out)\mu}\a_{-n}^{(\out)\mu}
\a_{m}^{(\inI)\nu}\a_{-m}^{(\inI)\nu}
\,\frac{1}{2\mu}
\left(\XII^2\right)\ket{\E}\nonumber\\
&&= 
-\frac{\omega_n^{(\out)}}{\mu\aout}(\Nt_{nn}^{\out\out}
-\Nt_{n-n}^{\out\out})\Nt_{m-m}^{\inI\inI}
-\frac{\omega_m^{(\inI)}}{\mu\aout}(\Nt_{mm}^{\inI\inI}
-\Nt_{m-m}^{\inI\inI})\Nt_{n-n}^{\out\out},
\end{eqnarray}
where the first term in the second line comes from 
the contraction of $\a_n^{(\out)\mu}$ and $\a_{-n}^{(\out)\mu}$ 
through the prefactor $\XII^2$ 
and of $\a_{m}^{(\inI)\nu}$ and $\a_{-m}^{(\inI)\nu}$
through the overlap $\ket{E}$,
while the second term comes form the opposite case.
Due to the property of the Neumann coefficients,
the second term  vanishes, and the first term reduces to
the field theory result.
It is useful to notice here that
the gauge theory result for vector impurities in the case 
(\ref{vector3}) is equal to
the scalar correspondent ($\mu, \nu 
\rightarrow i, j$) except for the overall sign, and  
to recall that the scalar case matches the duality relation (\ref{duality})
by using the vertex  $(1/2)\XI^2\ket{\E}$. 
In addition to this, we can easily check that the relation
\begin{equation}
 {}_{123}\bra{v}
\a_{n}^{(\out)\mu}\a_{-n}^{(\out)\mu}
\a_{m}^{(\inI)\nu}\a_{-m}^{(\inI)\nu}
\left(\XI^2+\XII^2\right)\ket{\E}={\cal O}\left(\frac{1}{\mu^2}\right)
\end{equation}
is satisfied in the present `flavor-changing' process.  Thus, 
$\XII^2\ket{\E}$ can be replaced with 
$-\XI^2\ket{\E}$ at the leading order in large $\mu$ limit,
confirming the validity of the duality relation. 
The same is true for the case of $m=0$.

The last case (\ref{vector4}) is the sum of the
above three cases (\ref{vector1}), (\ref{vector2}) 
and (\ref{vector3}) on the gauge theory side.
We can check that the same is true 
for the expressions on the string side: Indeed, 
\begin{equation}
{}_{123}\bra{v}
\a_{n}^{(\out)\mu}\a_{-n}^{(\out)\mu}
\a_{m}^{(\inI)\mu}\a_{-m}^{(\inI)\mu}
\,\frac{1}{2\mu}\left(\XII^2\right)\ket{\E},
\end{equation}
is given by 
summing all possible contractions of 
$\a_{n}^{(\out)\mu}$, $\a_{-n}^{(\out)\mu}$,
$\a_{m}^{(\inI)\mu}$ and $\a_{-m}^{(\inI)\mu}$
through the prefactor $\XII^2$ or the vertex$\ket{\E}$,
which coincide with the sum of
all the three amplitudes considered above on the 
string side in terms of the Neumann functions. 

\subsection{Mixed impurities}
Next, we consider the case with one vector and one scalar impurities, namely, 
$i_{m}\mu_{-m}+\vac\to i_n\mu_{-n}$. 
The gauge theory result for 
$(\Delta_\inI+\Delta_\inII-\Delta_\out)C_{123}$ is 
\cite{GKT} 
\begin{equation}\label{mixed}
\frac{\lambda'}{2}\C \frac{\sin^2(\pi n y)}
{y\pi^2\left(n^2-m^2/y^2\right)}
\left(n+\frac{m}{y}\right)^2.
\end{equation}
On the string theory side,
both the $\XI$ and $\XII$ parts in the prefactor in 
$|H_3\rangle_h$ 
contribute as 
\begin{eqnarray}
&&{}_{123}\bra{v}
\a_{n}^{(\out)i}\a_{-n}^{(\out)\mu}
\a_{m}^{(\inI)i}\a_{-m}^{(\inI)\nu}
\,\frac{1}{2\mu}\left(\sum_{i=5}^{8}(\XI^i)^2
 +\sum_{\mu=1}^{4}(\XII^\mu)^2\right)\ket{\E}\\
&&=\frac{1}{2\mu}
\left(\frac{\omega_n^{(\out)}}{\aout}+
 \frac{\omega_m^{(1)}}{\ainI}\right)
(\Nt^{\out\inI}_{m-n}+\Nt^{\out\inI}_{mn})\Nt^{\out\inI}_{mn}
-\frac{1}{2\mu}
\left(\frac{\omega_n^{(\out)}}{\aout}+
 \frac{\omega_m^{(\inI)}}{\ainI}\right)
(\Nt^{\out\inI}_{m-n}-\Nt^{\out\inI}_{mn})\Nt^{\out\inI}_{mn}\nonumber,
\end{eqnarray}
where the first part in the second line 
is the contribution from $(\XI^i)^2\ket{E}$ and 
the second form $(\XII^\mu)^2\ket{E}$, and the net result is
\begin{equation}
\frac{1}{\mu}\left(\frac{\omega_n^{(\out)}}{\aout}+
 \frac{\omega_m^{(\inI)}}{\ainI}\right)
\Nt^{\out\inI}_{mn}\Nt^{\out\inI}_{mn}.
\end{equation}
We can easily confirm that 
in the large $\mu$ limit this precisely reduces to the field theory result (\ref{mixed}).

\subsection{Fermionic impurities}
 We first explain the convention for representing 
spinor impurities. We essentially  follow the ref.\cite{GT}. Decompose the SU(4) R-symmetry group as SU(4) $=$ SO(4)$\times$U(1)
 $=$ SU(2)$\times$SU(2)$\times$U(1) 
with $U(1)$ being the subgroup corresponding to large 
orbital angular momentum $J$: 
\[
 {\bf 4} \to ({\bf 2},{\bf 1})_{+}+({\bf 1},{\bf 2})_{-}, 
\]
where the subscript $\pm$  represents the U(1) charge. 
With this decomposition of the R-charge index,
the correspondence between the fermionic fields
$\lambda^A_{\a}$ and $\bar\lambda^A_{\dot\a}$,
with $A$ and $\a$ being R-charge and Lorentz spinor
indices, and the string theory creation operators 
$b^{\dagger}_{\a_1\a_2}$ and $b^{\dagger}_{\dot\a_1\dot\a_2}$
with $SO(8) \, (=SO(4)\times SO(4)=SU(2)\times SU(2)\times SU(2)\times SU(2))$
indices is given by
\begin{eqnarray}
&\lambda^{A}_{\a}\to (\lambda_{r\a,1/2},\lambda_{\dot r\a,-1/2}) 
\sim (b^\dagger_{\a_1\a_2},b^\dagger_{\dot\a_1\a_2}),\\
&\bar \lambda^{A}_{\dot\a}\to 
(\bar\lambda_{r\dot\a,-1/2},\bar\lambda_{\dot r\dot\a,1/2})
\sim (b^\dagger_{\a_1\dot\a_2},b^\dagger_{\dot\a_1\dot\a_2}). 
\end{eqnarray}
The original SU(4) index $A$ of $\lambda_{\alpha}^A$ is represented by the set $(r, 1/2)+(\dot{r},-1/2)$ 
and similarly of $\overline{\lambda}^A_{\alpha}$ by its conjugate
$(r, -1/2)+(\dot{r},1/2)$. Note that the range of indices are 
$r=3,4$, $\dot r=1,2$, $\a_1,\a_2=1,2$, 
and $\dot\a_1, \dot\a_2=1,2$. On the right end of the above 
symbolic relation, the fermion oscillators are represented 
by the indices ($r\sim \alpha_1, \alpha \sim \alpha_2)$ etc., 
with $\alpha$ being originally 
the spinor indices of 4D target space 
on the boundary. 
We call the sector with the U(1) charge 
$J=1/2$ a BMN fermion
while the one with $J=-1/2$ an anti-BMN fermion,
and we will focus on the former. The SU(2) indices 
are contracted in the standard way   
by the $\epsilon$ symbol
 $\epsilon_{\a\b}$ and $\epsilon^{\a\b}\equiv\epsilon^{-1}_{\a\b}$.

As for two fermionic impurities,
it is sufficient  to consider the four types listed bellow,
accompanied with the corresponding gauge theory results 
for $(\Delta_\inI+\Delta_\inII-\Delta_\out)C_{123}$ 
which can be extracted from the work \cite{GT}:
\begin{eqnarray}
&&\lambda_{31m}\lambda_{32-m}+\vac\to \lambda_{31n}\lambda_{32-n}:
\qquad -\lambda'\C\frac{\sin^2(\pi n y)}{ y\pi^2}
\frac{n}{n-m/y}\label{fermion1},\\
&&\lambda_{31m}\lambda_{31-m}+\vac\to \lambda_{31n}\lambda_{31-n}: 
\qquad-\lambda'\C\frac{\sin^2(\pi n y)}{ y\pi^2}\frac{2nm/y}{n^2-m^2/y^2},\\
&&\lambda_{31m}\lambda_{42-m}+\vac\to \lambda_{31n}\lambda_{42-n}:
\qquad-\frac{\lambda'}{2}\C\frac{\sin^2(\pi n y)}{ y\pi^2}\frac{n+m/y}{n-m/y}\label{fermion3},\\
&&\lambda_{31m}\lambda_{41-m}+\vac\to \lambda_{31n}\lambda_{41-n}: 
\qquad-\lambda'\C\frac{\sin^2(\pi n y)}{ y\pi^2}\frac{m/y}{n-m/y}\label{fermion4}.
\end{eqnarray}

On the string theory side, the contribution from
the $|H_3\rangle_{SV}$ part in the interaction vertex is already calculated in \cite{GT} as 
\begin{eqnarray}
\mu^{-1}\bra{v}
\b_{-n}^{(1)12}\b_{n}^{(1)11}
\b_{11,m}^{(2)}\b_{12,-m}^{(2)}
|H_3\rangle_{SV}
\label{fermionic-SV-case1}
&=&-C_N \frac{\sin^2(\pi n y)}{(\mu\aout)^2\pi^2 y}, \\
\mu^{-1}\bra{v}
\b_{-n}^{(1)11}\b_{n}^{(1)11}\b_{11,m}^{(2)}\b_{11,-m}^{(2)}
|H_3\rangle_{SV}
&=&0\label{fermionic-SV-case2}, \\
\mu^{-1}\bra{v}
\b_{-n}^{(1)22}\b_{n}^{(1)11}\b_{11,m}^{(2)}\b_{22,-m}^{(2)}
|H_3\rangle_{SV}
&=&0\label{fermionic-SV-case3}, \\
\mu^{-1}\bra{v}
\b_{-n}^{(1)21}\b_{n}^{(1)11}\b_{11,m}^{(2)}\b_{21,-m}^{(2)}
|H_3\rangle_{SV}
&=&C_N \frac{\sin^2(\pi n y)}{(\mu\aout)^2\pi^2 y},
\label{fermionic-SV-case4}
\end{eqnarray}
where $\b_n$ is the exponential basis of each string defined
in a similar manner as $\a_n$:
\begin{equation}
\b_0=b_0,\qquad
\b_{n}=\frac{1}{\sqrt{2}}(b_n-ib_{-n}),\quad
\b_{-n}=\frac{1}{\sqrt{2}}(b_n+ib_{-n}).
\end{equation}
For these cases of pure fermionic external states 
with undotted indices,
the prefactor reduces to
\begin{equation}
 P_{SV}=\frac{\mu}{3}Y^4,
\end{equation}
where $Y^4$ is defined as $Y^4\equiv Y^2_{\a_1\b_2}Y^{2\a_1\b_2}=12Y_{11}Y_{12}Y_{21}Y_{22}$
with $Y^2_{\a1\b1}=Y_{\a1\gamma_2}Y_{\b1}^{\gamma_2}$. See the 
Appendix B for the definitions of these quantities.  
Using the explicit form of $Y^4$,
we can easily confirm  that the amplitudes on the 
string side  vanish in 
both case of (\ref{fermionic-SV-case2}) and (\ref{fermionic-SV-case3}), and that 
(\ref{fermionic-SV-case4}) is equal to 
the minus of (\ref{fermionic-SV-case1}).
With the definition of $Y^{\a_1\a_2}$ in (\ref{YZ}),
the string-matrix element (\ref{fermionic-SV-case1}) is given by
$-(\overline{G}_n^{(\out)}\overline{G}_m^{(\inI)})^2$, which reduces 
in the large $\mu$ limit 
to the right hand side of (\ref{fermionic-SV-case1})
using the asymptotic form
for $\overline{G}^{(r)}$ given in (\ref{asymptotic forms}).

Next turning to 
the contribution from the $|H_3\rangle_D$ part,  
we first find 
\begin{eqnarray}
\mu^{-1}\bra{v}
\b_{-n}^{(\out)11}\b_{n}^{(\out)11}\b_{11,m}^{(\inI)}\b_{11,-m}^{(\inI)}
|H_3\rangle_D &=&\left(\frac{2\omega_n^{(\out)}}{\mu\aout}+
 \frac{2\omega_m^{(\inI)}}{\mu\ainI}\right)\times
\nonumber \\
 &&\hspace{-1cm}
\left(
-\frac{1}{4}(Q_{nm}^{\out\inI}+Q_{mn}^{\inI\out})^2
+\frac{1}{4}(Q_{nm}^{\out\inI}-Q_{mn}^{\inI\out})^2
\right)\nonumber\\
&=&-C_N\frac{\sin^2(\pi n y)}{(\mu\aout)^2 y\pi^2}\,\frac{4nm/y}{n^2-m^2/y^2},
\end{eqnarray}
where we have used the relation $Q_{nm}^{\out\inI}=-N^{\out\inI}_{nm}$,
$Q_{mn}^{\inI\out}=-N^{\inI\out}_{-m-n}$ in the large $\mu$ limit,
which can be easily shown by the definition of $Q^{rs}_{nm}$, (\ref{fermionic Neumann}), and the asymptotic form
of $U^{(r)}$, (\ref{asymptotic forms}), given in the 
Appendix B. 
For other three cases  (\ref{fermion1}), (\ref{fermion3}) and (\ref{fermion4}), we find that 
the $|H_3\rangle_D$ contribution gives one and the same expression 
\begin{equation}\label{fermion1-result}
 \left(\frac{2\omega_n^{(\out)}}{\aout}+
 \frac{2\omega_m^{(\inI)}}{\ainI}\right)
\frac{1}{4}\left(Q_{mn}^{\inI\out}-Q_{nm}^{\out\inI}\right)^2
=-C_N \frac{\sin^2(\pi n y)}{(\mu\aout)^2 y \pi^2}\,\frac{n+m/y}{n-m/y}. 
\end{equation}
 Combining these two types of contributions in each case,
we obtain the matrix elements 
which precisely coincide with the gauge theory results 
(\ref{fermion1})-(\ref{fermion4}). 

\section{Concluding remarks}
Finally, we remark on some relevant 
 problems left in the present paper 
and on possible future directions. 
First of all, we have to emphasize again that
 our main predictions are not 
restricted to the impurity-preserving processes which almost all of other works have been  limited to. 
In the supergravity sector,  our holographic relation 
summarized in 
(\ref{clambdarelation1}) $\sim$ 
(\ref{clambdarelation3}) is  
valid by its construction for general processes. 
The extremal cases of the supergravity sector where conformal dimensions 
are preserved $\Delta_1=\Delta_2+ \Delta_3$ are generalized by lifting the degeneracy, owing to the 
higher-order effects in $\lambda'$, 
 to the impurity preserving processes including 
  string excitation modes. 
In the last two sections, we have confirmed that our relation 
is indeed valid in this case with nontrivial stringy effects. 
It is quite plausible that 
the relation should then be naturally extended to 
impurity non-preserving sectors. We can note, for example, 
 that the $\alpha'$-correction factor $f$ in the holographic relation is consistent with the fact that the 
CFT coefficients for such cases start from 0-th order 
in $1/\mu$ just as the impurity-preserving processes, while the 3-point string vertices for such cases in general start at most 
from the first order in $1/\mu$, because of 
 the large $\mu$ behavior 
of the Neumann functions. In fact, it is easy to confirm that 
our ansatz gives the correct results for a 
few simple cases. However, a systematic check 
of more general classes of 
impurity non-preserving processes is beyond the scope of the 
present paper and is left to a forthcoming work. 

There are many possible directions following the present work: 
For instance, in connection with the question of impurity nonpreserving 
processes, we should investigate the string-loop corrections. 
In most of the existing literature, impurity nonpreserving 
contributions have been ignored often without appropriate 
justification. 
It  would also be interesting to extend the 
discussions of section 2 and 3 to higher-point 
correlation functions, 
from both standpoints of supergravity limit and 
full string theory, and to see to what extent the structure suggested 
in our first work \cite{dsy} is realized in higher orders. 
Another important problem is to derive 
the holographic string-field theory in conjunction with  
our prescription of the holographic duality directly 
from the gauge-theory side. For such an attempt, the 
collective-field approach discussed in \cite{jev} seems to be 
suggestive. 

\vspace{0.5cm}
\noindent
{\large Acknowledgements}

We would like to thank H. Shimada for discussions at an 
early stage of the present work. 
The present work is supported in part by Grant-in-Aid for Scientific 
Research (No. 13135205 (Priority Areas) and No. 16340067 (B))  from the Ministry of  Education,
Science and Culture.

%\newpage
\appendix 
\section{Reduction to SO(4) basis}
\setcounter{equation}{0}
\renewcommand{\theequation}{\Alph{section}.\arabic{equation}}
\renewcommand{\thesubsection}{\Alph{section}.\arabic{subsection}}
First we summarize the definitions of overlap integrals 
of SO(6) harmonics, 
\begin{equation}
Y^I= {\cal C}_{i_1\cdots i_k}^I x^{i_1}\cdots x^{i_k}
\end{equation}
following the convention of 
ref. \cite{lmrs}. 
Using the formula for the integration over the 5-dimensional unit sphere $S^5$ ($\omega=\pi^3$ = the area of a unit 5-sphere), 
\begin{equation}
{1\over \omega_5}\int_{S^5}x^{i_1}\cdots x^{i_{2m}}
={2^{1-m}\over (m+2)!}\times 
(\mbox{All possiblle  contractions}), 
\end{equation}
we derive 
\[
\int_{S^5}Y^{I_1}Y^{I_2}=
{\cal C}_{i_1\cdots i_k}^{I_1}{\cal C}_{j_1\cdots j_k}^{I_2}
\int_{S^5}x^{i_1}\cdots x^{i_k} x^{j_1}\cdots x^{j_k}
\]
\begin{equation}
=\pi^3 {2^{1-k}\over (k+2)!}k! \langle {\cal C}^{I_1}{\cal C}^{I_2}\rangle 
=\pi^3{\delta^{I_1I_2} \over 2^{k-1} (k+1)(k+2)},  
\label{so6harmonorm}
\end{equation}
and 
\begin{equation}
\int_{S^5}
Y^{I_1}Y^{I_2}Y^{I_3}=\pi^3 
{2^{-(\Sigma -2)/2}
\over ({\Sigma\over 2} +2)!}
{k_1! k_2!k_3! \over \alpha_1 ! \alpha_2 ! \alpha_3!}
\langle {\cal C}^{I_1}{\cal C}^{I_2}{\cal C}^{I_3}\rangle
\equiv a(k_1,k_2,k_3) \langle {\cal C}^{I_1}{\cal C}^{I_2}{\cal C}^{I_3}\rangle,
\label{3pointso6harmo} 
\end{equation}
since in this case $m=\Sigma/2$ and ${k_1! k_2!k_3! \over \alpha_1 ! \alpha_2 ! \alpha_3!}
$ is equal to the number of same contractions 
occur, due to the total symmetry of the tensors ${\cal C}^I$. 

Now we convert these integrals on $S^5$  into a Gaussian 
average for the SO(4) directions by taking the 
large J limit. Such a calculation has 
previously done in ref. \cite{polman} for some special cases.  
First by expressing the $S^5$ harmonics 
in terms the $S^3$ harmonics $\tilde{Y}^{I_1}$, 
\begin{equation}
Y^{I} =2^{-J/2}\sqrt{{(J+\tilde{k})!\over J!\tilde{k}!} }
\e^{iJ\psi} \cos^J\theta \sin^{\tilde{k}_1}\theta \, \, \tilde{Y}^{I_1}
\end{equation}
\begin{equation}
Y^{\overline{I}} =2^{-J/2}\sqrt{{(J+\tilde{k})!\over J!\tilde{k}!} }
\e^{-iJ\psi} \cos^J\theta \sin^{\tilde{k}_1}\theta \, \, \tilde{Y}^{\overline{I}}
\end{equation}
Let us first check this formula by 
confirming the normalization integral. 
\begin{equation}
\int_{S^5}Y^{\overline{I}}Y^{I}
=2^{-J}{(J+\tilde{k})!\over J!\tilde{k}!}
\int_0^{2\pi}d\psi \int_{-\pi/2}^{\pi/2}d\theta 
\cos \theta |\sin\theta|^3
\cos^{2J} \theta \sin^{2\tilde{k}}\theta 
\int_{S^3}
\tilde{Y}^{\overline{I}}\tilde{Y}^{I}
\end{equation}
In the limit $J\rightarrow \infty$, this can be 
evaluated by making a change of integration variables  
from $\theta$ and the Cartesian coordinates 
$x^i$ on $S^3 \, (\sum_{i=1}^3 (x^i)^2=1) $ to a 
4-vector $y^0=\theta, y^i = \theta x^i$ as 
\begin{eqnarray}
\lim_{J\rightarrow \infty}\int_{S^5}Y^{\overline{I}}Y^{I}
&=&\lim_{J\rightarrow \infty}2^{-J}{(J+\tilde{k})!\over J!\tilde{k}!}
2\pi \int d^4 y \, \e^{-Jy^2}|y^{\tilde{k}}\tilde{Y}^I|^2 \\
&=&2^{-J}{(J+\tilde{k})!\over J!\tilde{k}!}2\pi 
{\pi^2 \over (J+{1\over 2})^2}
{\tilde{k}!\over (2(J+{1\over 2}))^{\tilde{k}}}
\langle \tilde{{\cal C}}^I\tilde{{\cal C}}^I
\rangle
={\pi^3 \over 2^{J+\tilde{k}-1}J^2}
\end{eqnarray}
which coincides with the (\ref{so6harmonorm}). 

Similarly,   the 3-point integral is found to be
\[
\int_{S^5}
Y^{\overline{I}_1}Y^{I_2}Y^{I_3}=2^{-J_1}
\Big[
{(J_1+\tilde{k}_1)! (J_2 +\tilde{k}_2)! (J_3 + \tilde{k}_3)! 
\over J_1!\tilde{k}_1!J_2!\tilde{k}_2!
J_3!\tilde{k}_3!}
\Big]^{1/2}\]
\[\times 
2\pi \int d\theta \cos\theta
|\sin\theta|^3 \cos^{J_1}\theta \sin^{\tilde{k}_1}\theta
\cos^{J_2}\theta \sin^{\tilde{k}_2}\theta\cos^{J_3}\theta \sin^{\tilde{k}_3}\theta
\]
\begin{equation}
\rightarrow 
\pi^3 {1\over 2^{-1+{\Sigma\over 2} } J_1^2}
\Big({J_2\over J_1}\Big)^{\tilde{k}_2/2}
\Big({J_3 \over J_1}\Big)^{\tilde{k}_3/2}
{\sqrt{\tilde{k}_1!\tilde{k}_2!\tilde{k}_3!}\over 
\tilde{\alpha}_1!\tilde{\alpha}_2!\tilde{\alpha}_3!}
\langle \tilde{{\cal C}}^{\overline{I}_1}
\tilde{{\cal C}}^{I_2}\tilde{{\cal C}}^{I_3}\rangle. 
\end{equation}
Equating this result with the large $J_1$ limit of (\ref{3pointso6harmo}) which is equal to 
\begin{equation}
\pi^3 {1\over 2^{-1+{\Sigma\over 2} } J_1^2}
{1\over  \tilde{\alpha}_1!}
\Big({J_2 J_3\over J_1}
\Big)^{\tilde{\alpha}_1}
\langle {\cal C}^{\overline{I}_1}{\cal C}^{I_2}
{\cal C}^{I_3}\rangle, 
\end{equation}
we find the relation between the 
SO(6) and SO(4) contractions ($\tilde{\alpha}_1=\alpha_1=
(\tilde{k}_2 +\tilde{k}_3 -\tilde{k}_1)/2$), 
\begin{equation}
\langle{\cal C}^{\overline{I}_1}{\cal C}^{I_2}
{\cal C}^{I_3}\rangle 
= \alpha_1!
\Big({J_1\over J_2 J_3}\Big)^{\tilde{\alpha}_1}
\Big({J_2\over J_1}\Big)^{\tilde{k}_2/2}
\Big({J_3 \over J_1}\Big)^{\tilde{k}_3/2}
 {\sqrt{\tilde{k}_1!\tilde{k}_2!\tilde{k}_3!}\over 
\tilde{\alpha}_1!\tilde{\alpha}_2!\tilde{\alpha}_3!}
\langle \tilde{{\cal C}}^{\overline{I}_1}
\tilde{{\cal C}}^{I_2}\tilde{{\cal C}}^{I_3}\rangle. 
\end{equation}
This is used in section 3 in order to express the 
3-point coupling in terms of the SO(4) variables explicitly. 

\section{Explicit form of string-field vertices}
\setcounter{equation}{0}
 In this appendix, we summarize various formulas  
which are necessary for our arguments in 
sections 5 and 6. We hope that this is useful 
to make the expositions of the present paper self-contained.  
\subsection{Preliminaries}
First, the Fourier mode expansions in terms of sin/cos basis
of  bosonic coordinate $x^{(r)}(\sigma_r)$
and  bosonic momentum $p^{(r)}(\sigma_r)$, as well as the
fermionic ones $\lambda^{(r)}(\sigma_r)$ and $\theta^{(r)}(\sigma_r)$,
are given by
\begin{eqnarray}
&&\hspace{-1cm} 
x^{(r)}(\sigma_r)=x^{(r)}_0+\sqrt{2}\sum_{n=1}^{\infty}
  \left(x^{(r)}_n \cos \frac{n\sigma_r}{|\ar|}
   +x^{(r)}_{-n}\sin\frac{n\sigma_r}{|\ar|}\right),\\
&&\hspace{-1cm} 
p^{(r)}(\sigma_r)=\frac{1}{2\pi|\ar|}
  \left[
   p^{(r)}_0+\sqrt{2}\sum_{n=1}^{\infty}
   \left(p^{(r)}_n \cos \frac{n\sigma_r}{|\ar|}
    +p^{(r)}_{-n}\sin\frac{n\sigma_r}{|\ar|}\right)
      \right],\\
&&\hspace{-1cm}
 \theta^{(r)}(\sigma_r)=\theta^{(r)}_0+\sqrt{2}\sum_{n=1}^{\infty}
  \left(\theta^{(r)}_n \cos \frac{n\sigma_r}{|\ar|}
   +\theta^{(r)}_{-n}\sin\frac{n\sigma_r}{|\ar|}\right),\\
&&\hspace{-1cm} \lambda^{(r)}(\sigma_r)=\frac{1}{2\pi|\ar|}
  \left[
   \lambda^{(r)}_0+\sqrt{2}\sum_{n=1}^{\infty}
   \left(\lambda^{(r)}_n \cos \frac{n\sigma_r}{|\ar|}
    +\lambda^{(r)}_{-n}\sin\frac{n\sigma_r}{|\ar|}\right)
      \right],
\end{eqnarray}
where\footnote{
The definition of the oscillators $a_{\pm n}^{(r)}$ 
is different from the usual one in the literature 
by a factor $-i$.
We use this definition since it is the 
appropriate one in the supergravity limit as we have 
discussed in section 3.}
\begin{eqnarray}
&&\hspace{-1cm}
x_n^{(r)}=\sqrt{\frac{\a'}{2\omega_n^{(r)}}}
\left(a_n^{(r)}+a_n^{(r)\dagger }\right),
\qquad
p_n^{(r)}=-i\sqrt{\frac{\omega_n^{(r)}}{2\a'}}
\left(a_n^{(r)}-a_n^{(r)\dagger }\right),\\
&&\hspace{-1cm}
\theta_0^{(r)}=\frac{\sqrt{\a'}}{2\sqrt{|\ar|}}
\left[
\frac{1+\Pi}{2}
\left\{(1+e(\ar))b_0^{(r)}
-(1-e(\ar)b_0^{(r)\dagger })\right\}\right.\nonumber\\
&&\hspace{3cm}+\left.\frac{1-\Pi}{2}
\left\{(1-e(\ar))b_0^{(r)}
+(1+e(\ar)b_0^{(r)\dagger })
\right\}\right],\\
&&\hspace{-1cm}
\theta_n^{(r)}=\frac{\sqrt{\a'}}{\sqrt{2|\ar|}}
\sqrt{\frac{n}{\omega_n^{(r)}}}
\left[
\frac{1+\Pi}{2}
\left\{U_{(r)n}^{-1/2}b_n^{(r)}
+e(\ar n)U_{(r)n}^{1/2}b_{-n}^{(r)\dagger})\right\}\right.\nonumber\\
&&\hspace{3cm}+\left.\frac{1-\Pi}{2}
\left\{U_{(r)n}^{1/2}b_n^{(r)}
+e(\ar n)U^{-1/2}_{(r)n}b_{-n}^{(r)\dagger})
\right\}\right] \quad (n\neq 0),\\
&&\hspace{-1cm}\lambda^{(r)}_n=\frac{|\ar|}{\a'}\theta_n^{(r)\dagger},
\end{eqnarray}
with the ordinal (anti-)commutation relations 
\begin{eqnarray}
[a_m^{(r)i},a_n^{(s)j\dagger}]=\delta^{rs}\delta^{ij}\delta_{mn},
\qquad
\{b^{(r)a}_m,b^{(s)b}_n\}=\delta^{rs}\delta^{ab}\delta_{mn}.
\end{eqnarray}
Here, $e(x)\equiv x/|x|$, 
$\omega_n^{(r)}\equiv\sqrt{n^2+\mu^2\ar^2}$,
$U_{(r)n}\equiv (\omega_{n}^{(r)}-\mu\ar)/|n|$,
and $\Pi\equiv \gamma^1\gamma^2\gamma^3\gamma^4$,
with $\gamma^i$ being SO(8) gamma matrices. 
With these Fock space basis, the free string Hamiltonian
for $r$-th string 
\begin{eqnarray}
 &&\hspace{-1cm}H^{(r)}=
\frac{1}{2}\int_0^{2\pi|\ar|}d\sigma\left[
2\pi\alpha' p^{(r)2}+\frac{1}{2\pi\alpha'}(\partial_\sigma x^{(r)})^2
+\frac{1}{2\pi\alpha'}\mu^2x^{(r)2})\right]\\
&&+\frac{1}{2}\int_{0}^{2\pi|\ar|}d\sigma\left[
-2\pi \a' \lambda^{(r)}\partial_\sigma \lambda^{(r)}+
\frac{1}{2\pi\a'}\theta^{(r)}\partial_\sigma\theta^{(r)}
+2\mu\lambda^{(r)}\Pi\theta^{(r)}\right]\nonumber
\end{eqnarray}
reduces to
\begin{eqnarray}
 H=\sum_{n=-\infty}^{\infty}
\frac{\omega_n^{(r)}}{|\ar|}
\left(
a_n^{(r)\dagger}a_n^{(r)}+b_n^{(r)\dagger}b_n^{(r)}
\right).
\end{eqnarray}
\subsection{{Overlap vertex}}
The overlap vertex takes the form 
\begin{eqnarray}
 \ket{\E}=\ket{E_a}\ket{E_b}\,\delta\left(\sum_{r=1}^3\ar\right),
\end{eqnarray}
where $\ket{E_a}$ and $\ket{E_b}$ are
the bosonic and fermionic overlap vertices which satisfy
\begin{eqnarray}
 \sum_{r=1}^{3}\tilde p^{(r)}(\sigma)\ket{E_a}=0,
\qquad \sum_{r=1}^{3}e(\ar)\tilde x^{(r)}(\sigma)\ket{E_a}=0,\\
 \sum_{r=1}^{3}\tilde \lambda^{(r)}(\sigma)\ket{E_b}=0,
\qquad \sum_{r=1}^{3}e(\ar)\tilde \theta^{(r)}(\sigma)\ket{E_b}=0.
\end{eqnarray}
Here,  $\tilde p^{(r)}(\sigma)\,\,(|\sigma|\le |\aout|)$
is defined as 
$\tilde p^{(r)}(\sigma)\equiv\Theta_r(\sigma)p^{(r)}(\sigma_r)$
with $\Theta_\inI(\sigma)=\theta(\pi\ainI-|\sigma|)$,
$\Theta_\inII(\sigma)=\theta(|\sigma|-\pi\ainI)$, and $ \Theta_\out=1$.
The parameter $\sigma_r$ is defined as
\begin{eqnarray}
 &&\hspace{-1cm}  \sigma_\inI=\sigma \hspace{3cm}~~ 
    -\pi\ainI\le\sigma\le\pi\ainI\\
 &&\hspace{-1cm} \sigma_\inII=
   \left\{
    \begin{array}{l}
     \sigma-\pi \ainI 
      \qquad\qquad~ \pi\ainI\le\sigma\le\pi(\ainI+\ainII)\\
     \sigma+\pi \ainI 
      \qquad\qquad -\pi(\ainI+\ainII)\le\sigma\le-\pi\ainI\\
    \end{array}
   \right.\\
 &&\hspace{-1cm} \sigma_\out=-\sigma \hspace{3cm}      
   -\pi(\ainI+\ainII)\le\sigma\le\pi(\ainI+\ainII).
\end{eqnarray}
The definitions of $\tilde x_{(r)}(\sigma)$, 
$\tilde \lambda_{(r)}(\sigma)$
and $\tilde \theta_{(r)}(\sigma)$ are  given in the same way.
We always assume $\ar(\equiv\a'p_{(r)}^+)$ satisfies the relation
$\ainI,\ainII>0,\,\,\aout<0$.

The explicit form of the overlap vertex is 
\begin{eqnarray}
 &&\hspace{-1cm}\ket{E_a}=\exp
 \left[
 -\frac12 
 \sum_{r,s=1}^{3} 
 a_m^{(r)\dagger} N_{mn}^{rs} a_n^{(s)\dagger}
 \right]\ket{v_a}_{123},\\
&&\hspace{-1cm}\ket{E_b}=
 \exp\left[
 \sum_{r,s=1}^{3}\sum_{m,n=0}^{\infty}
\left(
b_{-m}^{(r)\a_1\a_2\dagger}Q_{mn}^{rs}b_{n\a_1\a_2}^{(s)\dagger}
+b_{m}^{(r)\dot\a_1\dot\a_2\dagger}
Q_{mn}^{rs}b_{-n\dot\a_1\dot\a_2}^{(s)\dagger}
\right)\right]\ket{v_b}_{123}.
\end{eqnarray}
This overlap vertex is based on 
the ground states $\ket{v_a}_{123}$ and $\ket{v_b}_{123}$
which are defined by $a^{(r)}_{n}\ket{v_a}_{123}=0$ and
$b^{(r)}_{n}\ket{v_b}_{123}=0$ for $n\in Z$.
Note that $\ket{E_b}$ is constructed on
the Fock vacuum $\ket{v_b}$ \cite{CKPRT,CPRT},
not on the SO(8) vacuum
$\ket{0}$, defined as $a_n^{(r)}\ket{0}=0\,(n\in Z)$,
$b_n^{(r)}\ket{0}=0\,(n\neq0)$
and $\theta_0^{(r)}\ket{0}=0$,
on which the original fermionic interaction vertex 
\cite{SV2,Pan1,PS1} was constructed.
As for the fermionic sector, 
the SO(8) spinor indices have been decomposed as SO(8) $=$
SO(4)$\times$SO(4) $=$
SU(2)$\times$SU(2)$\times$SU(2)$\times$SU(2),
according to the works \cite{Pan2,PS2}.

%The details are referred to \cite{Pan2,PS2}.
The bosonic Neumann coefficients $N_{mn}^{rs}$, $N_m^r$ and
the fermionic Neumann coefficients $Q^{rs}_{mn}$, $Q^{r}_{m}$
are given by 
\begin{eqnarray}
 N_{00}^{r's'}&=&(1-4\mu \a K)
 \left(\delta^{r's'}+\frac{\sqrt{\arp\asp}}{\aout}\right),\\
 N_{00}^{r'\out}&=&\delta^{r'\out}-\sqrt{-\frac{\arp}{\aout}},\\
 N_{m0}^{rs'} 
 &=&  -\sqrt{2\mu\asp}\epsilon_{s't'}\atp(C_{(r)}^{1/2}N^{r})_{m},\\
 N^r_m&=&-(C^{-1/2}A^{(r)\rm T}\Gamma^{-1}B)_m,\\
 N_{mn}^{rs} 
 &=& \delta^{rs}\delta_{mn}
 -2 (C_{(r)}^{1/2}C^{-1/2}A^{(r)\rm T }
\Gamma^{-1}A^{(s)}C^{-1/2}C_{(s)}^{1/2})_{mn},\\
 N^{rs}_{-m-n}&=&-(U_{(r)}N^{rs}U_{(s)})_{mn}, \\
 Q_{mn}^{rs}&=&e(\ar)\sqrt{\frac{|\as|}{|\ar|}}\label{fermionic Neumann}
\left(U_{(r)}^{1/2}C^{1/2}N^{rs}C^{-1/2}U^{1/2}_{(s)}\right)_{mn},\\
Q_{m0}^{rr'}&=&-\epsilon_{r't'}\sqrt{\arp}\atp\frac{e(\ar)}{\sqrt{|\ar|}}
\left(U^{1/2}_{(r)}C_{(r)}^{1/2}C^{1/2}N^r\right)_m, \\
Q_{00}^{\out r'}&=&-Q_{00}^{r'\out}=\frac{1}{2}\sqrt{-\frac{\arp}{\aout}},\\
{\rm Otherwise}&=&0,
\end{eqnarray}
%Note that $Q^{rs}_{mn}\neq Q^{sr}_{nm}$ while
%$N^{rs}_{mn}= N^{sr}_{nm}$.
where $n, m > 0$, $r',s'\in\{2,3\}$, $r,s\in\{1,2,3\}$, 
$\a\equiv\aout\ainI\ainII$, and 
\begin{eqnarray}
&&\hspace{-1.5cm}
C_{mn}=m \delta_{mn},
\qquad C_{(r)mn}=\omega_{m(r)} \delta_{mn},\qquad
U_{(r)}=C^{-1}(C_{(r)}-\mu \ar),\\
&&\hspace{-1.5cm}K = -\frac{1}{4}B^{\rm T}\Gamma^{-1}B,
\qquad \Gamma=\sum_{r=1}^{3}A^{(r)}U_{(r)}A^{(r)\rm T},\\
&&\hspace{-1.5cm}A^{(\inI)}_{mn}=
  \frac{2\sqrt{mn}}{\pi}
  \frac{y (-1)^{n+1}}{n^2-y^2m^2}\sin(\pi m y),\quad
 A^{(\inII)}_{mn}=
  \frac{2\sqrt{mn}}{\pi}
  \frac{(1-y) }{n^2-(1-y)^2m^2}\sin \pi m y,\\
&&\hspace{-1.5cm}A^{(\out)}_{mn}=\delta_{mn},\qquad
 \hspace{3.9cm}B_m=\frac{2}{\pi y(1-y)\aout}
 m^{-3/2}\sin(\pi m y),
 \end{eqnarray}
with $y=-\ainI/\aout$ and $1-y=-\ainII/\aout$.

When we compare string amplitudes 
on the both sides of bulk and boundary in the plane-wave limit,
the appropriate Fock basis is the one spanned by 
the oscillators defined with 
the exponential Fourier mode basis
which corresponds directly to BMN operators:
\begin{equation}
\a_0=a_0,\qquad
\a_{n}=\frac{1}{\sqrt{2}}(a_n-ia_{-n}),\quad
\a_{-n}=\frac{1}{\sqrt{2}}(a_n+ia_{-n}).
\end{equation}
The Neumann coefficients 
in terms of the exponential oscillator basis, $\widetilde N_{mn}^{rs}$, is 
\begin{eqnarray}
\sum_{m,n=-\infty}^{\infty}a_{m}^{(r)\dagger}N_{mn}^{rs}a_{n}^{(s)\dagger}
=\sum_{m,n=-\infty}^{\infty}
\a_{m}^{(r)\dagger}\widetilde N_{mn}^{rs}\a_{n}^{(s)\dagger},
\end{eqnarray}
where
\begin{eqnarray}
&&\hspace{-1cm}
\widetilde N_{00}^{rs}= N_{00}^{rs},\quad
\widetilde N_{0m}^{rs}= \widetilde N_{m0}^{rs}
=\widetilde N_{0-m}^{rs}= \widetilde N_{-m0}^{rs}=\frac{1}{\sqrt{2}}
N_{0m}^{rs},\\
&&\hspace{-1cm}\widetilde N_{mn}^{rs}=\widetilde N_{-m-n}^{rs}= 
\frac{1}{2}(N_{mn}^{rs}-N_{-m-n}^{rs}),\quad
\widetilde N_{m-n}^{rs}=\widetilde N_{-mn}^{rs}= 
\frac{1}{2}(N_{mn}^{rs}+N_{-m-n}^{rs}).
\end{eqnarray}
We present the large $\mu$ behavior 
of $\widetilde N_{mn}^{rs}$'s below in the Appendix \ref{large mu}.

\subsection{Prefactors}
The prefactor which was first
constructed for the overlap vertex
based on the SO(8) vacuum $\ket{0}$ \cite{SV1,SV2,Pan1,PS1}
can be reformulated for the overlap vertex
$\ket{\E}$ which is based on the genuine Fock vacuum \cite{Pan2,PS2}. 
The form of this prefactor is\footnote{
This definition differs form the one
in (3.28) of \cite{Pan2} by a factor $-{2\a}/{\a'}$.
Note also 
the difference of the total factor of $K^i$ between here and there.}
\begin{eqnarray}\label{P_SV}
&&\hspace{-2cm}|H_3\rangle_{SV}=P_{SV}\ket{\E},\\
&&\hspace{-2cm}P_{SV}=\frac{1}{2}\left[
 \left(K^{i}\widetilde K^{j}+\mu%\frac{\mu\a}{\a'}
\delta^{ij}\right)V_{ij}
-\left(K^{\mu}\widetilde K^{\nu}+\mu%\frac{\mu\a}{\a'}
\delta^{\mu\nu}\right)V_{\mu\nu}
\right.\nonumber\\
&&\left.-K^{\dot\a_1\a_1}\widetilde K^{\dot\a_2\a_2}
 S_{\a_1\a_2}(\YI)S^{*}_{\dot\a_1\dot\a_2}(\YII)
-\widetilde K^{\dot\a_1\a_1} K^{\dot\a_2\a_2}
 S^{*}_{\a_1\a_2}(\YI)S_{\dot\a_1\dot\a_2}(\YII)
\right]
\end{eqnarray}
where $K^I$, $\widetilde K^I$ $\,(I=i,\mu)$ and
$Y^{\a_1\a_2}$, $Z^{\dot\a_1\dot\a_2}$ are 
bosonic and fermionic constituents of the prefactor defined as
\begin{eqnarray}
&&\hspace{-1cm}K^{J}=\XI^{J}+\XII^{J},\qquad \widetilde K^J=\XI^J-\XII^J,\\
&&\hspace{-1cm}\XI=i\sqrt{-\frac{\a'}{\a}}(1-4\mu\a K)^{1/2}
 \sum_{r=1}^{3}\sum_{n=0}^{\infty}F_n^{(r)}a_{n}^{(r)\dagger},\\
&&\hspace{-1cm}\XII=-\sqrt{-\frac{\a'}{\a}}(1-4\mu\a K)^{1/2}
\sum_{r=1}^{3}\sum_{n=1}^{\infty}
  U_{n(r)}F_n^{(r)}a_{-n}^{(r)\dagger},\\
&&\hspace{-1cm}Y^{\a_1\a_2}=
\sqrt{-\frac{\a'}{\a}}(1-4\mu\a K)^{1/2}
\sum_{r=1}^3\sum_{n=0}^{\infty}G_n^{(r)}b_n^{(r)\a_1\a_2\dagger}, \\
&&\hspace{-1cm}Z^{\dot\a_1\dot\a_2}=
\sqrt{-\frac{\a'}{\a}}(1-4\mu\a K)^{1/2}
\sum_{r=1}^3\sum_{n=0}^{\infty}
G_n^{(r)}b_{-n}^{(r)\dot\a_1\dot\a_2\dagger},\label{YZ}
\end{eqnarray}
with
\begin{eqnarray}
&&\hspace{-1cm} 
F^{(\inI)}_0=-\sqrt{\frac{2}{\a'}}\sqrt{\mu\ainI}\ainII,\quad
F^{(\inII)}_0=\sqrt{\frac{2}{\a'}}\sqrt{\mu\ainII}\ainI,\quad
F^{(\out)}_0=0,\\
&&\hspace{-1cm} 
F^{(r)}_n=-\frac{\a}{\sqrt{\a'}}\,\frac{1}{1-4\mu\a K}\,
\frac{1}{\ar}
(U_{(r)}^{-1}C_{(r)}^{1/2}CN^r)_{n} \quad(n>0),\\
&&\hspace{-1cm} 
G^{(\inI)}_0=-\sqrt{\frac{1}{\a'}}\sqrt{\ainI}\ainII,\quad
G^{(\inII)}_0=\sqrt{\frac{1}{\a'}}\sqrt{\ainII}\ainI,\quad
G^{(\out)}_0=0,\\
&&\hspace{-1cm} 
G^{(r)}_n=-\frac{\a}{\sqrt{\a'}}\,\frac{1}{1-4\mu\a K}\,
\frac{e(\ar)}{\sqrt{|\ar|}}
(U_{(r)}^{-1/2}C_{(r)}^{1/2}C^{1/2}N^r)_{n} \quad(n>0).
\end{eqnarray}
The other quantities in the prefactor is defined as 
\begin{eqnarray}
&&\hspace{-1cm}V_{ij}\equiv \delta_{ij}
 \left[
 1+\frac{1}{12}(Y^4+Z^4)+\frac{1}{144}Y^4 Z^4
 \right] \nonumber\\
&&\hspace{3cm}-\frac{i}{2}\left[
 \YI{}_{ij}^2(1+Z^4)-Z^2_{ij}(1+\frac{1}{12}Y^4)
 \right]
+\frac{1}{4}(Y^2Z^2)_{ij},\\
&&\hspace{-1cm}V_{\mu\nu}\equiv\delta_{\mu\nu}
 \left[
 1-\frac{1}{12}(Y^4+Z^4)+\frac{1}{144}Y^4Z^4
 \right] \nonumber\\
&&\hspace{3cm}-\frac{i}{2}\left[
 \YI{}_{\mu\nu}^2(1-Z^4)-Z^2_{\mu\nu}(1-\frac{1}{12}Y^4)
 \right]
+\frac{1}{4}(Y^2Z^2)_{\mu\nu},\\
&&\hspace{-1cm}S(Y)\equiv Y+\frac{i}{3}Y^3,
\end{eqnarray}
with
\begin{eqnarray}
&& K^{\dot\a_r\b_r}\equiv K^i\sigma_i^{\dot\a_r\b_r},\quad
\widetilde K^{\dot\a_r\b_r}\equiv K^i\sigma_i^{\dot\a_r\b_r},\qquad
(r=1,2)\\
&&Y^2_{\a_1\b_1}\equiv Y_{\a_1\a_2}Y^{~~\a_2}_{\b_1},\quad
Y^2_{\a_2\b_2}\equiv Y_{\a_1\a_2}Y^{\a_1}_{~~~\b_2},\\
&&Y^3_{\a_1\b_2}\equiv Y^2_{\a_1\b_1}Y^{\b_1}_{~~~\b_2},\quad
Y^4\equiv Y^2_{\a_1\b_1}Y^{2\a_1\b_1},\\
&&Y^{2ij}\equiv Y^{2\a_1\b_1}\sigma^{ij}_{\a_1\b_1},\quad
Z^{2ij}\equiv Z^{2\dot\a_1\dot\b_1}\sigma^{ij}_{\dot\a_1\dot\b_1},\quad
(Y^2Z^2)^{ij}\equiv Y^{2k(i}Z^{2j)k},
\end{eqnarray}
We refer the reader to the ref.\cite{Pan2} for more details.

%We can easily see that for pure bosonic external states
%$P_{SV}$ takes the form 
%\begin{eqnarray}\label{effective PSV for bosons}
%P_{SV}=\frac{1}{2}
%\left[
%\sum_{i=5}^{8}
%\left(\left(\XI^i\right)^2-\left(\XII^{i}\right)^2\right)
%-\sum_{\mu=1}^{4}
%\left(\left(\XI^{\mu}\right)^2-\left(\XII^{\mu}\right)^2\right)
%\right]
%\end{eqnarray}
%while for pure fermionic external states considered in section (\ref{}) 
%takes the form
%\begin{eqnarray}
%P_{SV}=\frac{\mu}{3}Y^4.
%\end{eqnarray}
%Note the relative minus sign between  
%cos and sin modes as well as between scalar and vector directions
%in (\ref{effective PSV for bosons}).

On the other hand,
the interaction vertex presented in \cite{DPPRT}, which is of the form
\begin{eqnarray}
|H_3\rangle_{D}= \sum_{r=1}^{3}\sum_{m=-\infty}^{\infty}
\frac{\omega_m^{(r)}}{\ar}
\left(\sum_{I=1}^{8}
a_m^{(r)I\dagger}a_{m}^{(r)I}
+\sum_{a=1}^{8}
b_m^{(r)a\dagger}b_{m}^{(r)a}
\right)\ket{\E}
\end{eqnarray} can be, using the factorization formula,  written as
\begin{eqnarray} \label{PD}
&&\hspace{-1cm}|H_3\rangle_{D}=P_{D}\ket{\E},\\
&&\hspace{-1cm}P_{D}=\frac{1}{4}\left(K^2+\widetilde K^2\right)  
-Y^{\a_1\a_2}\widetilde Y_{\a_1\a_2}
-Z^{\dot\a_1\dot\a_2} \widetilde Z_{\dot\a_1\dot\a_2},
\end{eqnarray}
where
\begin{eqnarray}
\widetilde Y^{\a_1\a_2}
=\sum_{r=1}^3\frac{n}{\ar}G_n^{(r)}b_n^{(r)\a_1\a_2\dagger},\qquad
\widetilde Z^{\dot\a_1\dot\a_2}
=\sum_{r=1}^3\frac{n}{\ar}G_n^{(r)}b_{-n}^{(r)\a_1\a_2\dagger}.
\end{eqnarray}
For the derivation of (\ref{PD}), see  the Appendix \ref{Factorization
formula} below. 
%For pure bosonic external states $P_D$ reduces to
%\begin{eqnarray}\label{effective PSV for bosons}
%P_{D}=\frac{1}{2}
%\left[
%\sum_{i=5}^{8}
%\left(\left(\XI^i\right)^2+\left(\XII^{i}\right)^2\right)
%+\sum_{\mu=1}^{4}
%\left(\left(\XI^{\mu}\right)^2+\left(\XII^{\mu}\right)^2\right)
%\right].
%\end{eqnarray}

\subsection{Large $\mu$ behavior {\rm\cite{He et. al.}}}\label{large mu}
The large $\mu$ behavior of $\widetilde N_{mn}^{rs}$
is given,
for $(m,n)\neq(0,0)$, by
\begin{eqnarray}\label{Neumann for large mu}
&&\hspace{-1cm}\widetilde N_{mn}^{\inI\inI}=\frac{(-1)^{m+n}}{4\pi\mu |\aout|y },
\quad
 \widetilde N_{mn}^{\inI\inII}=\frac{(-1)^{m+1}}
{4\pi\mu|\aout| \sqrt{y(1-y)}}\\
&&\hspace{-1cm}\widetilde N_{mn}^{\inII\inII}=\frac{1}{4\pi\mu |\aout|(1-y)},
\quad
 \widetilde N_{mn}^{\out\out}=\frac{(-1)^{m+n+1}\sin(\pi m y)\sin(\pi n y)}
{\pi \mu|\aout|}\\
&&\hspace{-1cm} \widetilde N_{mn}^{\inI\out}
=\frac{(-1)^{m+n+1}\sin(\pi n y)}
{\pi \sqrt{y}(n-m/y)},\quad
\widetilde N_{mn}^{\inII\out}=\frac{(-1)^{n}\sin(\pi n y)}
{\pi \sqrt{1-y}(n-m/(1-y))}
\end{eqnarray}
and, for $m=n=0$, by
\begin{eqnarray}
&&\hspace{-1cm}\widetilde N_{00}^{\out\out}=0,\quad
 \widetilde N_{00}^{\out\inI}=-\sqrt{y},\quad
 \widetilde N_{00}^{\out\inII}=-\sqrt{1-y},\\
&&\hspace{-1cm}\widetilde N_{00}^{\inI\inII}
=-\frac{1}{4\pi\mu|\aout| \sqrt{y(1-y)}},
\quad
\widetilde N_{00}^{\inI\inI}=\frac{1}{4\pi\mu|\aout| y},
\quad
 \widetilde N_{00}^{\inII\inII}=\frac{1}{4\pi\mu|\aout| (1-y)}
\end{eqnarray}

When we compute string amplitudes for fermions, the large $\mu$ behavior 
of $F_n^{(r)}$, $G_n^{(r)}$, $U_m^{(r)}$ and $1-4\mu\a K$ are also
useful\footnote{
Note that the definition of the Neumann vector $N^{r}_{n}$ here,
with which $F_n^{(r)}$ and $G_n^{(r)}$  are
 defined, differs by 
$C_{(r)}^{-1/2}U_{r}$ from that of the ref.\cite{He et. al.}.
}:
\begin{eqnarray}\label{asymptotic forms}
 &&\hspace{-1.5cm}F_{n}^{(\inI)}=(-1)^{n+1}\frac{2|\aout|}{\sqrt{\a'}}
\sqrt{\mu |\aout|y}\,(1-y),\qquad
 F_{n}^{(\inII)}=\frac{2|\aout|}{\sqrt{\a'}}\sqrt{\mu|\aout|(1-y)}\,y,\nonumber \\
 &&\hspace{-1.5cm}F_{n}^{(\out)}=(-1)^{n+1} \frac{2|\aout|}{\sqrt{\a'}}
\frac{y(1-y)}{\sqrt{\mu|\aout|}}\,\,n \sin(\pi n y), \nonumber \\
&&\hspace{-1.5cm}\overline{G}^{(\inI)}_n=\frac{(-1)^{n+1}}{\sqrt{2\pi \mu |\aout| y}},\quad
\overline{G}^{(\inII)}_n=\frac{1}{\sqrt{2\pi \mu |\aout| (1-y)}},\quad
\overline{G}^{(\out)}_n=\frac{(-1)^{n+1}\sqrt{2} \sin(\pi n y)}{\sqrt{\pi \mu |\aout|}}\\
&&\hspace{-1.5cm}U_n^{(\inI)}=\frac{n}{2 \mu |\aout| y},\qquad
U_n^{(\inII)}=\frac{n}{2 \mu |\aout| (1-y)},\qquad
U_n^{(\out)}=\frac{2\mu |\aout| }{n}, \nonumber \\
&&\hspace{-1.5cm} 1-4\mu \a K = \frac{1}{4\pi \mu |\aout| y(1-y)},  \nonumber 
\end{eqnarray}
where we have defined 
\EQ
\overline{G}^{(r)}_n=\sqrt{-{\alpha'\over \alpha}}
(1-4\mu \alpha K)^{1/2}G^{(r)}_n. 
\EN

\subsection{Factorization formula}\label{Factorization formula}

We first prove the formula
\begin{equation}\label{formula1}
 \sum_{r=1}^3 \sum_{n=0}^{\infty}\frac{\omega_n^{(r)}}{\ar}
a_n^{(r)\dagger}a_n^{(r)}\ket{\E}
=\frac{1}{2}\XI^2\ket{\E},
%\quad
% \sum_{r=1}^3 \sum_{n=1}^{\infty}\frac{\omega_{n}^{(r)}}{\ar}
%a_{-n}^{(r)\dagger}a_{-n}^{(r)}\ket{\E}
%=\frac{1}{2}\XII^2\ket{\E},
\end{equation}
using the factorization formula obtained in \cite{Schwarz, Pan1}.
Operating the annihilation operator $a_n$ on the vertex $\ket{\E}$,
the left hand side of (\ref{formula1}) can be written as
\begin{eqnarray}
&&\hspace{-2cm}\sum_{r=1}^3 \sum_{n=0}^{\infty}\frac{\omega_n^{(r)}}{\ar}
a_n^{(r)\dagger}a_n^{(r)}\ket{\E}
=
 -\sum_{r,s=1}^3 \left(
\frac{\omega_0^{(r)}}{\ar}N^{rs}_{00}
a_0^{(r)\dagger}a_0^{(s)\dagger}
+\sum_{n=1}^{\infty}
\frac{\omega_0^{(r)}}{\ar}N^{rs}_{0n}
a_0^{(r)\dagger}a_n^{(s)\dagger}\right.\nonumber\\
&&\hspace{+2.5cm}\left.+\sum_{n=1}^{\infty}
\frac{\omega_n^{(r)}}{\ar}N^{rs}_{n0}
a_n^{(r)\dagger}a_0^{(s)\dagger}
+\sum_{n,m=1}^{\infty}
\frac{\omega_m^{(r)}}{\ar}N^{rs}_{mn}
a_m^{(r)\dagger}a_n^{(s)\dagger}
\right)\ket{\E}.
\end{eqnarray}
By the definition of Neumann matrices, the first term
in the right hand side becomes
\begin{eqnarray}
\sum_{r,s=1}^{3}\frac{\omega_0^{(r)}}{\ar}N^{rs}_{00}
a_0^{(r)\dagger}a_0^{(s)\dagger}=\frac{1}{2}X_0^2,
\end{eqnarray}
and the sum of the second and the third terms reduces to
\begin{eqnarray}
&&\sum_{r,s=1}^3\sum_{n=1}^{\infty}
\frac{\omega_0^{(r)}}{\ar}N^{rs}_{0n}
a_0^{(r)\dagger}a_n^{(s)\dagger}
+\sum_{r,s=1}^3\sum_{n=1}^{\infty}
\frac{\omega_n^{(r)}}{\ar}N^{rs}_{n0}
a_n^{(r)\dagger}a_0^{(s)\dagger}
=X_0X_+,
\end{eqnarray}
where $X_0$ and $X_+$ is defined as
zero-mode and positive-mode parts of $\XI$, 
such as
\begin{eqnarray}
&&\hspace{-1cm}\XI=i
\sqrt{-\frac{\a'}{\a}}(1-4\mu\a K)^{1/2}
 \left(\sum_{r=1}^{3}F_0^{(r)}a_{0}^{(r)\dagger}
+ \sum_{r=1}^{3}\sum_{n=1}^{\infty}F_n^{(r)}a_{n}^{(r)\dagger}\right)
\equiv X_0+X_+.
\end{eqnarray}
Using the property of the Neumann matrix, $N_{nm}^{rs}=N_{mn}^{sr}$, 
and the factorization formula \cite{Schwarz,Pan1},
\begin{eqnarray}
 N^{rs}_{nm}=-\frac{\a}{1-4\mu\a K}
\frac{1}{\ar\omega_m^{(s)}+\as\omega_n^{(r)}}
(U_{(r)}^{-1}C_{(r)}^{1/2}CN^r)_n(U_{(r)}^{-1}C_{(s)}^{1/2}CN^s)_m,
\end{eqnarray}
the fourth term can be written as
\begin{eqnarray}
\sum_{r,s=1}^3\sum_{n,m=1}^{\infty}
\frac{\omega_n^{(r)}}{\ar}N^{rs}_{nm}
a_n^{(r)\dagger}a_m^{(s)\dagger}
&=& \frac{1}{2}
\sum_{r,s=1}^3\sum_{n,m=1}^{\infty}
\left(\frac{\omega_n^{(r)}}{\ar}N^{rs}_{nm}
+N^{rs}_{nm}\frac{\omega_m^{(s)}}{\as}\right)
a_n^{(r)\dagger}a_m^{(s)\dagger}\nonumber\\
&=&\frac{1}{2} X_+^2 \, .
\end{eqnarray}
Combining  all the results above, we obtain
the formula (\ref{formula1}).

Noticing that the Neumann coefficient with negative Fourier modes
is given by
\begin{eqnarray}
 N_{-m-n}^{rs}=-(U_{(r)}N^{rs}U_{(s)})_{mn},\quad (m,n > 0)
\end{eqnarray}
and the definition of $\XII$, which has 
the extra $iU_{(r)}$ factor compared with $\XI$, 
we can easily see that 
the similar formula for negative modes,
\begin{eqnarray}
 \sum_{r=1}^3 \sum_{n=1}^{\infty}\frac{\omega_{n}^{(r)}}{\ar}
a_{-n}^{(r)\dagger}a_{-n}^{(r)}\ket{\E}
=\frac{1}{2}\XII^2\ket{\E},
\end{eqnarray}
can also hold.

With the definition of the fermionic Neumann coefficient $Q_{mn}^{rs}$,
it is not difficult to prove the formula 
\begin{eqnarray}
&&\hspace{-1cm}\sum_{r=1}^3\sum_{n\in Z}\frac{\omega_n^{(r)}}{\ar}
\left(
b_{n\a_1\a_2}^{(r)\dagger}b_{n}^{(r)\a_1\a_2}
+b_{n\dot\a_1\dot\a_2}^{(r)\dagger}b_{n}^{(r)\dot\a_1\dot\a_2}
\right)\ket{\E_b}\nonumber\\
&&\hspace{5cm}=-\left(Y_{\a_1\a_2}\widetilde Y^{\a_1\a_2}
  + Z_{\dot\a_1\dot\a_2}\widetilde Z^{\dot\a_1\dot\a_2}\right)
\ket{\E_b}
\end{eqnarray}
in the same manner as the bosonic case.

%\newpage

\end{document}